\documentclass[%
superscriptaddress,
footinbib,
bibnotes,
 amsmath,amssymb,
twocolumn,
floatfix,
]{revtex4}

\usepackage{graphicx}
\usepackage{bm}
\usepackage{gensymb}
\usepackage{color}
\usepackage{lipsum}
\usepackage{siunitx}

\begin{document}

\title{Simulations of LiCl--Water: The Effects of Concentration and Supercooling
}

\author{Philip H. Handle}
\affiliation{%
 Institute of Physical Chemistry,  University of Innsbruck, Innsbruck, Austria
}
\affiliation{%
Institute of General, Inorganic and Theoretical Chemistry, University of Innsbruck, Innsbruck, Austria
}

\date{\today}
\begin{abstract}
Aqueous solutions of LiCl are probably the most studied electrolyte solutions related to the complexity of liquid water at low temperatures.
Despite the large amount of available experimental data hardly any computational studies were performed on LiCl solutions in this context.
In this study, we present molecular dynamics simulations of LiCl--water at ambient and
supercooled conditions spanning a large concentration range.
The molecular insight gained provides information on how the presence of the ions impacts the hydrogen bond network.
It is found that this influence changes appreciably  when supercooled states are considered.
While the local structure of water molecules beyond the first hydration shells barely changes with concentration  at room temperature, a change is found for those molecules at low temperature.
Additionally, we scrutinize the possibility of a phase separation in this system as indicated by several experimental studies.
Our analyses do not show signs of such a phase separation at \SI{240}{\kelvin}, but are consistent with a possible separation at even lower temperatures.

\end{abstract}
             
\maketitle

\section{Introduction}

Aqueous solutions of simple electrolytes are ubiquitous on earth~\cite{franks2000water, finney2015water} and play crucial roles in maintaining life as we know it~\cite{kahler20-scirep}.
Relevant for the present work is the study of electrolyte solutions as a means of elucidating pure water's behaviour at low temperatures~\cite{angell02-crev,bachler19-pccp}.
As water is supercooled it shows an increasingly complex behaviour~\cite{debenedetti03-jpcm} for whose explanation several scenarios have been discussed~\cite{mishima98-nature, angell04-arpc, angell08-science,gallo16-crev,handle2017supercooled,anisimov18-prx}.
A particularly intriguing scenario proposes a low temperature liquid-liquid phase transition (LLPT) ending in a liquid-liquid critical point (LLCP).
This scenario originated from computer simulations~\cite{poole92-nature} of the ST2 water model~\cite{stillinger74-jcp} suggesting the appearance of a low-density (LDL) and a high-density liquid (HDL) in deeply supercooled water.
Over the years this scenario gained support from further simulation studies employing a variety of molecular models of water~\cite{poole05-jpcm,palmer14-nature,paschek08-cpc,corradini10-jcp,abascal10-jcp,wikfeldt2011spatially, sumi13-rscadv, yagasaki14-pre, biddle17-jcp,handle18-jcp,palmer18-crev}.
The LLCP scenario is also consistent with recent experimental results obtained for supercooled water at positive~\cite{kim17-science} and negative pressures~\cite{holten2017compressibility}.
It furthermore provides a rather compelling explanation for the apparent solid polyamorphism documented in several experimental studies~\cite{mishima84-nature,mishima85-nature,mishima94-jcp,loerting01-pccp, klotz05-prl, winkel08-jcp, loerting11-pccp, handle2018experimental1, handle2018experimental2, mariedahl2018x}.
Yet, decisive experimental results are still missing, since the $T$-$P$ region in which the LLPT is predicted has so far not been accessible to experiments.
Therefore, the experimental evidence for this scenario relies on consistency arguments, extrapolations, or out-of-equilibrium~\footnote{Out-of-equilibrium here specifically denotes not equilibrated systems. That is, systems in metastable equilibrium (e.g., supercooled water) are \emph{not} considered to be out-of-equilibrium.} experiments.

One such indirect approach is the study of aqueous solutions~\cite{angell02-crev,bachler19-pccp}.
In this realm LiCl is the most prominent electrolyte used in experiments~\cite{angell68-jcp, angell70-jcp, kanno87-jcp, suzuki00-prl,suzuki02-jcp, mishima04-jcp, mishima05-jcp, mishima07-jcp, mishima11-jpcb, suzuki11-jcp, bove11-prl, winkel11-jcp, bove13-jcp, suzuki13-jcp,  ruiz14-pccp, ruiz18-pccp}.
Already the first studies by Angell and Sare~\cite{angell68-jcp, angell70-jcp} suggest that aqueous LiCl solutions might separate into a water-rich and a salt-rich liquid before vitrifying.
Similar findings were reported by Kanno~\cite{kanno87-jcp} and Suzuki and Mishima~\cite{suzuki00-prl}.
Recent transient grating experiments furthermore suggest that LiCl solutions with a mole fraction of LiCl $x_\text{LiCl}$ lower than \SI{14.3}{\percent} become heterogeneous at the nanoscale below \SI{190}{\kelvin}~\cite{bove13-jcp}.
It was proposed that the suspected phase separation occurs due to an immiscibility dome in the $T$-$P$-$x_\text{LiCl}$ diagram, which is continuously connected to the hypothetical LLPT of pure water~\cite{suzuki13-jcp}.
However, the exact nature of the separation is not clear.
While some studies suggest the water-rich liquid to be of low density~\cite{suzuki00-prl, suzuki13-jcp} other studies find that a high-density water-rich liquid separates from the solution~\cite{kanno87-jcp,suzuki02-jcp}.
Currently, none of these views can be excluded and it might very well be that both are correct.
In the latter case the different separations occur at different $T$-$P$-$x_\text{LiCl}$ conditions entailing a more complex behaviour of supercooled LiCl solutions~\cite{bachler19-pccp}.
Moreover, it is not clear how those separations are connected to a possible LLPT.

Surprisingly, there are barely any computational studies of LiCl--H$_2$O in this context.
One work~\cite{bove11-prl} consists of a combined experimental and computational study of a vitrified LiCl solution of $x_\text{LiCl}=\SI{14.3}{\percent}$ subjected to pressure changes.
The performed molecular dynamics (MD) and \emph{ab initio} MD simulations indicate that the coordination number of the Li ion increases by one (either a water or a chloride) when the system is compressed from ambient $P$ to \SI{3}{\giga\pascal}, giving a microscopic explanation of the experimentally observed densification.
Another study~\cite{camisasca18-jcp} investigated LiCl solutions of the same concentration in the temperature range from \SIrange{200}{300}{\kelvin} utilising MD calculations.
Here the simulations revealed that the dynamical cross-over, a feature that has been linked to the LLCP scenario~\cite{xu2005relation}, is absent in the studied temperature range.
A third study~\cite{le2011nanophase} made use of the coarse-grained mW water model~\cite{molinero2009water} in conjunction with a generic solute S~\cite{le2011nanophase}.
In this study solutions below $x_\text{S}=\SI{20}{\percent}$ were found to form a nano-segregated glass upon cooling, where the water-rich phase is in a low-density state.
Interestingly, it was also shown that the generic solute S bears some resemblance to LiCl.

Motivated by the scarcity of numerical data we provide novel simulations of supercooled LiCl solutions.
Building on the work of Aragones et al.~\cite{aragones14-jpcb}, who studied aqueous LiCl solutions at ambient conditions, we both extend the concentration range and explore supercooled states.
We also consider very dilute systems probing the connection to phenomena present in pure supercooled water.

\section{Simulation Details}

For this study four different force fields to describe LiCl--H$_2$O were used.
The first consists of the TIP4P-Ew water model~\cite{horn04-jcp} together with the Joung-Cheatham LiCl parameters~\cite{joung08-jpcb}, which were specifically designed for the use with TIP4P-Ew.
For the second description the parameters of the model were slightly modified following the suggestion of Aragones et al.~\cite{aragones14-jpcb}.
That is, the combination rules governing the Lennard-Jones cross-interaction between lithium and chloride were altered.
This modification was shown to enhance the description of the cation-anion structure making it comparable to experiments~\cite{aragones14-jpcb}.
The relevant equations are:
\begin{align}
 \sigma_{ij}=\eta\frac{\sigma_i+\sigma_j}{2}\label{eq:sigma},\\
 \epsilon_{ij}=\chi\sqrt{\epsilon_i\epsilon_j}\label{eq:epsilon}.
\end{align}
Here, $i$ and $j$ represent the two interaction sites considered, which can be Li, Cl, or O.
Please note, that the charges are omitted in the notation throughout the manuscript and that only the oxygen is a Lennard-Jones interaction site in TIP4P-type water models.
The parameters $\epsilon$ and $\sigma$ control the potential depth and diameter of the particles, respectively.

For the first parameter set all Lennard-Jones cross-interactions are calculated using $\eta=\chi=1$ corresponding to the standard Lorentz-Berthelot (LB) combination rules.
The modifications introduced by Aragones et al.~\cite{aragones14-jpcb} are $\eta=0.932$ and $\chi=1.88$ for the Li--Cl cross-interaction and $\eta=\chi=1$ in all other cases.
To refer to these \emph{modified} Lorentz-Berthelot combination rules, MLB will be used as a shorthand.
Both variants discussed so far are used in conjunction with the TIP4P-Ew water model and thus will be abbreviated as Ew-LB and Ew-MLB, respectively.

The third and fourth force fields used are based on the TIP4P/2005 water model~\cite{abascal05-jcp}.
The use of TIP4P/2005 is desired, since it is considered to be a very accurate rigid water model~\cite{vega11-pccp}.
It also displays polyamorphism~\cite{wong15-jcp,handle19-jcp} and it likely exhibits an LLPT~\cite{abascal10-jcp,wikfeldt2011spatially, sumi13-rscadv, yagasaki14-pre, biddle17-jcp,handle18-jcp}.
Conveniently, it was shown that the Joung-Cheatham LiCl parameters are transferable to TIP4P/2005~\cite{aragones14-jpcb, moucka12-jpcb}.
For this combination Aragones et al.~\cite{aragones14-jpcb} again introduced an MLB variant ($\eta=0.934$, $\chi=1.88$), which we will also use.
These two descriptions of LiCl--H$_2$O will be abbreviated as 2005-LB and 2005-MLB, respectively.

The systems studied contain 1000 water molecules and the amount of LiCl is given by the desired concentration.
In total eleven different concentrations between $x_\text{LiCl}=\SI{0.1}{\percent}$ and $x_\text{LiCl}=\SI{33.3}{\percent}$  were investigated (see Tab.~S-I in the Electronic Supplementary Information -- ESI --  for a complete listing).
All MD simulations were performed in the $N\!PT$ ensemble.
The pressure was set to \SI{1}{\bar} in all cases and two different temperatures (\SI{240}{\kelvin} and \SI{298}{\kelvin}) were considered.
When compared to experimental data the systems at \SI{240}{\kelvin} are supercooled with respect to ice formation for $x_\text{LiCl}<\SI{8.3}{\percent}$ and supercooled with respect to hydrate formation for $x_\text{LiCl}>\SI{18.1}{\percent}$~\cite{monnin02-jced}.
For $\SI{8.3}{\percent}<x_\text{LiCl}<\SI{18.1}{\percent}$ the solution is still in the stable domain, which reaches its minimum at $T\approx\SI{199}{\kelvin}$ for the eutectic concentration $x_\text{LiCl}\approx\SI{12.5}{\percent}$~\cite{monnin02-jced}.
At \SI{298}{\kelvin} the systems were simulated with all four force fields (i.e., Ew-LB, Ew-MLB, 2005-LB, and 2005-MLB),
at \SI{240}{\kelvin}  only Ew-MLB and 2005-MLB were used for all systems while their LB counterparts were studied for $x_\text{LiCl}=\SI{2.4}{\percent}$ and $x=\SI{14.3}{\percent}$ to probe the influence of the combination rule modification.
For comparison also pure water systems (TIP4P/2005 and TIP4P-Ew) were studied at both temperatures.

The simulations were performed utilising GROMACS 5.1.4~\cite{vanderspoel05-jcc}.
The cubic simulation boxes were treated with periodic boundary conditions and the equations of motions were integrated using the leap-frog algorithm with a time step of \SI{2}{\femto\second}.
Temperature and pressure are controlled using a Nos\'{e}-Hoover thermostat~\cite{nose84-mp, hoover85-pra} and a  Parinello-Rahman barostat~\cite{parrinello81-jap}, respectively.
The Coulombic interactions were calculated using the particle mesh Ewald method~\cite{essmann95-jcp} with a Fourier spacing of \SI{0.1}{\nano\meter}.
An identical cut-off of \SI{0.95}{\nano\meter} was used for both the Lennard-Jones and the real space Coulomb interactions.
Lennard-Jones interactions beyond the cut-off distance were included assuming a uniform fluid density.
The bond constraints were maintained using the LINCS (Linear Constraint Solver) algorithm~\cite{hess08-jctc} of 6$\text{th}$ order with one iteration to correct for rotational lengthening.
Most systems were simulated for \SI{500}{\nano\second}, where the first \SI{10}{\nano\second} were not used for analysis.
At \SI{240}{\kelvin} the two highest concentrations ($x_\text{LiCl}=\SIlist{25.0;33.3}{\percent}$) were simulated for up to \SI{3}{\micro\second}.

\section{Results}

The obtained numerical data for LiCl--H$_2$O is  analysed in a variety of ways.
We evaluate thermodynamic quantities (Section~\ref{ssec:thd}), diffusion coefficients (Section~\ref{ssec:diff}), and studied the structure using radial distribution functions (Section~\ref{ssec:rdf}).
Furthermore, we evaluated the coarse-grained density field as introduced by Testard et al.~\cite{testard2014intermittent} (Section~\ref{ssec:cg}), the structural order parameter introduced by Russo and Tanaka~\cite{russo14-natcomm} (Section~\ref{ssec:rt}), and we analyse the overlap of different first hydration shells (Section~\ref{ssec:ohs}).
The latter three methods were specifically applied to look for signs of a phase separation as indicated by experimental studies~\cite{angell68-jcp, angell70-jcp,kanno87-jcp,suzuki00-prl, bove13-jcp, suzuki13-jcp}.
We note that also other order parameters were successfully applied to study pure water~\cite{tanaka2019revealing}.
However, quantities like the tetrahedral order parameter $q$~\cite{errington2001relationship}, which relies on the four nearest neighbours, or $g_5(r)$~\cite{saika2000computer, cuthbertson11-prl}, which relies on the fifth nearest neighbour, were not considered, because ambiguities would arise once the water molecules are close to ions.
On similar grounds we refrain from analysing partial RDFs based on low-density water molecules only or on high-density water molecules only~\cite{martelli19-jcp}.

In the following we will only show data obtained using the MLB combination rules (i.e., Ew-MLB and 2005-MLB).
Differences to calculations using the LB combination rules are discussed in the ESI.

\begin{figure}
 \centering
\includegraphics[width=0.49\textwidth]{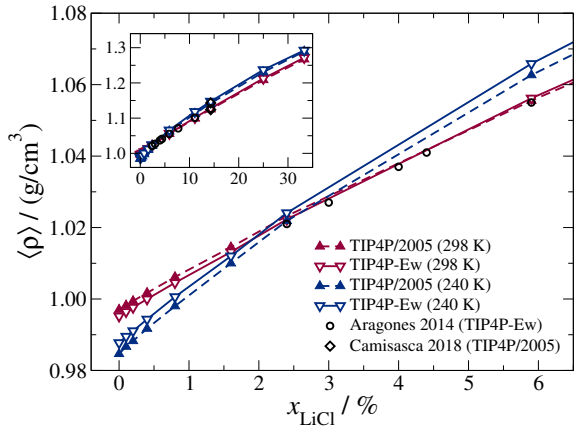}
 \caption{Density $\rho$ as a function of LiCl mole fraction $x_\text{LiCl}$. The filled upward triangles show the data for the TIP4P/2005 water model, the open downward triangles the data for the TIP4P-Ew water model.
 The red data correspond to simulations at \SI{298}{\kelvin} the blue data to simulations at \SI{240}{\kelvin}.
 Black open circles indicate the data for the TIP4P-Ew water model as reported by Aragones et al.~\cite{aragones14-jpcb} and black diamonds indicate the data for the TIP4P/2005 water model as reported by Camisasca et al.~\cite{camisasca18-jcp, camisasca20erratum}.
 The main figure shows only the low concentration part, while the inset shows the whole studied concentration range.
 }
 \label{fig:thd}
\end{figure}

\subsection{Thermodynamics}
\label{ssec:thd}

The thermodynamical parameters evaluated here are the averages of the potential energy $\left<U\right>$ and density $\left<\rho\right>$.
The complete data set is collected in the ESI (Tab.~S-I) for reference.
For all studied force fields an increase in LiCl content decreases $\left<U\right>$.
For pure TIP4P/2005 at \SI{298}{\kelvin} \SI{-47.87}{\kilo\joule\per\mole} is found, which changes to $\approx\SI{-245}{\kilo\joule\per\mole}$ for $x_\text{LiCl}=\SI{33.3}{\percent}$.
If the TIP4P-Ew water model is used the potential energy is slightly lower at low concentrations, but becomes highly similar to the values obtained for TIP4P/2005 at high concentrations.
At \SI{240}{\kelvin} $\left<U\right>$ is lower than at \SI{298}{\kelvin} over the whole concentration range and it also decreases as  $x_\text{LiCl}$ is increased.

The behaviour of $\left<\rho\right>$ is shown in Fig.~\ref{fig:thd}.
At \SI{298}{\kelvin} the density increases as $x_\text{LiCl}$ increases.
TIP4P-Ew produces slightly lower densities at low concentrations, but slightly higher densities at high concentrations when compared to TIP4P/2005.
As the temperature is decreased to \SI{240}{\kelvin} TIP4P/2005 shows lower densities than TIP4P-Ew over the whole concentration range.
In addition, the temperature decrease gives rise to an intriguing feature.
At low concentrations the density decreases as a consequence of temperature change, while it increases at high concentrations.
The cross-over takes place between $\SI{2.4}{\percent}$ and $\SI{5.9}{\percent}$ for TIP4P/2005 and between $\SI{1.6}{\percent}$ and $\SI{2.4}{\percent}$ for TIP4P-Ew (see Fig.~\ref{fig:thd}).
This merits a closer look.
For pure water the density increase upon cooling is the result of an increasing population of tetrahedrally ordered water molecules~\cite{handle19-jcp}.
This phenomenology is still present at low concentrations.
However, at higher LiCl concentrations the density increases upon temperature decrease, which is similar to the behaviour of regular systems.
This indicates that the ions disrupt the hydrogen bonds (HBs) enough to prevent the formation of a tetrahedral low-density network.

\subsection{Diffusion Coefficients}
\label{ssec:diff}

\begin{figure}
 \centering
    \includegraphics[width=0.48\textwidth]{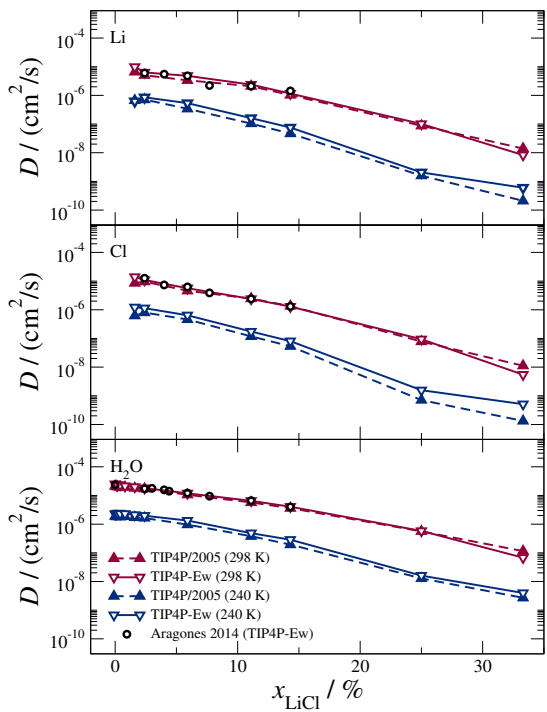}
 \caption{Diffusion coefficients $D$ for Li (top panel), Cl (middle panel) and H$_2$O (bottom panel) as a function of LiCl mole fraction $x_\text{LiCl}$. The filled upward triangles show the data for the TIP4P/2005 water model, the open downward triangles the data for the TIP4P-Ew water model.
 The red data correspond to simulations at \SI{298}{\kelvin} the blue data to simulations at \SI{240}{\kelvin}.
 For the ions only $x_\text{LiCl}\geq\SI{1.6}{\percent}$ allow for statistically significant data.
 The black open circles indicate the data  for the TIP4P-Ew water model as reported by Aragones et al.~\cite{aragones14-jpcb}.
 }
 \label{fig:diff}
\end{figure}

The diffusion coefficient of all three species, i.e., Li, Cl, and water, are shown in Fig.~\ref{fig:diff}.
In all cases an increase in the LiCl mole fraction leads to a decrease in diffusivity.
While the two ionic species show a drop of about three orders of magnitude over the whole concentration range, the water diffusivity decreases by only two orders of magnitude.
The obtained data for \SI{298}{\kelvin} agree well with previously obtained diffusion coefficients for TIP4P-Ew~\cite{aragones14-jpcb}.
As the temperature is changed from \SI{298}{\kelvin} to \SI{240}{\kelvin} the diffusion coefficients of all species are shifted down by approximately one order of magnitude.
This shift is slightly larger for higher concentrations.
We also note that the two different water models studied behave very similarly.
Only at the highest concentrations the results for the two water models differ.
This difference is more pronounced as the temperature is lowered.

\subsection{Structure}
\label{ssec:rdf}

\begin{figure}[b!]
 \centering
\includegraphics[width=0.48\textwidth]{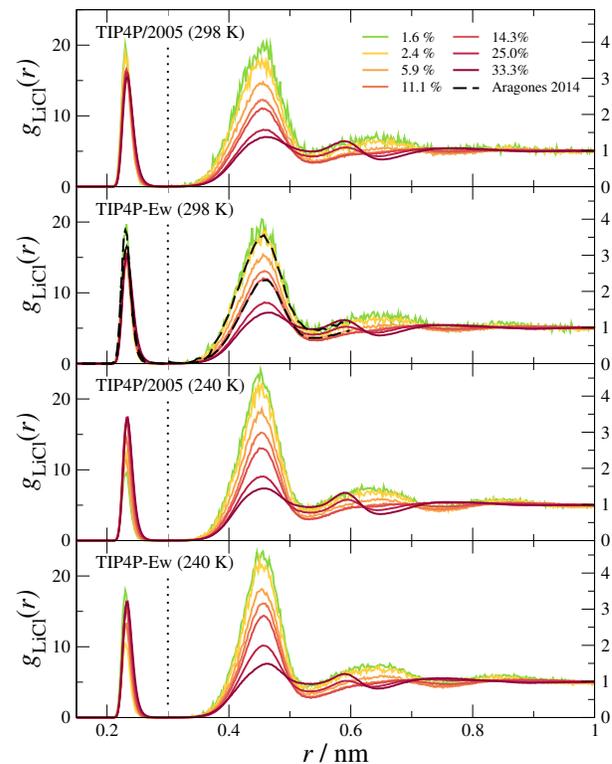}
 \caption{Li-Cl RDFs  as a function of concentration. The two upper panels show the behaviour at \SI{298}{\kelvin} for TIP4P/2005 and TIP4P-Ew, respectively, whereas the bottom two panels show the behaviour at \SI{240}{\kelvin}. The different colours represent the different concentrations as indicated in the legend. 
 Note that the main peak is plotted on a different scale (left axis) than the rest of the RDF (right axis).
 The dashed black lines indicate the RDFs for $x_\text{LiCl}=0.024$ and $x_\text{LiCl}=0.143$ as reported by Aragones et al.~\cite{aragones14-jpcb}.}
 \label{fig:licl-rdf}
\end{figure}

\begin{figure*}
 \centering
\includegraphics[width=0.48\textwidth]{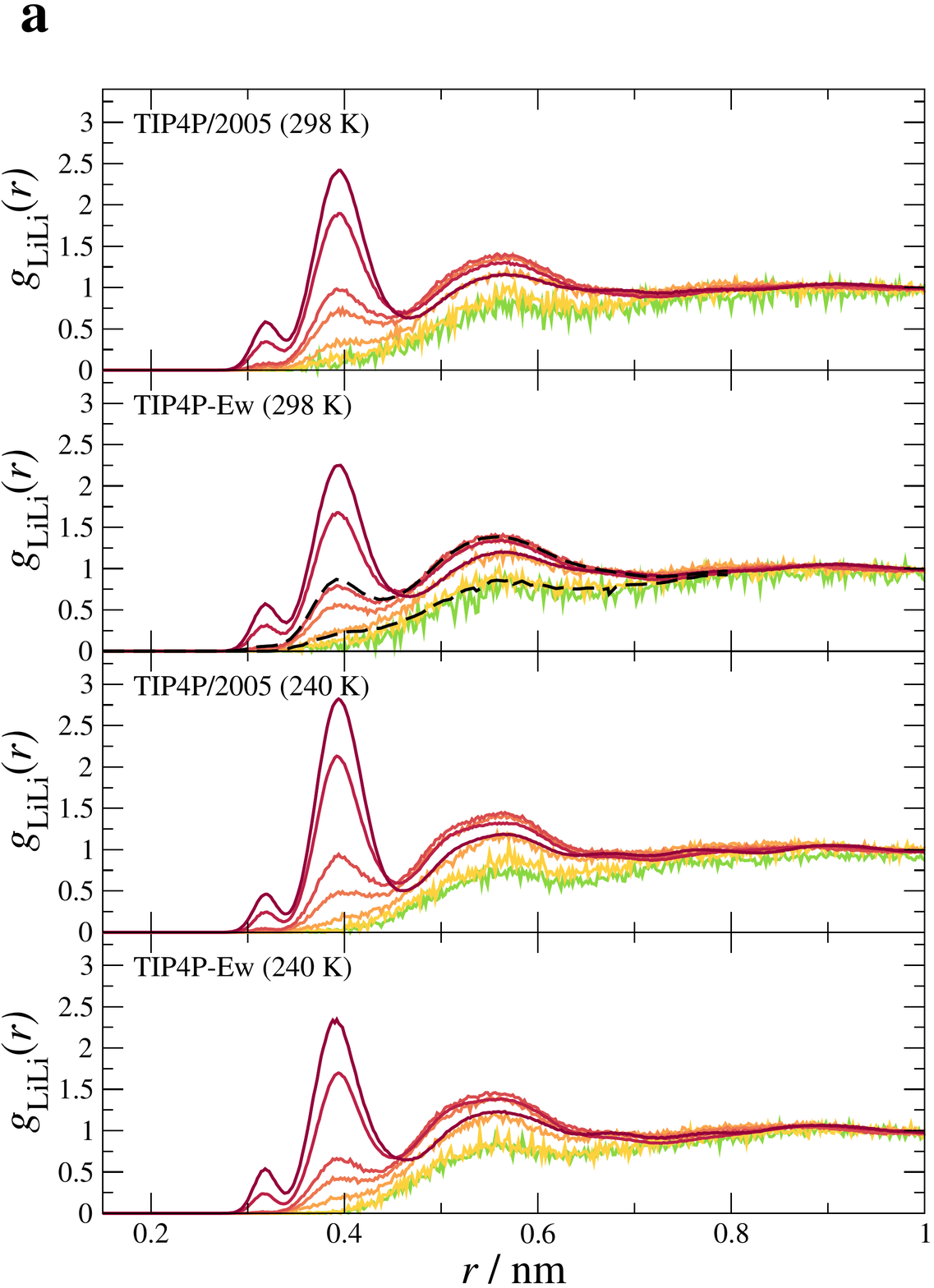}
\includegraphics[width=0.48\textwidth]{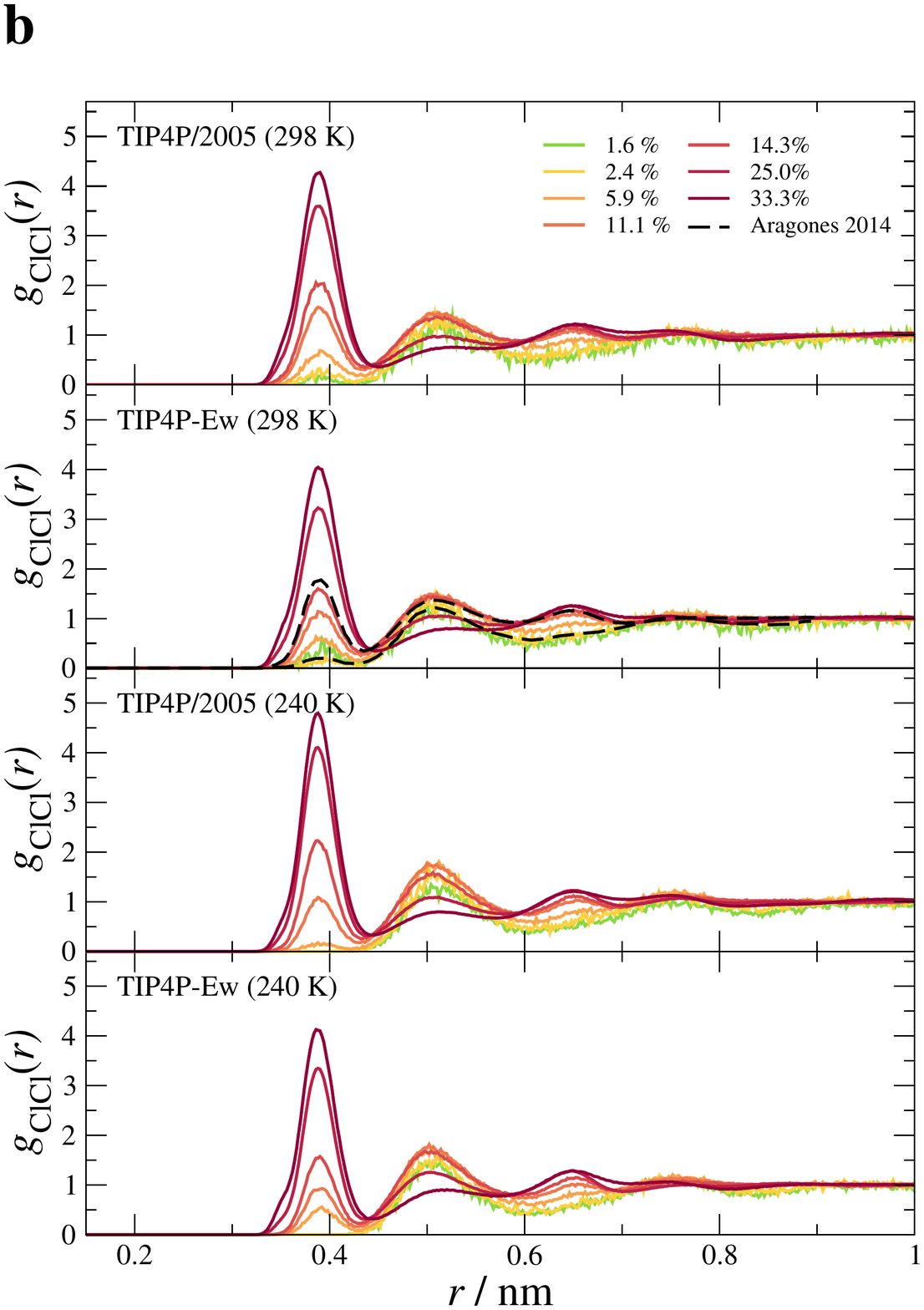}
 \caption{Li-Li (a), and Cl-Cl (b) RDFs  as a function of concentration. In each subfigure the two upper panels show the behaviour at \SI{298}{\kelvin} for TIP4P/2005 and TIP4P-Ew, respectively, whereas the bottom two panels show the behaviour at \SI{240}{\kelvin}. The different colours represent the different concentrations as indicated in the legend. 
 The dashed black lines indicate the RDFs for $x_\text{LiCl}=0.024$ and $x_\text{LiCl}=0.143$ as reported by Aragones et al.~\cite{aragones14-jpcb}.}
 \label{fig:ion-rdf}
\end{figure*}

The four atom types (Li, Cl, O, H) present allow for the calculation of ten different partial radial distribution functions RDFs.
The three different ion-ion structures for both water models used and both temperatures considered are shown in Figs.~\ref{fig:licl-rdf} and \ref{fig:ion-rdf}.
Here, the minimum concentration shown is $x_\text{LiCl}=\SI{1.6}{\percent}$, since the small number of ions at lower concentrations does not allow to obtain statistically significant RDFs.

In Fig.~\ref{fig:licl-rdf} we report the Li-Cl RDFs. 
The two water models produce similar results: the first two peaks are well separated at all concentrations studied reflecting a clear separation of the first and second cation--anion coordination shells.
For both temperatures the first and second peak are changing their intensity as the concentration is increased, but they stay at the same distances.
At \SI{298}{\kelvin} the first and second peak decrease as the concentration is increased, while at \SI{240}{\kelvin} the first peak's intensity increases with concentration and the intensity of the second peak decreases.
A more complex change is observed for the higher coordination shells where an increase in concentration contracts the third and fourth shell.

In Fig.~\ref{fig:ion-rdf}a the Li-Li RDFs are shown.
For both studied temperatures an increase in the LiCl molar fraction initially increases the main peak at $r\approx\SI{0.55}{\nm}$.
At high concentrations, however, this peak decreases again, while two pre-peaks grow,
a strong one at $r\approx\SI{0.4}{\nm}$ and a smaller one at $r\approx\SI{0.32}{\nm}$.
This indicates that as the concentration is increased more and more Li ions come into close proximity.
Also, they occupy more localised shells as is manifested by the rather sharp pre-peaks contrasting the broad main maximum present at low concentrations.

The Cl-Cl structure is shown in Fig.~\ref{fig:ion-rdf}b.
Similar to the Li-Li case a strong pre-peak develops as the concentration is increased.
Here only one pre-peak is evolving, though, and it seems to grow at the expense of the main peak located at $r\approx\SI{0.5}{\nm}$.
At the same time also the region between \num{0.6} and \SI{0.7}{\nano\metre} becomes more populated.

For both the Li-Li and the Cl-Cl structure it is found, that the choice of water model has barely an influence.
Only the strong pre-peaks are slightly larger and sharper in TIP4P/2005 than in TIP4P-Ew.
Moreover, also the temperature change has almost no effect on the two structures, with the exception of a slight increase in the strong pre-peak's height for TIP4P/2005.

\begin{figure*}
 \centering
    \includegraphics[width=0.48\textwidth]{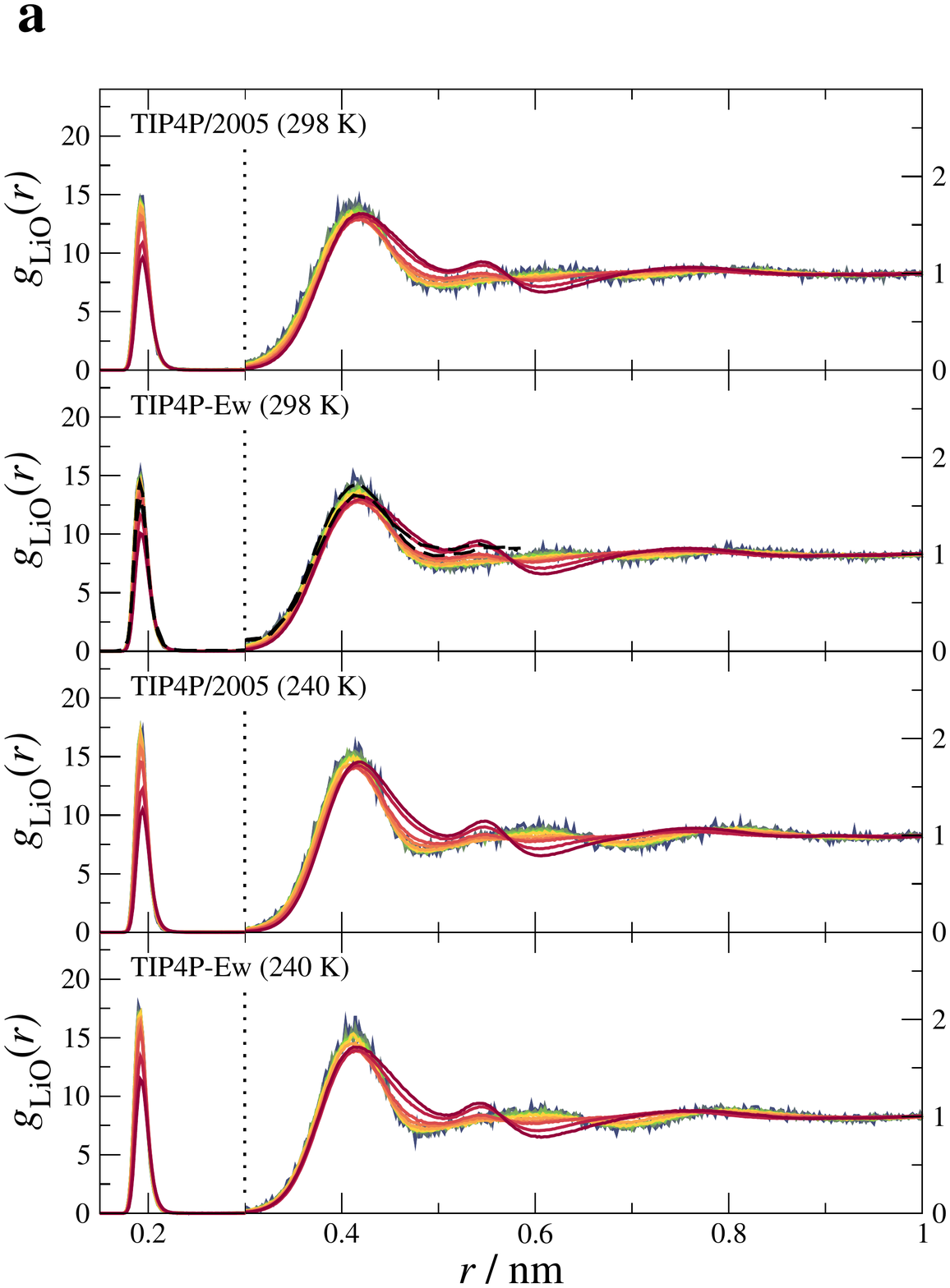}
\includegraphics[width=0.48\textwidth]{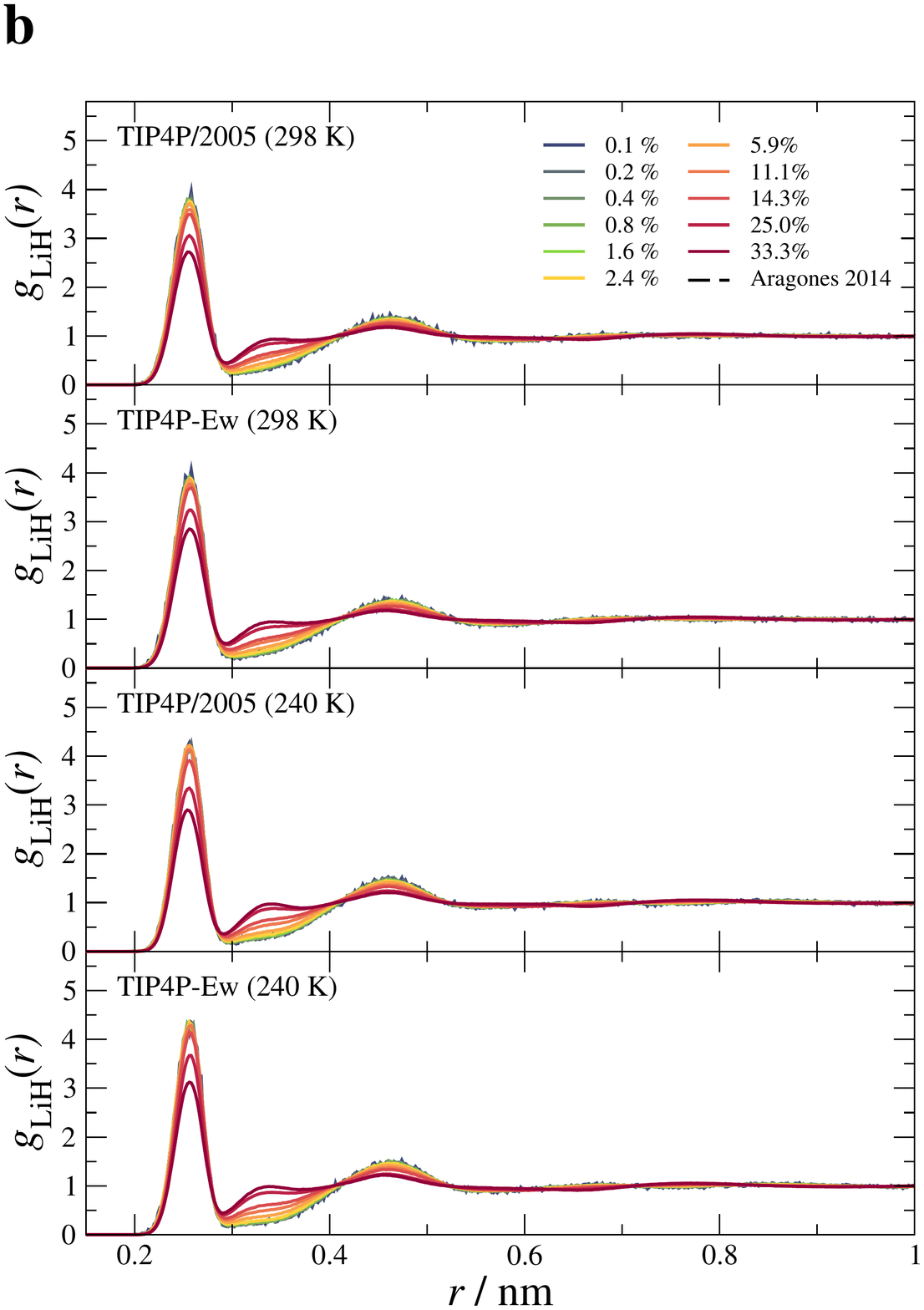}
 \caption{Li-O (a) and Li-H (b) RDFs as a function of concentration. In each subfigure the two upper panels show the behaviour at \SI{298}{\kelvin} for TIP4P/2005 and TIP4P-Ew, respectively, whereas the bottom two panels show the behaviour at \SI{240}{\kelvin}. The different colours represent the different concentrations as indicated in the legend.
 Note that in part a the main peak is plotted on a different scale (left axis) than the rest of the RDF (right axis).
 The dashed black lines indicate the RDFs for $x_\text{LiCl}=0.024$ and $x_\text{LiCl}=0.143$ as reported by Aragones et al.~\cite{aragones14-jpcb}.}
 \label{fig:li-rdf}
\end{figure*}

\begin{figure*}
 \centering
\includegraphics[width=0.48\textwidth]{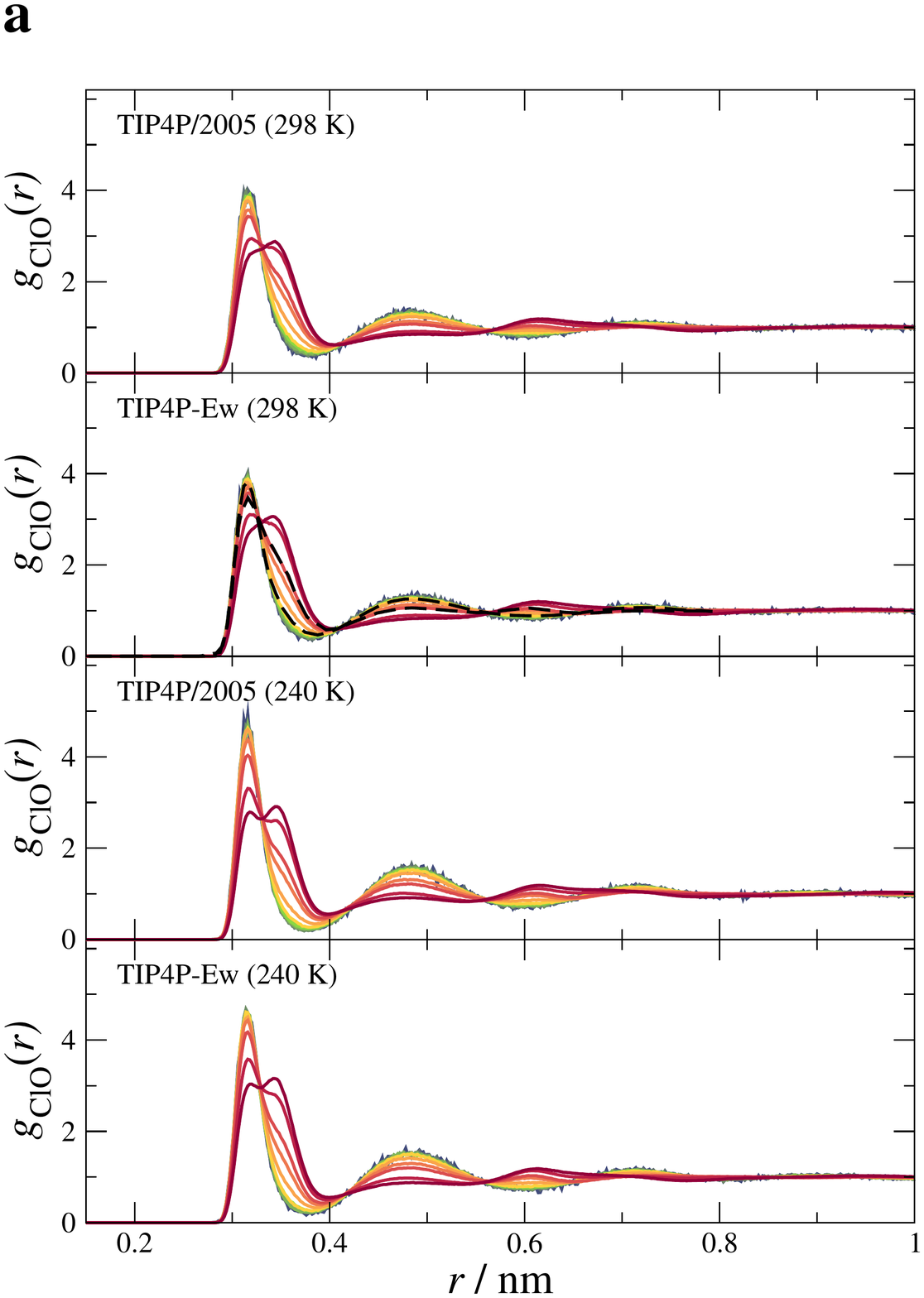}
\includegraphics[width=0.48\textwidth]{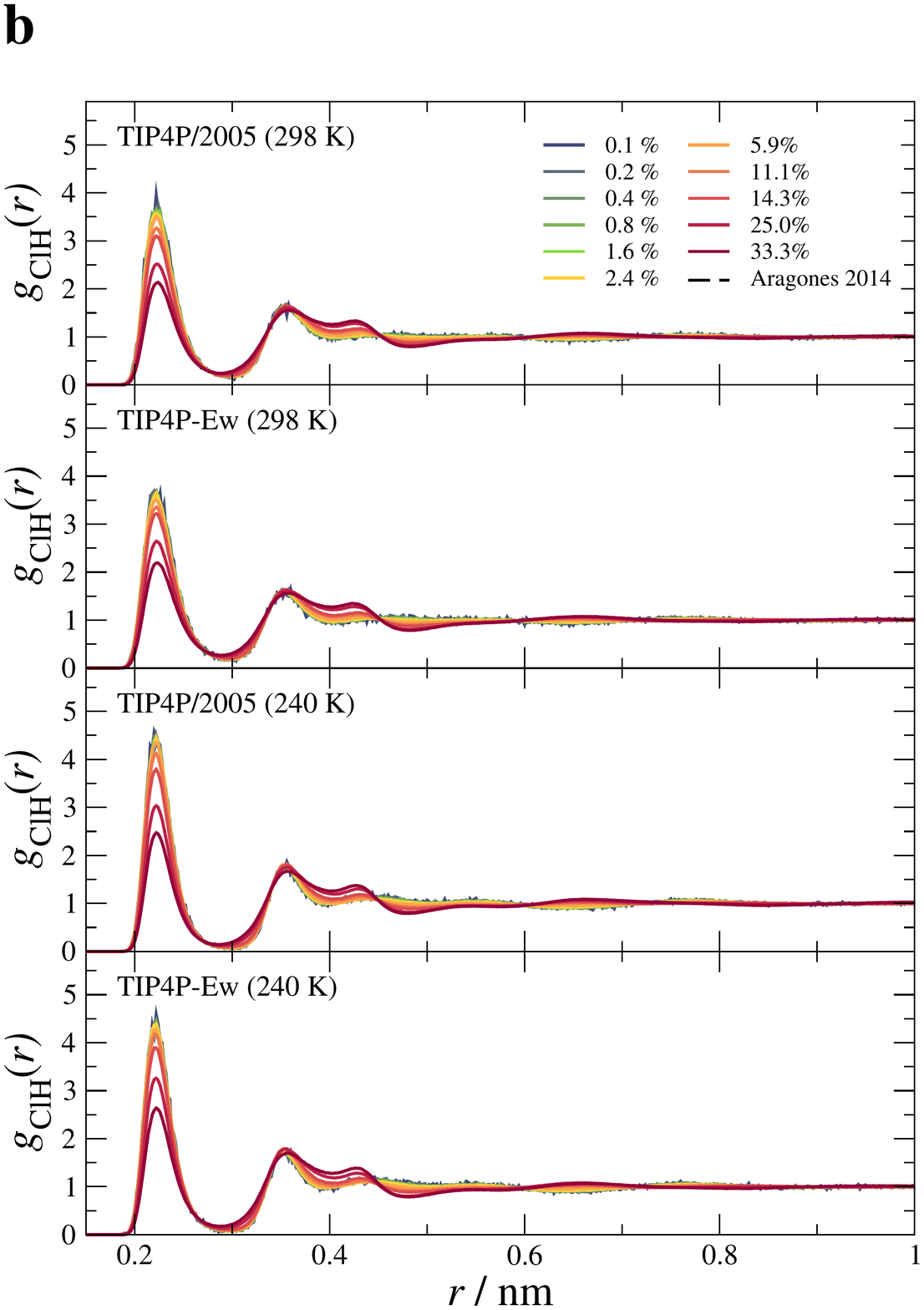}
 \caption{Cl-O (a) and Cl-H (b) RDFs as a function of concentration. In each subfigure the two upper panels show the behaviour at \SI{298}{\kelvin} for TIP4P/2005 and TIP4P-Ew, respectively, whereas the bottom two panels show the behaviour at \SI{240}{\kelvin}. The different colours represent the different concentrations as indicated in the legend.
 The dashed black lines indicate the RDFs for $x_\text{LiCl}=0.024$ and $x_\text{LiCl}=0.143$ as reported by Aragones et al.~\cite{aragones14-jpcb}.}
 \label{fig:cl-rdf}
\end{figure*}

The hydratisation structures of the two ionic species are shown in Figs.~\ref{fig:li-rdf} and ~\ref{fig:cl-rdf}.
Here, the respective RDFs can be obtained for the full concentration range.
For all shown RDFs the two water models again yield similar results.
The Li hydration structure is shown in Fig.~\ref{fig:li-rdf}.
Part a shows the Li-O RDFs, where the height of the first peak decreases with increasing LiCl concentration, while its position is unaltered.
The second shell is well separated from the first shell at all concentrations.
As the concentration is increased the low $r$ side of the second shell does not change, while a shoulder grows to towards larger $r$.
At the same time the higher shells become contracted.

The Li-H RDFs shown in Fig.~\ref{fig:li-rdf}b exhibit a decrease of the first maximum as the concentration is increased.
Moreover, the hydrogen atoms seem to progressively populate regions between the first and second shell.
In combination with the concentration independence of the second Li-O shell at low $r$ this indicates that the water molecules of the second hydration shell do not move towards the first shell as the concentration is increased, but they rotate such that the hydrogens are more likely to point towards the first hydration shell of the Li.
For both the Li-O and the Li-H structure it is found that the change in temperature has little effect (see Fig.~\ref{fig:li-rdf}).

The Cl hydration structure is depicted in Fig.~\ref{fig:cl-rdf}, with
part a showing the Cl-O RDFs.
It evident that the Cl-O structure is more sensitive to changes in LiCl concentration than the Li-O structure.
As the concentration increases a shoulder towards larger distances of the main peak develops.
For the two highest concentrations considered this shoulder is so intense, that it forms a second peak.
At the same time the population of the second shell diminishes and the region between the second and third shell becomes more populated.
If the temperature is lowered to \SI{240}{\kelvin} both the main and the second peak of the dilute systems grow and become slightly more separated.
In contrast, the RDFs at higher concentrations barely change as the temperature is lowered.
This leads to a more pronounced change in the RDFs at \SI{240}{\kelvin} upon concentration increase.

\begin{figure*}
 \centering
\includegraphics[width=0.48\textwidth]{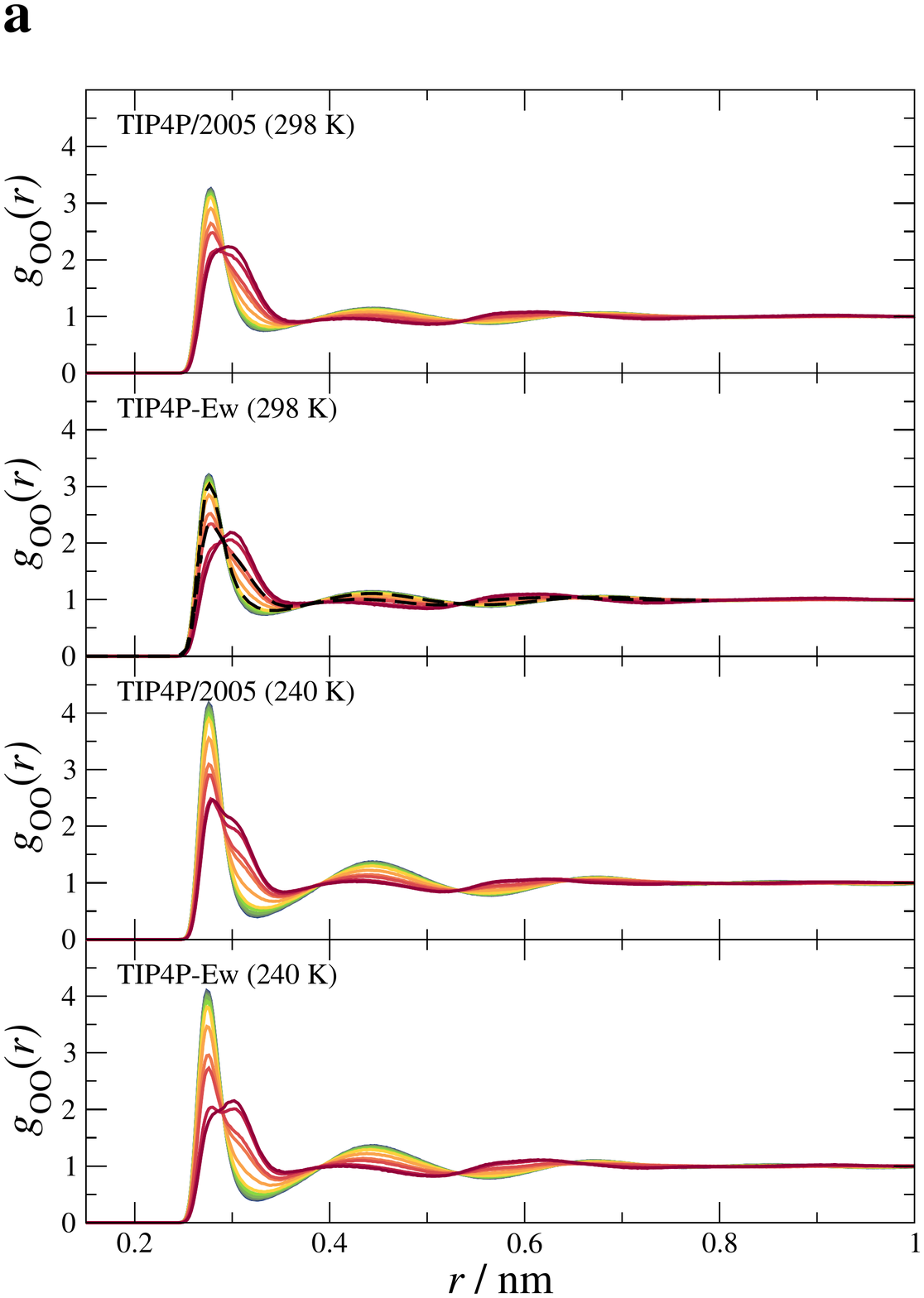}
\includegraphics[width=0.48\textwidth]{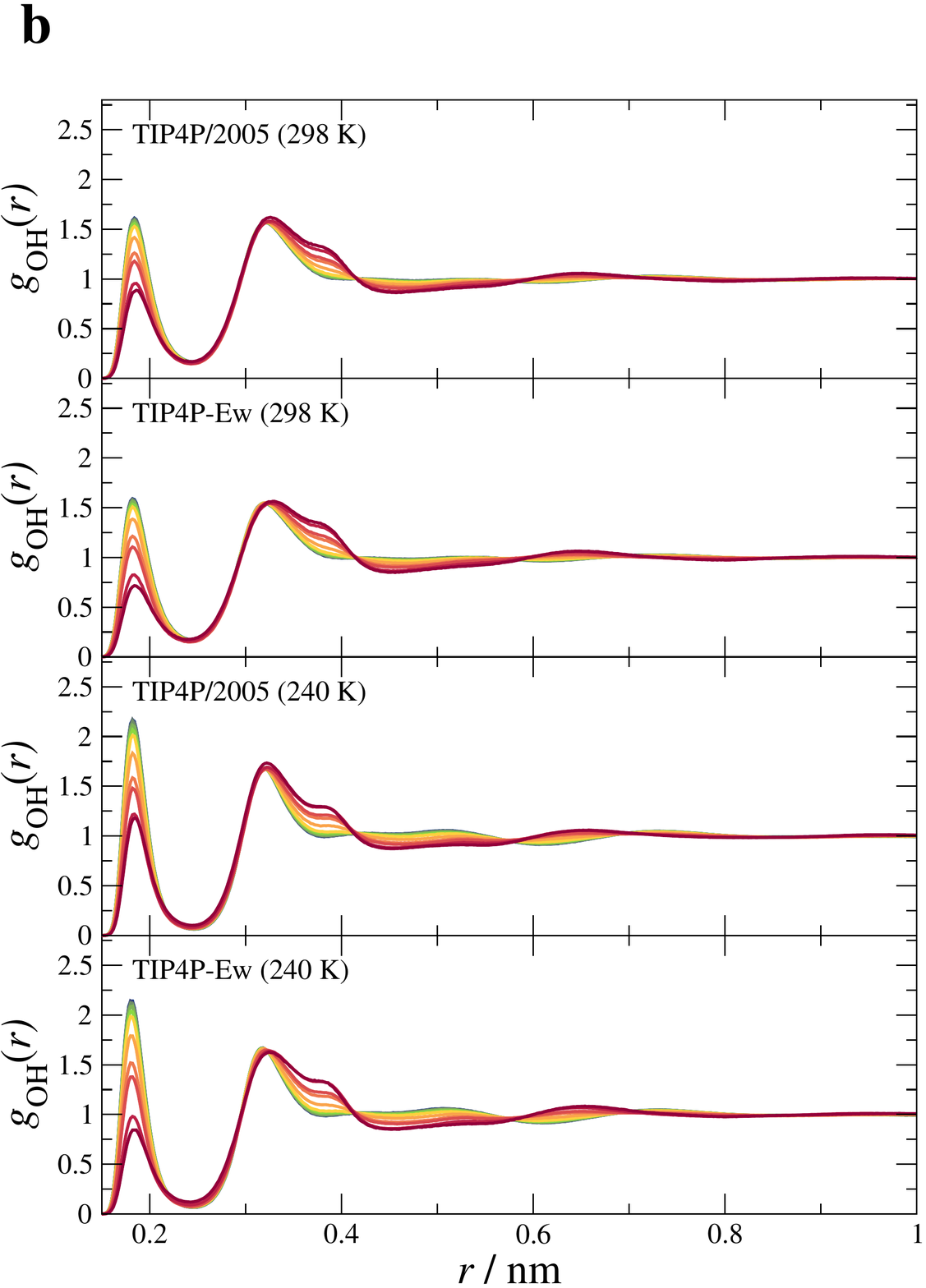}
\includegraphics[width=0.48\textwidth]{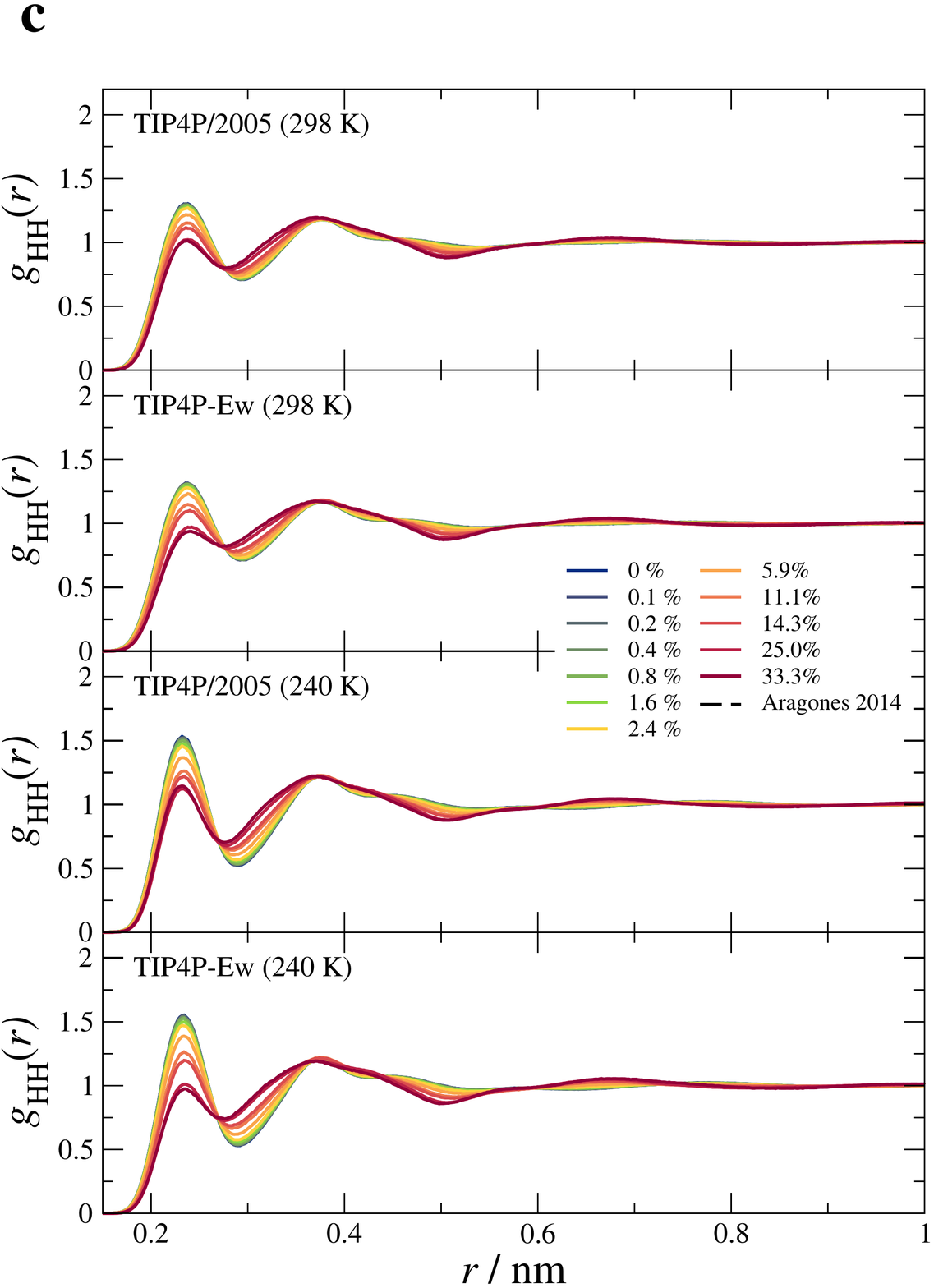}
 \caption{O-O (a), O-H (b), and H-H (c) RDFs as a function of concentration. In each subfigure the two upper panels show the behaviour at \SI{298}{\kelvin} for TIP4P/2005 and TIP4P-Ew, respectively, whereas the bottom two panels show the behaviour at \SI{240}{\kelvin}. The different colours represent the different concentrations as indicated in the legend.
 The dashed black lines indicate the RDFs for $x_\text{LiCl}=0.024$ and $x_\text{LiCl}=0.143$ as reported by Aragones et al.~\cite{aragones14-jpcb}.}
 \label{fig:water-rdf}
\end{figure*}

The Cl-H structure is shown in Fig.~\ref{fig:cl-rdf}b.
Here, the main peak is decreasing with increasing concentration, while the second peak shows little change.
The third shell, however, becomes quite populated at high concentrations and it is located close to the second shell.
Also the shells beyond the third become more contracted.
The rather intense third shell at high concentrations is likely related to the double-peak feature appearing in the Cl-O structure.
Since the second Cl-O maximum appearing at high concentrations is located at $r\approx\SI{0.34}{\nano\meter}$ and the third maximum in the Cl-H RDFs appears at $r\approx\SI{0.42}{\nano\meter}$  their distance is $\approx\SI{0.08}{\nano\meter}$, slightly less than the O-H bond length of TIP4P-type models ($r_\text{OH}=\SI{0.09572}{\nano\meter}$)~\cite{horn04-jcp, abascal05-jcp}.
Thus, these two signals could stem from water molecules close to a chloride, but pointing with the O towards the Cl and with the Hs in the opposite direction.
This unfavourable arrangement, O and Cl electrostatically repel each other, explains the larger Cl-O distance.
Also, it could indicate that the respective water molecules are sandwiched between two or more chlorides, forcing them into unfavourable positions with respect to some chlorides.

Fig.~\ref{fig:water-rdf} shows the three water-water RDFs.
In this figure the RDFs for the pure water models are also shown.
It is again visible, that both water models show similar results.
Moreover, it is found that temperature has a larger effect than in all other RDFs discussed so far.
The O-O RDFs depicted in Fig.~\ref{fig:water-rdf}a show that the increasing LiCl concentration produces a shoulder in the first peak populating the interstitial region.
Typically such a phenomenology is related to a decrease in tetrahedral order~\cite{russo14-natcomm}.
At the same time the first and second shell become less populated and the third shell contracts.
For \SI{240}{\kelvin} the trend is similar, but the first and second shell are more separated for the low concentrations so that the change in the interstitial region is more pronounced.
Similar to the Cl-O RDFs this entails that at low concentrations the O-O RDFs are more affected by the temperature decrease, while the higher concentrations barely change.
Only a slight increase of the main peak is visible at the highest concentrations in TIP4P/2005.

The O-H structure is shown in Fig.~\ref{fig:water-rdf}b.
At \SI{298}{\kelvin} the addition of LiCl leads to a decrease of the first peak, signalling a disruption of the HB network.
The second peak is again barely affected, but a shoulder grows towards larger $r$ at the expense of the third shell and also the fourth shell contracts.
Since the shoulder of the first peak in the O-O RDF and the shoulder of the second peak in the O-H RDF are $\approx\SI{0.08}{\nano\meter}$ apart, this again suggests that the same water molecules are responsible for the two features.
As the temperature is lowered to \SI{240}{\kelvin} the O-H RDFs change only slightly.
One difference is that the first peak is enhanced at \SI{240}{\kelvin} when compared to \SI{298}{\kelvin} indicating an increase in hydrogen bonding.
These changes are more pronounced for the low concentrations, in line with the respective changes of the O-O and Cl-O RDFs.
In addition, the second peak is slightly sharper at \SI{240}{\kelvin} and clearer separated from the first peak.

The H-H RDFs, are shown in Fig.~\ref{fig:water-rdf}c.
Here the first peak decreases as the concentration is increased and the second peak becomes broader.
The lowering of the temperature again affects the lower concentrations more, where the first peak becomes enhanced and more separated from the second one.
We also note small differences between the two water models as the increase of LiCl content has more effect on the RDFs of TIP4P-Ew.

\begin{figure}
 \centering
\includegraphics[width=0.49\textwidth]{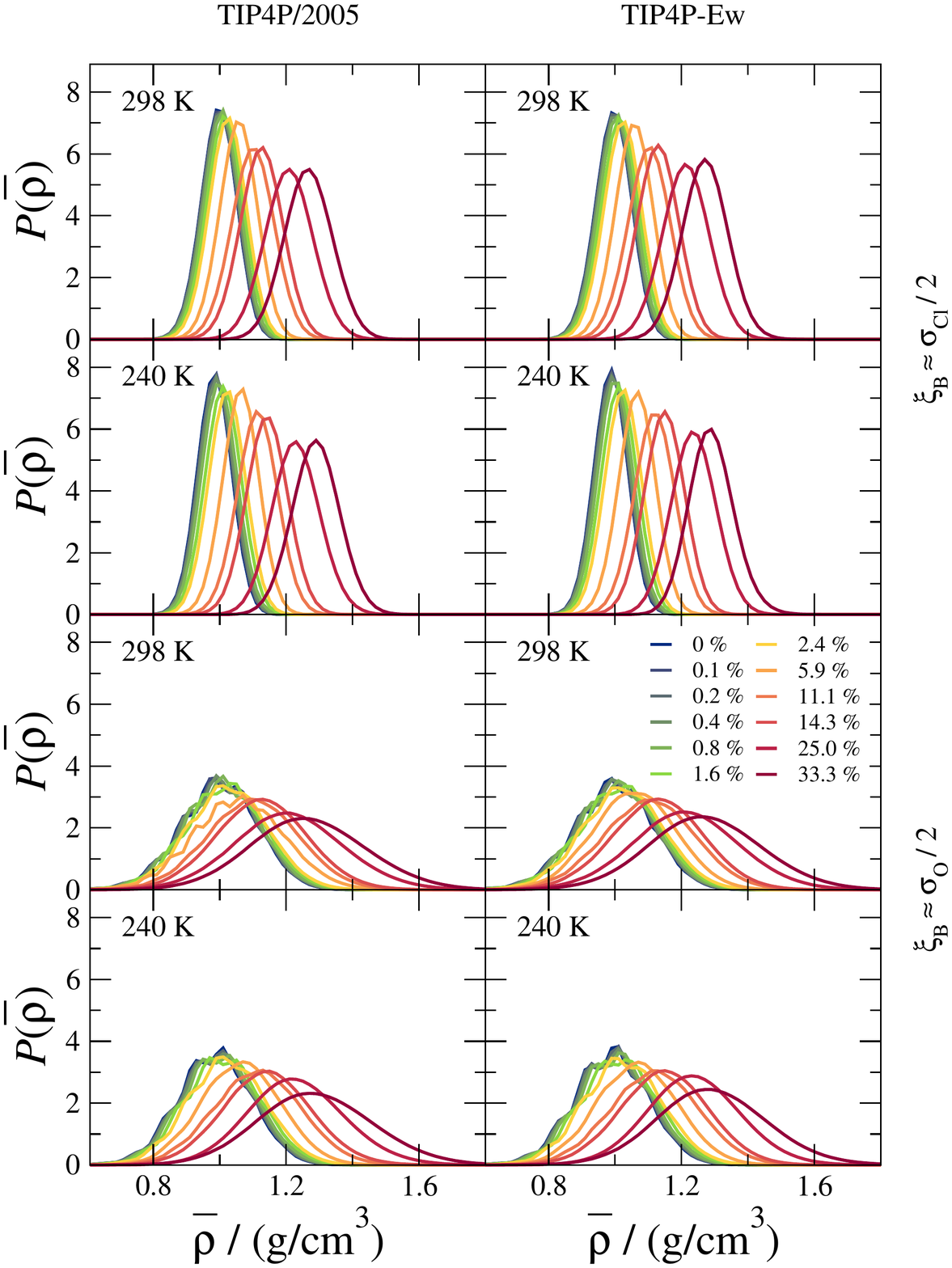}
 \caption{Distribution of the coarse-grained density for both water models and both temperatures studied. The left column shows the data for TIP4P/2005 and the right column for TIP4P-Ew. The top two rows show the results for $\xi_\text{b}\approx\sigma_\text{Cl}/2$ and the bottom two rows show the results for $\xi_\text{b}\approx\sigma_\text{O}/2$. The different colours indicate the different concentrations as indicated in the legend.
 }
 \label{fig:cg}
\end{figure}

\subsection{Coarse-Grained Density}
\label{ssec:cg}

As discussed in the introduction several experimental studies indicate the presence of a low-temperature phase separation in LiCl--H$_2$O at low temperatures~\cite{angell68-jcp, angell70-jcp,kanno87-jcp,suzuki00-prl, bove13-jcp, suzuki13-jcp}.
In particular, it is expected that a water-rich phase separates form a salt-rich phase.
To investigate this we make use of the coarse-grained density field as defined by Testard et al.~\cite{testard2014intermittent}.
Testard et al.~\cite{testard2014intermittent} studied the spinodal decomposition of a binary Lennard-Jones mixture~\cite{kob95-pre}, where the phase separation manifested itself as a bimodality in the coarse-grained density distribution~\cite{testard2014intermittent}. 
Therefore, this methodology is highly suitable to investigate whether a spatial inhomogeneous density field is present in the solutions studied here.

For this analysis the cubic simulation boxes are subdivided into voxels of sidelength $\xi_\text{b}$.
Then, each voxel is assigned a local density $\rho(\mathbf{r})$.
The local density is calculated within a sphere of radius $\xi_\text{s}$ centred at the voxel's centre $\mathbf{r}$.
More formally this is expressed through the following relation:
\begin{equation}
\rho(\mathbf{r})=\frac{3}{4\pi\xi_\text{s}^3}\sum_{i=1}^{N}m_i \theta(\xi_\text{s}-\left|\mathbf{r}-\mathbf{r}_i \right|).
\end{equation}
Here $\theta$ represents the Heaviside step function and $m_i$ is the mass of particle $i$ located at $\mathbf{r}_i$.
The sum includes all species present, i.e., $N=N_\text{Li}+N_\text{Cl}+N_{\text{H}_2\text{O}}$.
In contrast to Testard et al.~\cite{testard2014intermittent}, the density calculated here is a mass density and not a particle density.

In a second step each voxel is assigned a coarse-grained density $\overline{\rho}(\mathbf{r})$, which takes the local density of the voxel and its six immediate neighbours into account:
\begin{align}
\overline{\rho}(\mathbf{r})=&\frac{1}{8}\left[2 \rho(\mathbf{r}) +  \rho(\mathbf{r} +  \xi_\text{b}\mathbf{e}_\text{x})+  \rho(\mathbf{r} +  \xi_\text{b}\mathbf{e}_\text{y}) +  \rho(\mathbf{r} +  \xi_\text{b}\mathbf{e}_\text{z})\right.\nonumber\\
&+\left.  \rho(\mathbf{r} -  \xi_\text{b}\mathbf{e}_\text{x}) +  \rho(\mathbf{r} -  \xi_\text{b}\mathbf{e}_\text{y}) +  \rho(\mathbf{r} -  \xi_\text{b}\mathbf{e}_\text{z}) \right] .
\end{align}
Here $\mathbf{e}_\text{x}$, $\mathbf{e}_\text{y}$, and $\mathbf{e}_\text{z}$ denote unit vectors in the respective direction.

To perform this analysis values for $\xi_\text{b}$ and $\xi_\text{s}$ have to be selected.
After a series of tests~\cite{testard2011etude} Testard et al.~\cite{testard2014intermittent} used $\xi_\text{b}=\sigma_\text{AA}/2$ and $\xi_\text{s}=\sigma_\text{AA}$ for their calculations, where $\sigma_\text{AA}$ corresponds to the diameter of the larger Lennard-Jones component.
Since, three atom-types interacting via a Lennard-Jones term are present here, i.e., Li, Cl, and O, we performed two separate calculations.
One based on $\sigma_\text{Cl}$, the largest species present, and one based on $\sigma_\text{O}$, the second largest species present.

Furthermore, it had to be taken into account that the box size varies in 
the $N\!PT$-MD calculations performed here.
Thus, we did not fix $\xi_\text{b}$, but the number of intervals $I$ into which the box is divided in each direction.
$I$ is specified from the first frame of each trajectory:
\begin{equation}
 I=\left\lfloor\frac{2L_0}{\sigma_i}\right \rfloor.
\end{equation}
Here, $i$ is either Cl or O, $L_0$ is the box length of the first frame, and the brackets indicate the floor function.
From this $\xi_\text{b}$ and $\xi_\text{s}$ are calculated via
\begin{equation}
 \xi_\text{b}=\frac{L}{I},\label{eq:xib}
 \end{equation}
and
\begin{equation}
 \xi_\text{s}=2\xi_{b},
\end{equation}
where $L$ is the box length of the frame considered.
For the calculations we used $\sigma_\text{Cl}\approx\SI{0.5}{\nano\meter}$ and $\sigma_\text{O}\approx\SI{0.32}{\nano\meter}$.
The exact values are $\sigma_\text{Cl}=\SI{0.49178}{\nano\meter}$~\cite{joung08-jpcb}, $\sigma_\text{O}=\SI{0.31589}{\nano\meter}$ (TIP4P/2005)~\cite{abascal05-jcp}, and $\sigma_\text{O}=\SI{0.316435}{\nano\meter}$ (TIP4P-Ew)~\cite{horn04-jcp}, but since the fluctuating box length induces variations in $\xi_\text{b}$ already (cf. Eq.~\ref{eq:xib}), the approximations have been used.

The results of this analysis are shown in Fig.~\ref{fig:cg}.
The left column shows the results for TIP4P/2005 and the right column shows the results for TIP4P-Ew.
For $\xi_\text{b}\approx\sigma_\text{Cl}/2$ the results are shown in the top two rows.
It is evident, that neither a change in water model nor a change in temperature has a strong effect on the coarse-grained density distribution $P(\overline{\rho})$.
The distribution is unimodal in all cases and it simply shifts to higher densities as the concentration is increased consistent with the density of the solution (cf. Tab.~S-I).
For $\xi_\text{b}\approx\sigma_\text{O}/2$ the results are shown in the bottom two rows.
These distributions are more spread out than in the $\xi_\text{b}\approx\sigma_\text{Cl}/2$ case.
Other than that the picture barely changes:
the distributions are unimodal and shift to higher densities as the concentration is increased.

None of the obtained coarse-grained density distributions indicate a phase separation.
That is, the coarse-grained local density field is homogeneous for all studied conditions.
This includes also the different combination rules, which do not alter the result (cf. Fig.~\ref{fig:thd} and Fig.~S12 in the ESI).

\subsection{Structural Order Parameter}
\label{ssec:rt}

As a second method to investigate the possibility of a low-temperature phase separation we consider the structural order parameter $\zeta$ introduced by Russo and Tanaka~\cite{russo14-natcomm}.
This order parameter is calculated for every water molecule as
\begin{equation}
 \zeta = d_{\neg\text{HB}} - d_\text{HB},
 \label{eq:zeta}
\end{equation}
where $d_\text{HB}$ is the distance to the \emph{farthest hydrogen bonded} neighbour and $d_{\neg\text{HB}}$ is the distance to the \emph{nearest non-hydrogen bonded} neighbour.
For a neighbouring molecule to be considered hydrogen bonded two criteria have to be satisfied:
i) the O-O distance has to be less than \SI{0.35}{\nano\meter} and
ii) the HOO angle has to be less than \SI{30}{\degree}.
In other words $\zeta$ is a measure of the distance between the first and the second coordination shell of water.
Positive values of $\zeta$ indicate well separated first and second hydration shells and thus a rather well developed tetrahedral HB-network.
Such low-density states have been termed S-states by Russo and Tanaka~\cite{russo14-natcomm}.
Values close to \SI{0}{\nano\metre} and below indicate that water molecules from the second shell penetrate the first shell distorting the HB network.
In Russo and Tanaka's terms this high-density local structure is called the $\rho$-state~\cite{russo14-natcomm}.
Pure S- and $\rho$-liquids would correspond to LDL and HDL in the LLCP scenario, respectively.

The distribution of the structural order parameter $P(\zeta)$ shows a bimodal behaviour for several water models~\cite{russo14-natcomm,shi18-jcp}, including TIP4P/2005, consistently revealing the presence of two distinct local environments.
We use this order parameter here to look for signs of a possible phase separation in LiCl--H$_2$O.

In addition to the regular distribution $P(\zeta)$ calculated for all water molecules in the system, we also consider partial distributions based on subsets of water molecules.
The distribution $P(\zeta | \text{Li})$ is calculated for water molecules which are exclusively part of a first coordination shell of Li, i.e., within \SI{0.23}{\nano\meter} of a Li \emph{and not} within \SI{0.38}{\nano\meter} of a Cl.
These two cut-off distances are based on the minima between the first and second peak in the Li-O (cf. Fig.~\ref{fig:li-rdf}a) and Cl-O (cf. Fig.~\ref{fig:cl-rdf}a) RDFs of dilute systems.
Note that these restrictions apply only to the central water molecule.
Its partners, i.e., the farthest hydrogen bonded neighbour and the nearest non-hydrogen bonded neighbour, can be any other water molecule in the system.
Constructed in this way $P(\zeta | \text{Li})$ encodes the local structural order for water molecules being exclusively part of the first hydration shell of Li.

Analogously, $P(\zeta | \text{Cl})$ is calculated for water molecules which are exclusively part of a first hydration shell of Cl (i.e., within \SI{0.38}{\nano\meter} of a Cl \emph{and not} within \SI{0.23}{\nano\meter} of a Li).
Additionally, $P(\zeta|\text{Both})$ is calculated for molecules being part of both a first hydration shell of Li and Cl (i.e., within \SI{0.23}{\nano\meter} of a Li \emph{and} within \SI{0.38}{\nano\meter} of a Cl), and
$P(\zeta|\text{Bulk})$ is calculated for all water molecules not being part of any first hydration shell (i.e., \emph{not} within \SI{0.23}{\nano\meter} of a Li \emph{and not} within \SI{0.38}{\nano\meter} of a Cl).
Using these definitions $P(\zeta)$ can be expressed as a linear combination of the four partial distributions:
\begin{align}
 P(\zeta)=&P(\text{Li})P(\zeta | \text{Li})+P(\text{Cl})P(\zeta | \text{Cl})\nonumber\\
 &+P(\text{Both})P(\zeta | \text{Both})+P(\text{Bulk})P(\zeta | \text{Bulk}).
\end{align}
Here, $P(\text{Li})$, $P(\text{Cl})$, $P(\text{Both})$, and $P(\text{Bulk})$ are the fractions of water molecules in the respective subset.
These are simply calculated as
\begin{equation}
P(i)=\frac{N^i_{\text{H}_2\text{O}}}{N_{\text{H}_2\text{O}}},
\end{equation}
where $i$ is Li, Cl, Both, or Bulk.
$N^i_{\text{H}_2\text{O}}$ is the number of water molecules in the respective subset, and $N_{\text{H}_2\text{O}}$ is the total number of water molecules in the system ($N_{\text{H}_2\text{O}}=1000$ here).

\begin{figure*}
 \centering
\includegraphics[width=0.48\textwidth]{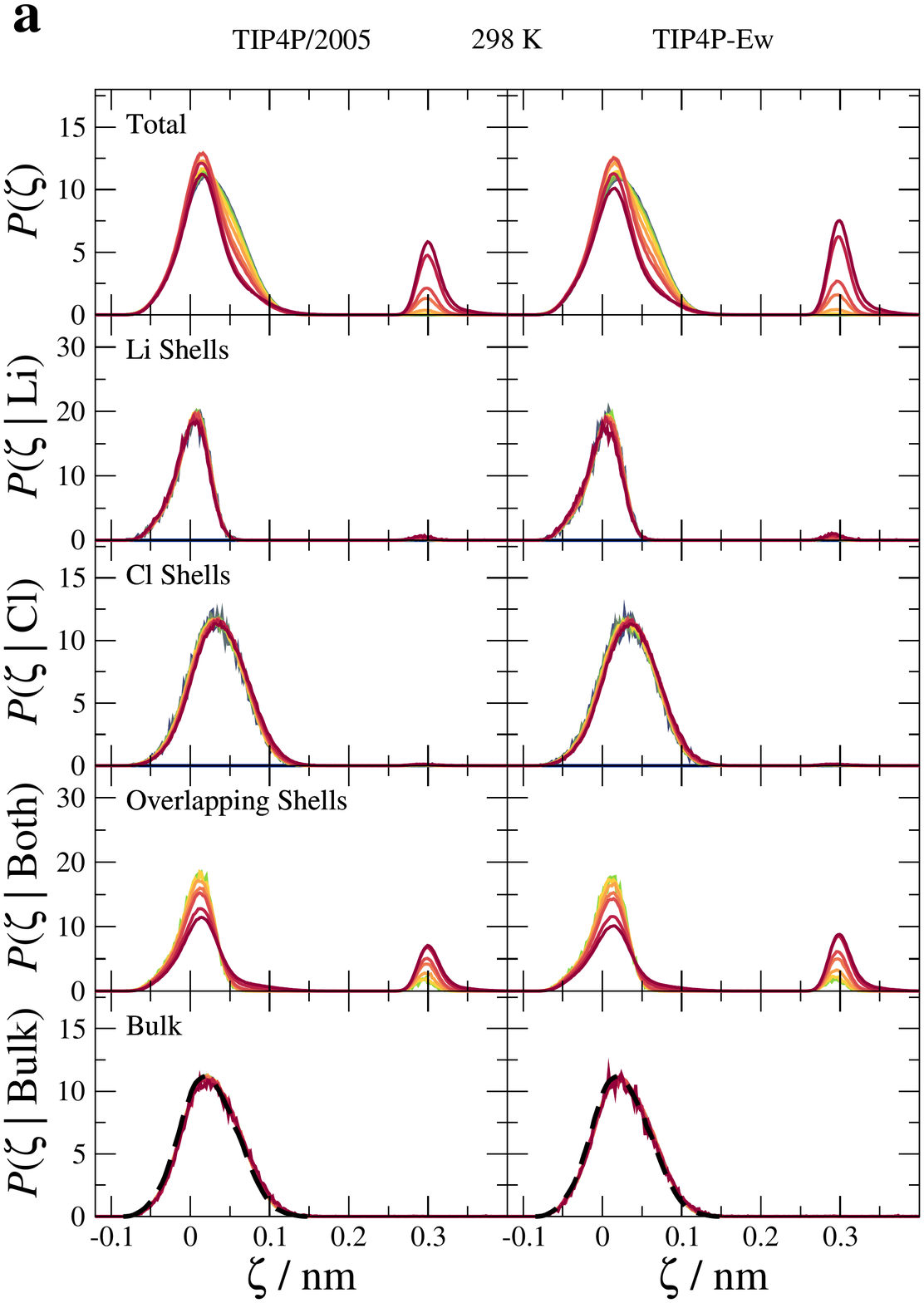}
\includegraphics[width=0.48\textwidth]{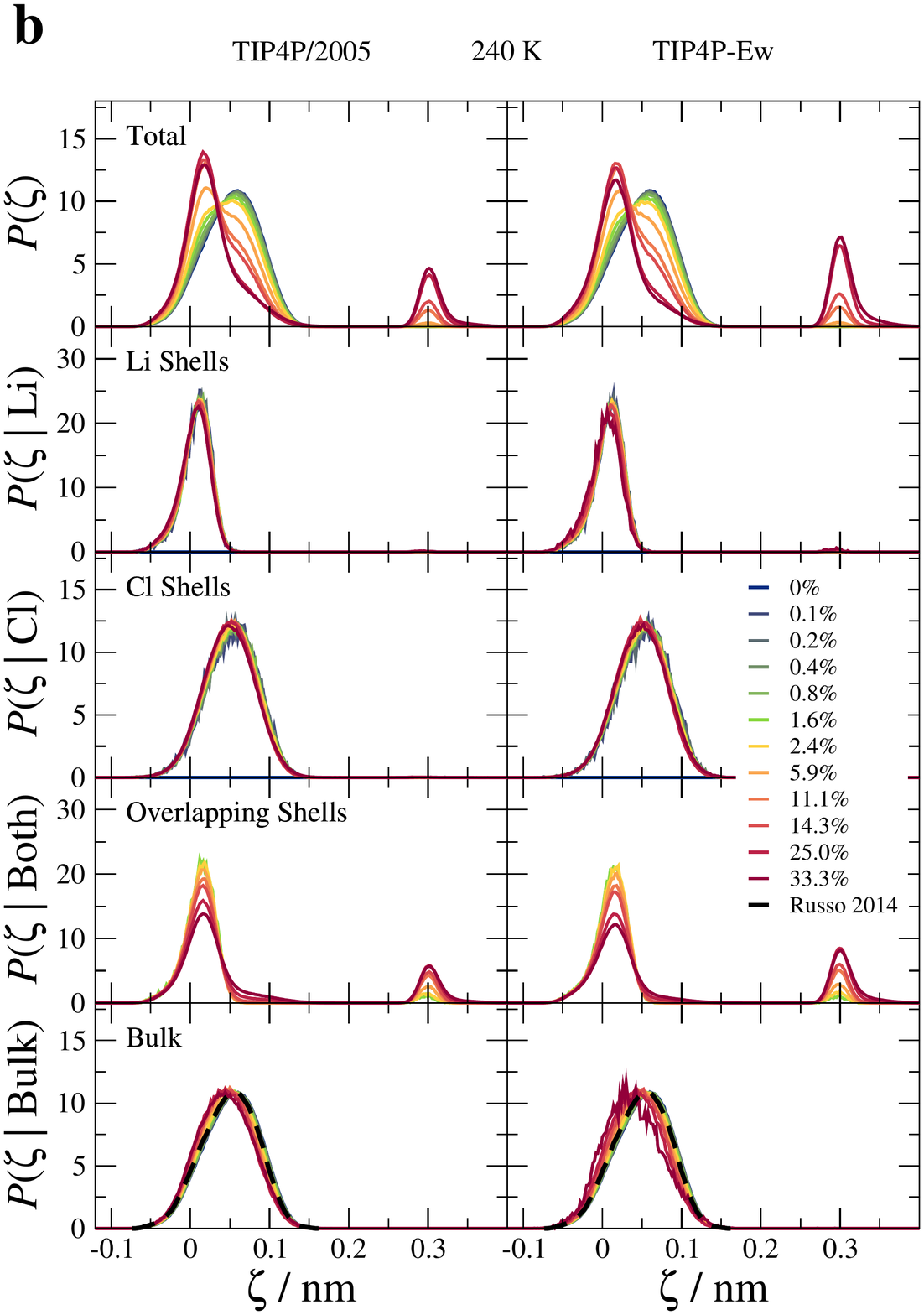}
\includegraphics[width=0.48\textwidth]{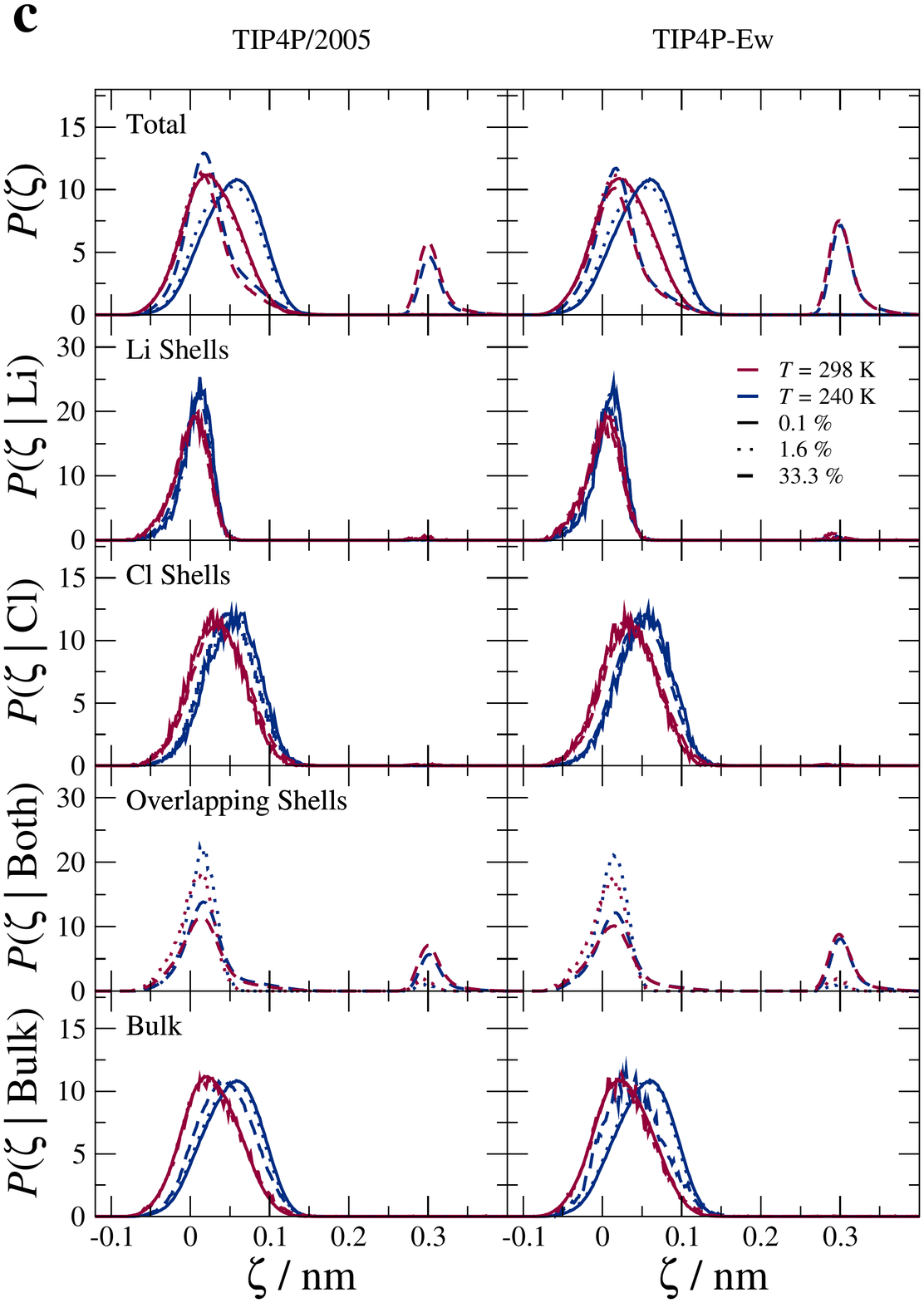}
 \caption{Distributions of the structural order parameter $\zeta$ at \SI{298}{\kelvin} (a) and \SI{240}{\kelvin} (b).
 The different colours indicate the different concentrations as indicated in the legend.
 Part c illustrates the effect of temperature change for selected concentrations (see dedicated legend).
 In all figures the left column shows the data for TIP4P/2005 and the right column for TIP4P-Ew.
  Rows from top to bottom show the data for all water molecules, molecules in first hydration shells of Li only, molecules in first hydration shells of Cl only, molecules in first hydration shells of both Li and Cl, and the bulk contribution, i.e, water not part of any first hydration shell.
  The dashed black lines indicate the data from Russo and Tanaka obtained for pure TIP4P/2005~\cite{russo14-natcomm}.}
 \label{fig:rt}
\end{figure*}

\begin{figure}
 \centering
\includegraphics[width=0.48\textwidth]{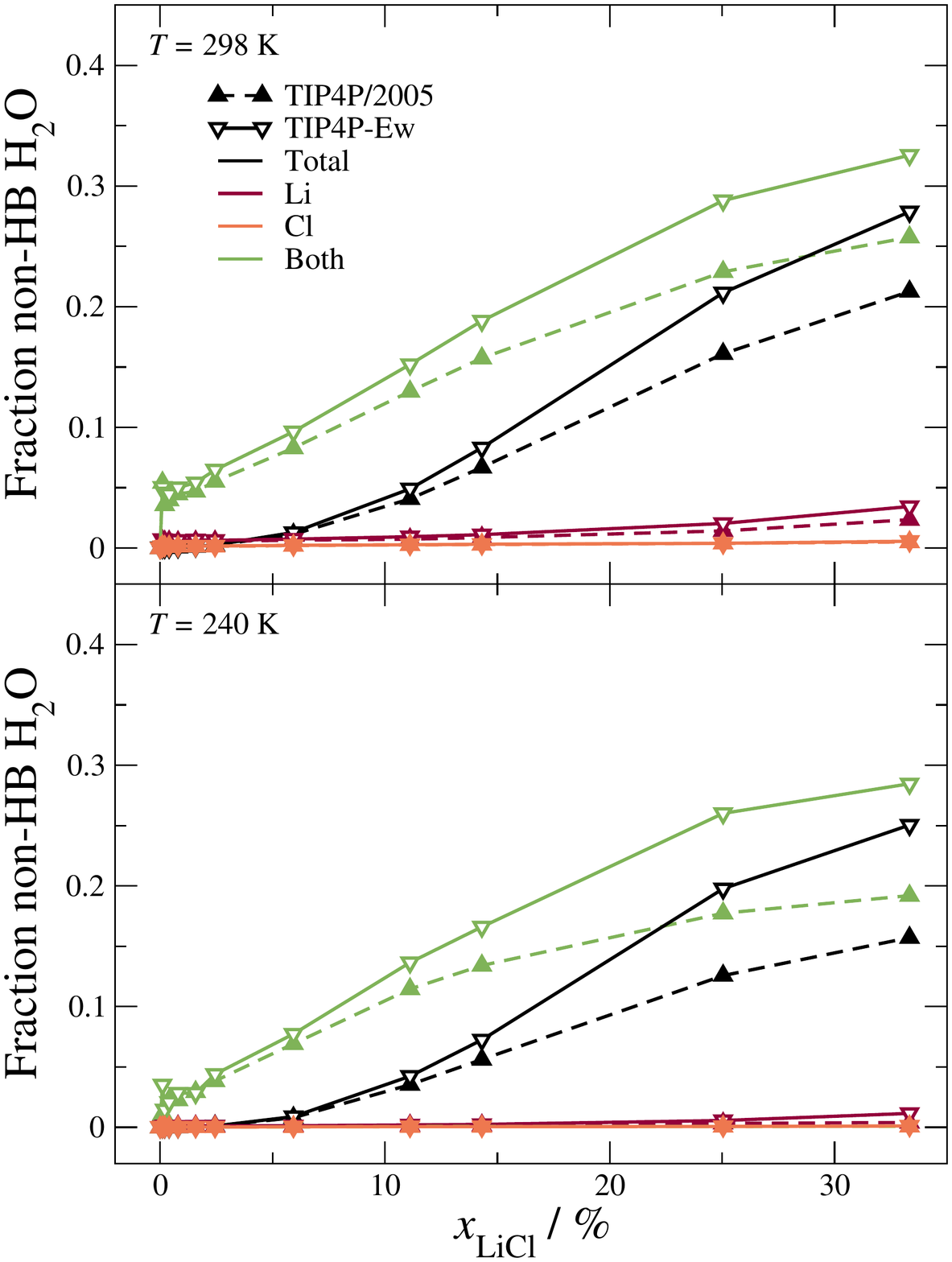}
 \caption{Fraction of water molecules with no HB to another water molecule as a function of LiCl mole fraction $x_\text{LiCl}$. The top panel shows the data for \SI{298}{\kelvin} the bottom panel shows the data for \SI{240}{\kelvin}.
 The data obtained with TIP4P/2005 are shown as filled upward triangles and the data obtained with TIP4P-Ew are shown as open downward triangles.
 The different subsets are shown in different colours as indicated in the legend.
 }
 \label{fig:rt-hb}
\end{figure}

The obtained $\zeta$-distributions are shown in Fig.~\ref{fig:rt}.
Part a shows the distributions for both water models at \SI{298}{\kelvin}.
The distribution for the full system $P(\zeta)$ is shown in the top row.
Here two peaks are visible.
The first peak is located slightly above $\SI{0}{\nano\meter}$, while the second peak is located close to $\SI{0.3}{\nano\meter}$.
For pure water only the main peak close to $\SI{0}{\nano\meter}$ is present.
As the LiCl concentration is increased the main peak becomes more symmetric and shifts even closer to $\SI{0}{\nano\meter}$.
The shift indicates that the HB network becomes more distorted as the concentration is increased.
At the same time the peak at $\zeta\approx\SI{0.3}{\nano\meter}$ grows.
This peak originates from water molecules that have no hydrogen bonded neighbours according to the criteria used.
In this case $d_{\text{HB}}$ is set to $\SI{0}{\nano\meter}$, which yields $\zeta=d_{\neg\text{HB}}$ based on the definition in Eqn.~\ref{eq:zeta}.
In other words, $\zeta$ reflects the next-neighbour distance for these water molecules.
While this disagrees with the idea of $\zeta$ being a measure for the distance of the first and second coordination shell, it visualises the amount of water molecules not properly included into the HB network.
Hence, this peak will be referred to as the non-HB peak in the following.
It is obvious that non-hydrogen bonded water molecules become more prevalent as the concentration of LiCl is increased.
This is clearly visible in Fig.~\ref{fig:rt-hb} where the black data in the top panel give the fraction of non-HB water molecules.
Almost all water molecules are integrated into the HB network until $x_\text{LiCl}=\SI{5.9}{\percent}$ where the number of non-HB water molecules start to increase significantly.
Note that in TIP4P/2005 the fraction of non-HB water molecules is always lower than in TIP4P-Ew.

To analyse the origin of the changes in $P(\zeta)$, the partial distributions have to be discussed.
These are shown in the bottom four rows of Fig.~\ref{fig:rt}a.
One can clearly see that only  $P(\zeta | \text{Both})$, the distribution calculated for molecules being part of first hydration shells of both Li and Cl, changes as the concentration is increased.
Interestingly, both peaks in $P(\zeta | \text{Both})$ do not shift, but the main peak shrinks while the non-HB peak grows with increasing LiCl amount.
For all concentrations the main peak is centred around $\SI{0}{\nano\meter}$ indicating a quite distorted environment of the corresponding water molecules.
The change in intensity of the two peaks is also reflected in the top panel of Fig.~\ref{fig:rt-hb} (green data) which shows that the fraction of non-HB water molecules in this subset increases to about \SI{30}{\percent} at the highest concentration.

The other three partial distributions do not change significantly as the concentration is increased.
$P(\zeta | \text{Li})$ is centred at $\approx\SI{0}{\nano\meter}$, is rather sharp, and bears similarities to the main peak of $P(\zeta | \text{Both})$.
$P(\zeta | \text{Li})$ is followed by $P(\zeta | \text{Bulk})$ slightly above $\SI{0}{\nano\meter}$ being markedly broader, and $P(\zeta | \text{Cl})$ centred at the largest $\zeta$ and similar in shape to $P(\zeta | \text{Bulk})$.
Only small non-HB peaks appear in both $P(\zeta | \text{Li})$ and $P(\zeta | \text{Cl})$ at high concentrations.
Fig.~\ref{fig:rt-hb} again shows the amount of water molecules with no HB in the first coordination shells of Li (red data) and Cl (orange data).
In the latter subset molecules without HBs amount to less than \SI{1}{\percent} even at the highest concentrations.
The locations of the main peaks suggests that independent of concentration Li favours a quite distorted environment, while the Cl hydration shell is more ordered.
The bulk component is located in between these two cases.

\begin{figure}
 \centering
\includegraphics[width=0.48\textwidth]{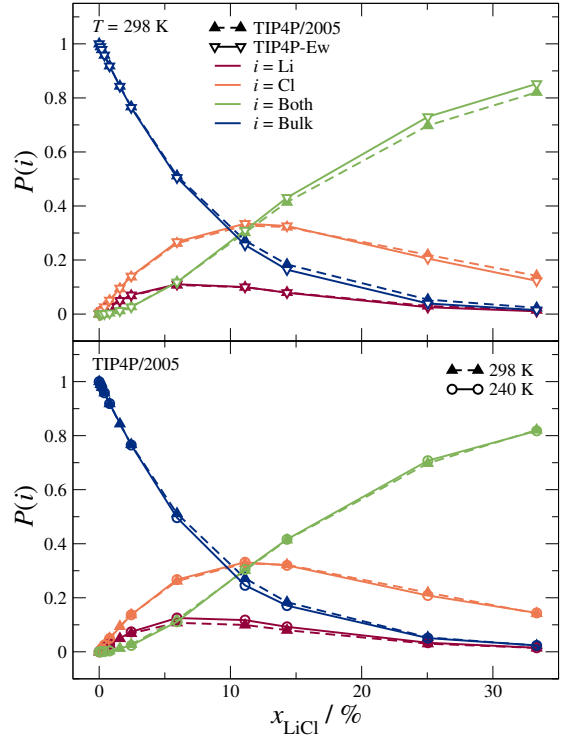}
 \caption{Relative contribution of the four partial distributions of the structural order parameter $\zeta$ to the full distribution $P(\zeta)$ as a function of LiCl mole fraction $x_\text{LiCl}$. The top panel shows a comparison of the two different water models TIP4P/2005 (filled upward triangles) and TIP4P-Ew (open downward triangles) for \SI{298}{\kelvin}.
 The different partial distributions are coloured as shown in the legend.
 The bottom panel shows a comparison of the two different temperatures \SI{298}{\kelvin} (filled upward triangles) and \SI{240}{\kelvin} (open downward triangles) for TIP4P/2005.
 }
 \label{fig:rt-amounts}
\end{figure}

As discussed above the full distribution $P(\zeta)$ exhibits a shift as $x_\text{LiCl}$ is increased.
In contrast, the partial distributions do not shift with changing $x_\text{LiCl}$ indicating no change in the different local environments.
Therefore, a change in the relative contributions of the partial distributions, i.e., the $P(i)$s, is responsible for the overall shift observed in $P(\zeta)$.
The $P(i)$s are shown in the top panel of Fig.~\ref{fig:rt-amounts}.
The bulk water partial distribution dominates $P(\zeta)$ until $x_\text{LiCl}=\SI{5.9}{\percent}$.
As the concentration increases further the first hydration shells of Li and Cl become more important.
Between \SI{5.9}{\percent} and \SI{11.1}{\percent} their influence reaches a maximum and declines again.
At $x_\text{LiCl}>\SI{14.3}{\percent}$ $P(\zeta)$ is dominated by water molecules being part of both a first hydration shell of Li and a first hydration shell of Cl.
The trend in $P(\zeta)$ indicates a shift to an even more distorted HB network than already present in pure water.
This is not due to a change in the bulk water component, but due to the introduction of first hydration shells of ions.
The shift in $P(\zeta)$ is towards $P(\zeta | \text{Li})$ and $P(\zeta | \text{Both})$, but away from $P(\zeta | \text{Cl})$.
That is, the ordering potential of Chloride is neutralised by the disorder of overlapping hydration shells. 

Another key finding is that the bulk contribution does not change as LiCl is added, i.e., the local structure of water beyond the first hydration shells is not altered.
Note that $P(\zeta | \text{Bulk})$ of both water models agrees well with the data of Russo and Tanaka obtained for pure TIP4P/2005 at \SI{300}{\kelvin}~\cite{russo14-natcomm} (cf. Fig.~\ref{fig:rt}a).
According to their analysis this distribution indicates that $\approx\SI{85}{\percent}$ of the water molecules are in the $\rho$-state, while only $\approx\SI{15}{\percent}$ are in the S-state.
This explains the small influence of the ions on the bulk component, since pure water at $\SI{298}{\kelvin}$ and $\SI{1}{\bar}$ already consists of a quite distorted HB network, reducing the effect distorting hydration shells.

\begin{figure*}
 \centering
    \includegraphics[width=0.86\textwidth]{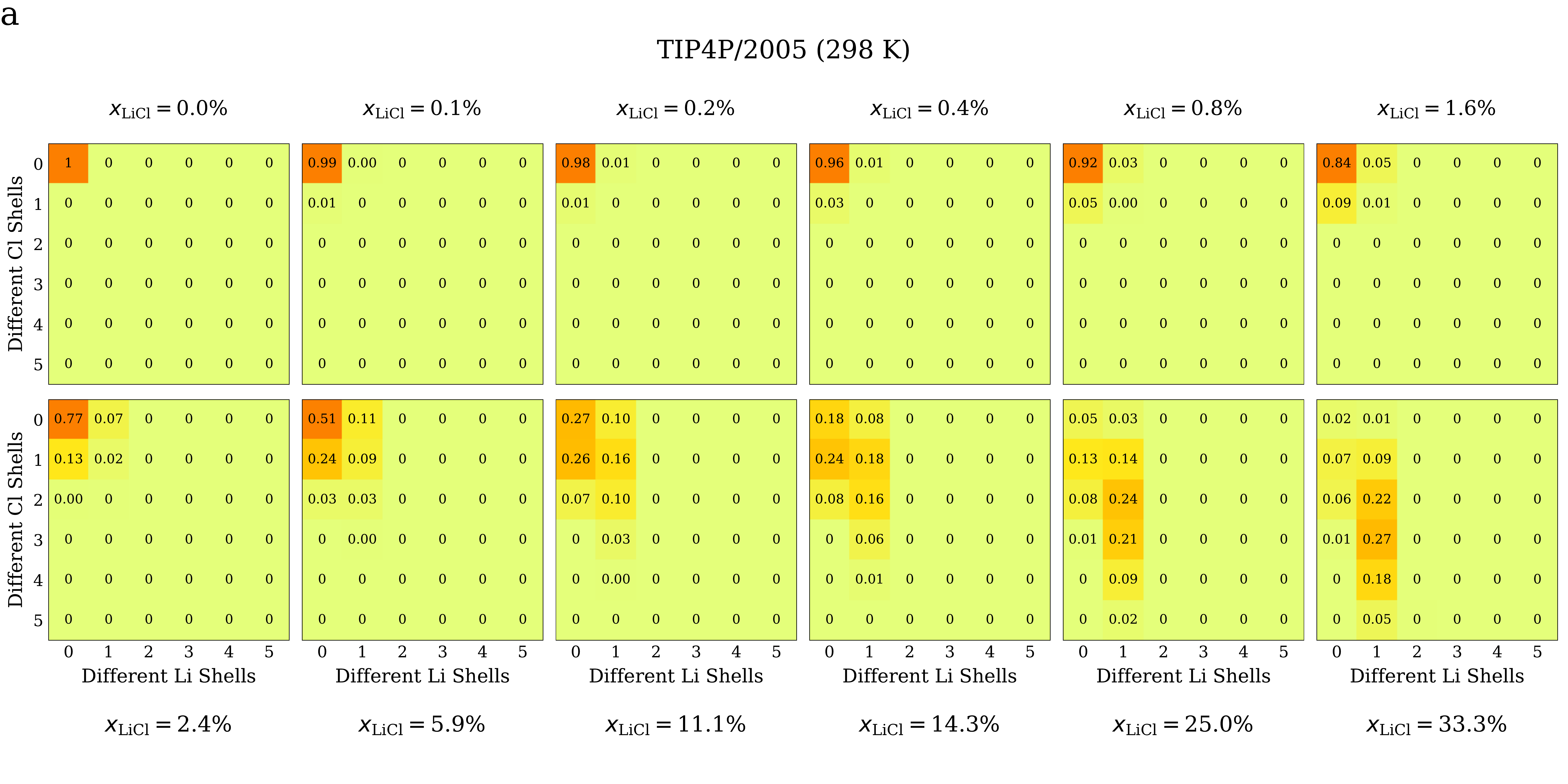}
    \includegraphics[width=0.86\textwidth]{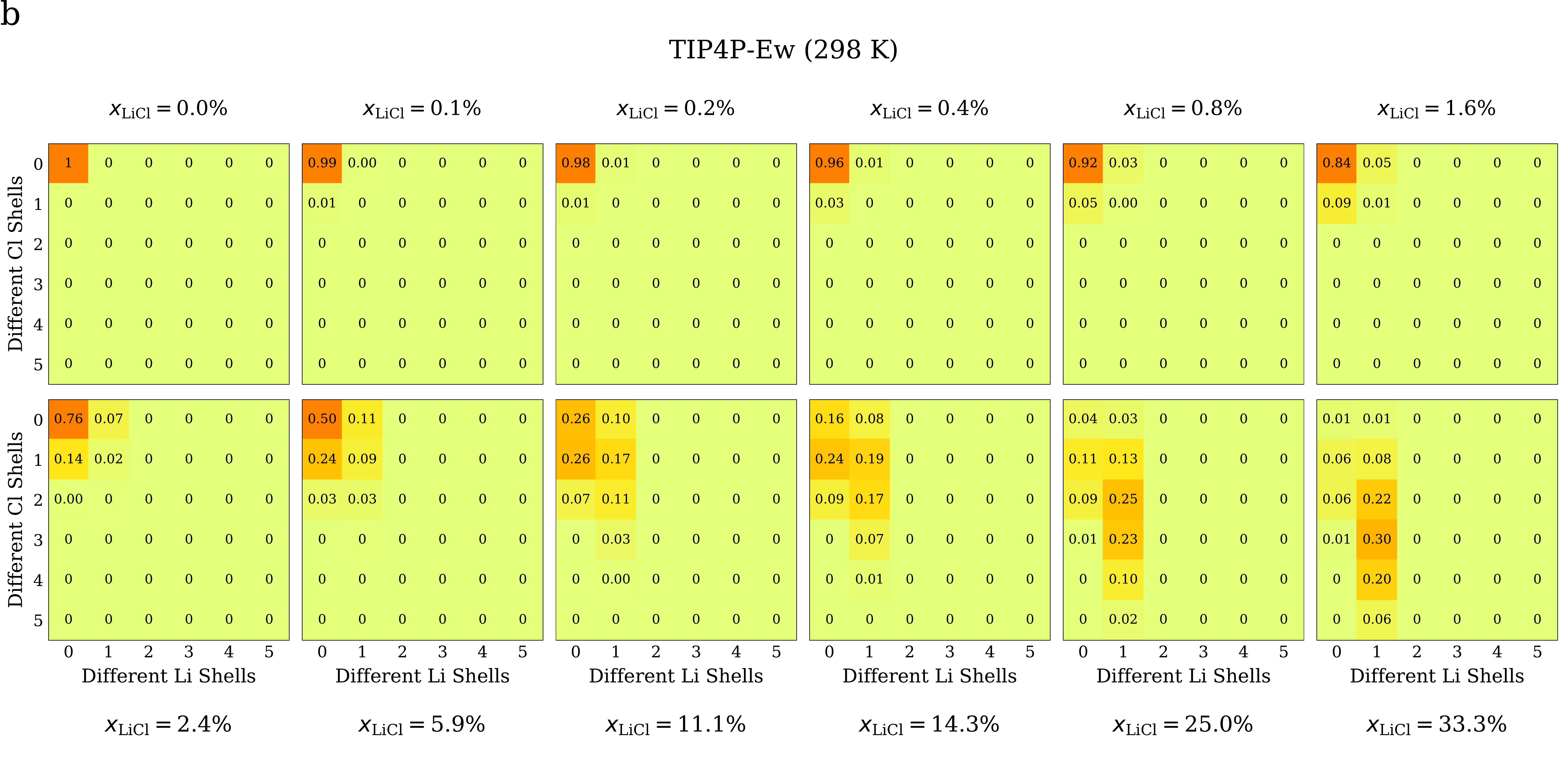}
 \caption{Relative amount of water molecules being part of different first hydration shells at \SI{298}{\kelvin}. The number of different first Li shells increases along the rows, the number of different Cl shells along the columns.
 Each matrix shows the results for a given concentration. Part a shows the results for the TIP4P/2005 water model and part b for the TIP4P-Ew water model.}
 \label{fig:shells-298}
\end{figure*}

The structural order parameter distributions obtained for $T=\SI{240}{\kelvin}$ are shown in Fig.~\ref{fig:rt}b.
For pure water $P(\zeta)$ shifts from $\approx\SI{0}{\nano\meter}$ to $\approx\SI{0.05}{\nano\meter}$ when decreasing the temperature from $\SI{298}{\kelvin}$ to $\SI{240}{\kelvin}$.
This indicates more tetrahedral order at $\SI{240}{\kelvin}$. 
As LiCl is added the distribution again shifts towards zero, while at the same time the non-HB peak starts to grow.
This is also evidenced by an increasing amount of non-HB water molecules in the bottom panel of Fig.~\ref{fig:rt-hb} (black data).
Fig.~\ref{fig:rt}b also reveals that in contrast to \SI{298}{\kelvin}, all partial distributions except $P(\zeta | \text{Li})$ change with concentration.
$P(\zeta | \text{Li})$ does also not change significantly with temperature (see Fig.~\ref{fig:rt}c).
The only effect the temperature decrease has is decreasing the non-HB peak.
This can be seen in Fig.~\ref{fig:rt-hb} (red data), where the increase in non-HB molecules in the Li shells is less at \SI{240}{\kelvin} (bottom panel) than at \SI{298}{\kelvin} (top panel).

\begin{figure*}
 \centering
    \includegraphics[width=0.86\textwidth]{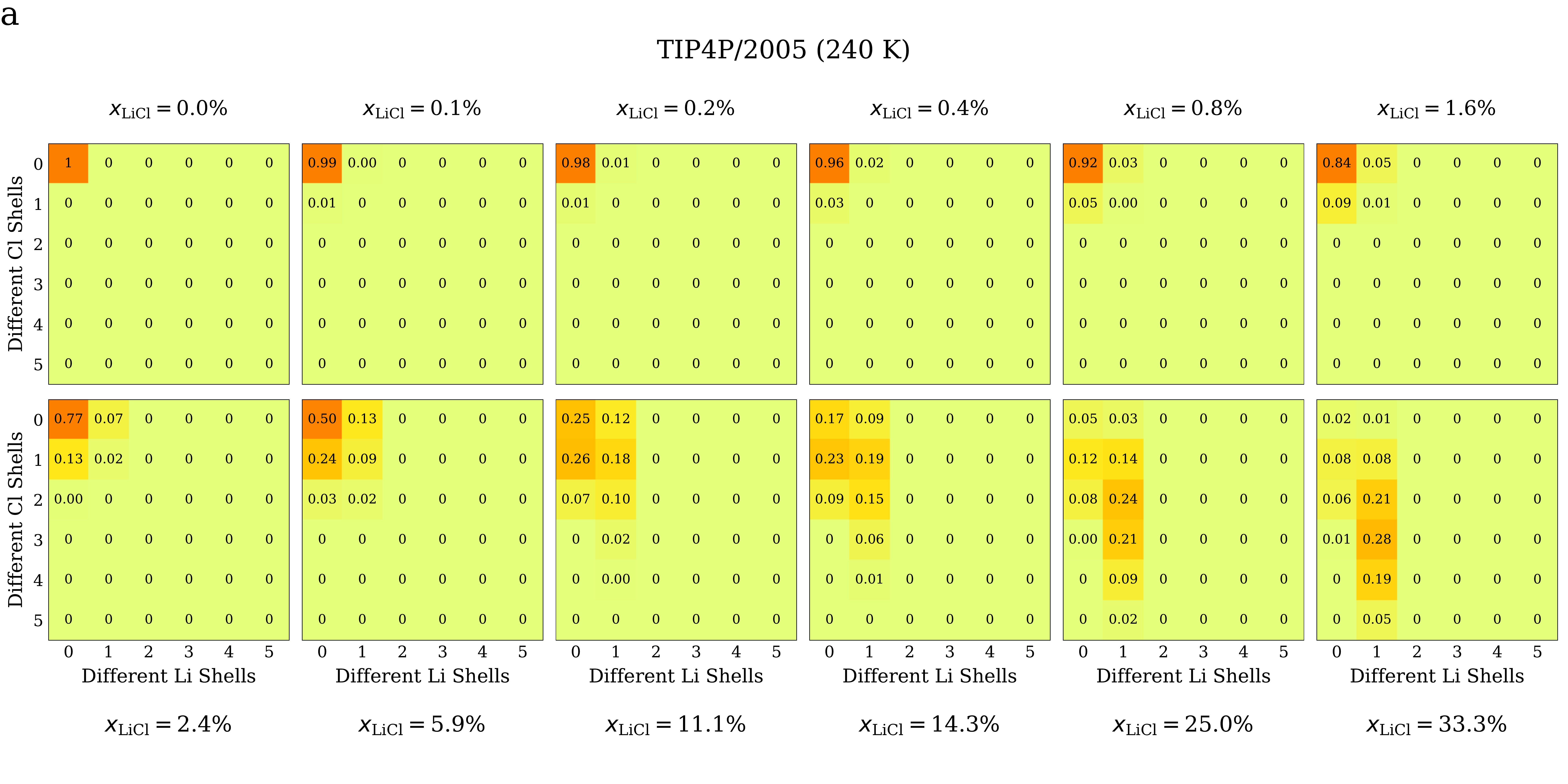}
    \includegraphics[width=0.86\textwidth]{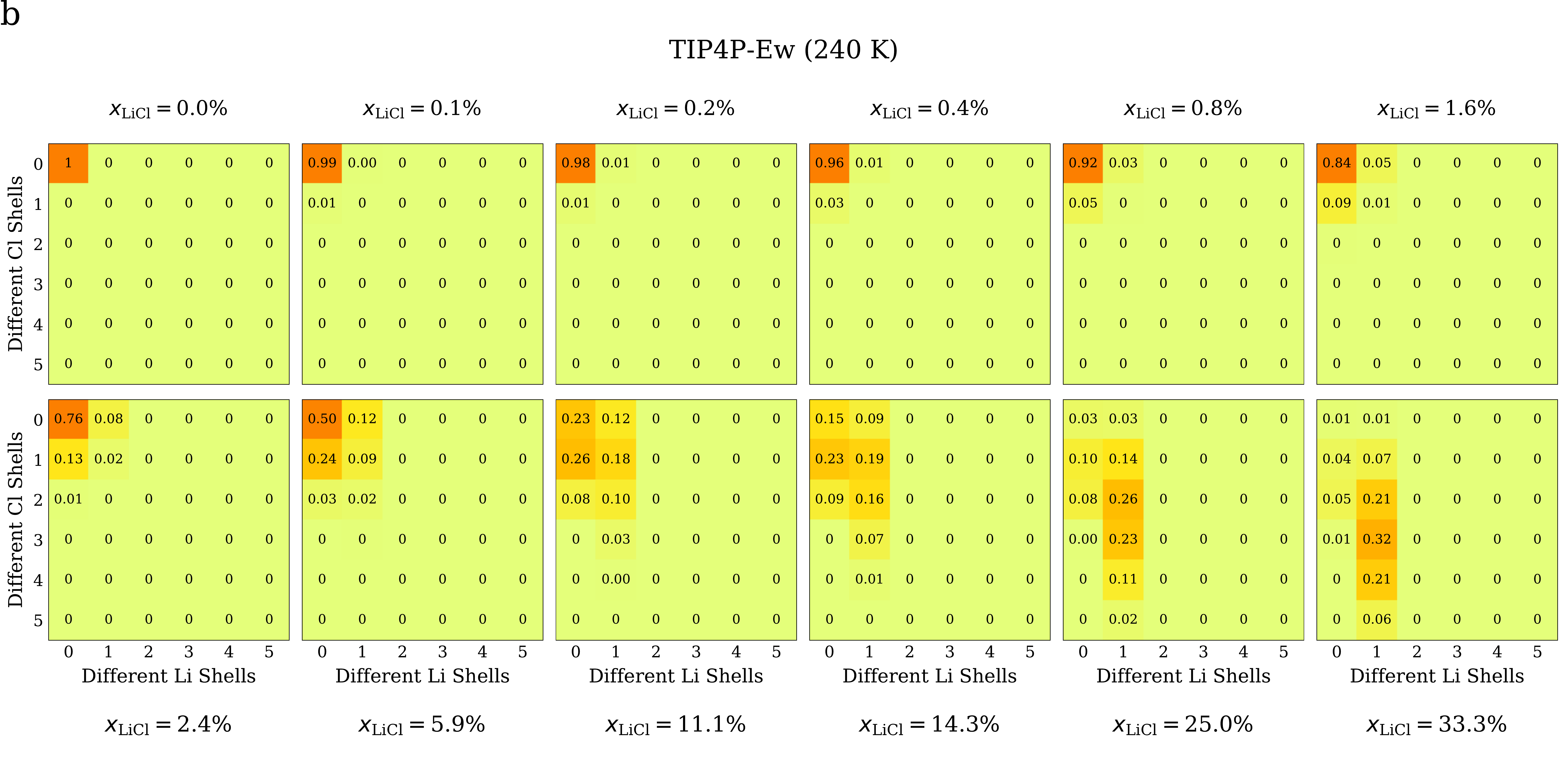}
 \caption{Relative amount of water molecules being part of different first hydration shells at \SI{240}{\kelvin}. The number of different first Li shells increases along the rows, the number of different Cl shells along the columns.
 Each matrix shows the results for a given concentration. Part a shows the results for the TIP4P/2005 water model and part b for the TIP4P-Ew water model.}
 \label{fig:shells-240}
\end{figure*}

$P(\zeta | \text{Cl})$ on the other hand slightly shifts to smaller $\zeta$ as LiCl is added (see Fig.~\ref{fig:rt}b).
Note that these changes are small compared to the changes introduced through the temperature decrease.
At $\SI{240}{\kelvin}$ all $P(\zeta | \text{Cl})$ distributions are shifted to higher $\zeta$ when compared to $\SI{298}{\kelvin}$ indicating a more ordered environment (see Fig.~\ref{fig:rt}c).
As is visible in Figs.~\ref{fig:rt}b and c $P(\zeta | \text{Both})$ behaves very similar as for $T=\SI{298}{\kelvin}$.
The two peaks present show no shifts and just exchange intensity as the  concentration increases. 
This is again reflected in the bottom panel of Fig.~\ref{fig:rt-hb} (green data).

Despite the changes in all partial distributions with concentration the overall changes in $P(\zeta)$ are still dominated by the changes in the $P(i)$s (cf. Fig.~\ref{fig:rt-amounts}b).
$P(\zeta)$ again shifts towards distorted environments, i.e., towards $P(\zeta | \text{Li})$ and $P(\zeta | \text{Both})$, and away from the more ordered $P(\zeta | \text{Cl})$.

Figs~\ref{fig:rt}b shows that for both water models $P(\zeta | \text{Bulk})$ of pure water and low concentrations agrees with the data of Russo and Tanaka obtained for TIP4P/2005 at \SI{240}{\kelvin}~\cite{russo14-natcomm}.
This indicates that $\approx\SI{55}{\percent}$ of the water molecules are in the $\rho$-state, while $\approx\SI{45}{\percent}$ are in the S-state~\cite{russo14-natcomm}.
As $x_\text{LiCl}$ is increased $P(\zeta | \text{Bulk})$ shifts to lower values of $\zeta$.
Such a shift was not visible at \SI{298}{\kelvin}.
An increase in LiCl concentration at \SI{240}{\kelvin} therefore leads to effects on water also beyond the first hydration shell.
This can be explained by the larger ratio of S-state molecules present in pure water at \SI{240}{\kelvin}.
Apparently, S-state molecules also beyond the hydration shells are forced towards the $\rho$-state by the presence of ions.

Fig.~\ref{fig:rt}b also reveals that the shift in $P(\zeta | \text{Bulk})$ is smaller for TIP4P/2005 than for TIP4P-Ew.
For the latter water model $P(\zeta | \text{Bulk})$ almost coincides with $P(\zeta | \text{Bulk})$ for \SI{298}{\kelvin}, while for TIP4P/2005 even at the highest concentrations $P(\zeta | \text{Bulk})$ is centered at higher $\zeta$ than at \SI{298}{\kelvin} (see Fig.~\ref{fig:rt}c).
Other than that the choice of water model has little influence on the structural order parameter.
The results for $P(\zeta)$ are highly similar (see Fig.~\ref{fig:rt}a) and even the relative contributions of the partial distributions to $P(\zeta)$ are almost identical (see top panel of Fig.~\ref{fig:rt-amounts}).

While the partial distributions themselves change, their relative contribution to $P(\zeta)$ is almost independent of temperature as is shown for TIP4P/2005 in the bottom panel of Fig.~\ref{fig:rt-amounts}.
This does not indicate the onset of a phase separation.
If water was about to separate from the solution, the relative contribution of the bulk component to $P(\text{Bulk})$ is expected increase when cooling the system from \SI{298}{\kelvin} to \SI{240}{\kelvin}.

\subsection{Overlapping Hydration Shells}
\label{ssec:ohs}

To provide additional information on the hydration structure we analysed how many water molecules are part of how many first hydration shells.
For \SI{298}{\kelvin} the results are shown in Fig.~\ref{fig:shells-298} and for \SI{240}{\kelvin} the results are shown in Fig.~\ref{fig:shells-240}.
We visualise the results as a $6\times6$ matrix for each $x_\text{LiCl}$ studied.
The columns indicate the number of different first Cl hydration shells the water molecules are part of, while
the rows indicate the number of different first Li hydration shells the water molecules are part of.
In each cell the fraction of respective water molecules is given.
In both figures part a shows the results for TIP4P/2005 and part b the results for TIP4P-Ew.
At both temperatures the two water models yield similar results.
For the trivial pure water case Fig.~\ref{fig:shells-298} shows that all water molecules are part of no hydration shell.
As the concentration is increased to $x_\text{LiCl}=\SI{0.4}{\percent}$ water molecules are either part of the bulk (the majority), or part of a single first hydration shell (either Li or Cl).

Between $x_\text{LiCl}=\SI{0.8}{\percent}$ and $\SI{1.6}{\percent}$ water molecules appear that are part of a first hydration shell of a single Li and a single Cl, signalling the presence of solvent shared or solvent separated ion pairs~\cite{marcus06-crev}.
These water molecules however, amount to less than $\SI{1}{\percent}$ of the total number of water molecules.
At $x_\text{LiCl}\geq\SI{2.4}{\percent}$ some water molecules are part of multiple Cl shells, but none are found that are part of more than one Li shell.
At $x_\text{LiCl}\geq\SI{11.1}{\percent}$ most water molecules are located in hydration shells and at $x_\text{LiCl}\geq\SI{25}{\percent}$ most molecules are part of multiple Cl shells and a single Li shell.   

Notably, the decrease in temperature does not significantly alter this phenomenology.
This reinforces the notion that no phase separation is about to occur at \SI{240}{\kelvin} (see Fig.~\ref{fig:shells-240}).
If this were the case the temperature decrease should lead to an increase in the amount of molecules being part of no hydration shell as well as the amount of molecules being part of multiple hydration shells.

\section{Discussion and Conclusion}

In this study we have presented $N\!PT$-MD simulations of aqueous LiCl solutions.
We considered mole fractions $x_\text{LiCl}$ between \num{0.1} and \SI{33.3}{\percent} and simulated the systems at \SI{298}{\kelvin} and \SI{240}{\kelvin}.
The ions modelled by the Joung-Cheatham parameters~\cite{joung08-jpcb} were hydrated in two different water models, namely TIP4P/2005~\cite{abascal05-jcp} and TIP4P-Ew~\cite{horn04-jcp}.

It was found that at all conditions studied the average potential energy $\left<U\right>$ becomes more negative with increasing amount of LiCl, while the average density $\left<\rho\right>$ increases.
Interestingly, the temperature change from \SI{298}{\kelvin} to \SI{240}{\kelvin} decreases the density at low concentrations, but it increases the density at high concentrations.
Since for pure water a decrease in temperature increases the local tetrahedral order and thereby decreases the density~\cite{russo14-natcomm,handle19-jcp}, this suggests that in the dilute solutions the water molecules can still form a fairly open HB network.
At higher concentration this seems to be prevented by the presence of the ions.
This is in line with the work of Camisasca et al.~\cite{camisasca18-jcp}, who found that high-density structures are enhanced in a $x_\text{LiCl}=\SI{14.3}{\percent}$ solution.
It also resonates well with the often quoted similarity between pressure and concentration (cf., e.g., Ref.~\onlinecite{bachler19-pccp}).

The ion-ion structure shows almost no difference for the two temperatures studied, but the LiCl concentration has quite a large influence.
Clear changes were visible in the second shell of the Li-Cl structure, and 
pre-peaks appeared in both the Li-Li and Cl-Cl structures indicating a tendency towards clusterisation.
This is also manifested as an increase in the number of overlapping first hydration shells with increasing concentration.
The hydratisation structures of the two ionic species behave quite differently.
While the Li-O RDFs are rather insensitive to both temperature and concentration change (especially the first and second shell), the Cl-O RDFs are influenced by both.
Here the concentration increase leads to a shoulder of the first peak and eventually to a double maximum.
Upon temperature change from \SI{298}{\kelvin} to \SI{240}{\kelvin} these changes are even more pronounced.
The reason for this is the significant temperature dependence of the dilute Cl-O RDFs, which show a clearer separation of the first and second peak at \SI{240}{\kelvin}.
Similar to the RDFs, the structural order parameter $\zeta$ of Russo and Tanaka~\cite{russo14-natcomm} indicates a significant difference in the hydration shells of the two ions.
The $\zeta$ distribution of the Li hydration shells again show almost no dependence on temperature and concentration, while the $\zeta$ distribution of the Cl hydration shells show a significant temperature and a slight concentration dependence.
It is also found that Li favours a distorted local environment when compared to the bulk component, while Cl prefers a slightly more ordered local environment.

The water-water structure is influenced by both $x_\text{LiCl}$ and temperature.
Especially the O-O RDFs showed a similar trend as the Cl-O RDFs: the appearance of a shoulder in the first peak and a strong temperature dependence of the dilute systems.
This again indicates that the tetrahedral order is only able to increase in the systems at low concentrations.
The behaviour of the O-H RDFs is consistent with this analysis, since the first peak (representing the HBs) decreases with concentration indicating that the number of hydrogen bonded water molecules decreases.
Here again the temperature decrease leads to an enhancement of the first peak at low LiCl content, while barely showing a change at high concentrations.

The structural order parameter $\zeta$ corroborates the analysis that the addition of ions decreases the tetrahedral order of the system.
Interestingly, it was found that this change is dominated by the introduction of first hydration shells at \SI{298}{\kelvin}.
That is, water beyond the first hydration shells does not significantly alter its local structure.
We note that this finding for \SI{298}{\kelvin} is consistent with femtosecond mid-infrared spectroscopy data that indicated that water beyond the first hydration shell is not influenced by the presence of ions~\cite{omta03-science}, a result that was later contested~\cite{marcus09-crev}.

However, at \SI{240}{\kelvin} also the local water structure beyond the first shells was influenced by the presence of the ions.
This is consistent with the larger effect of LiCl addition found for both the density and the RDFs at \SI{240}{\kelvin}.
The reason for this is that pure water at \SI{240}{\kelvin} is more ordered compared to \SI{298}{\kelvin} as indicated by the different $\zeta$ distributions.
Consequently, LiCl addition, which distorts the HB network, has a larger impact at lower temperature.

At both temperatures the decrease in tetrahedral order is induced by the hydration shells of Li or by overlapping hydration shells.
It is not induced by isolated Cl hydration shells, which appear to favour more ordered surroundings.

Despite the indications for ion-ion clustering the simulation boxes are homogeneous at all studied conditions and none of the quantities analysed show signs of an imminent phase separation.
The coarse-grained density distribution, that exhibits a bimodality during a phase separation in a binary Lennard-Jones mixture~\cite{testard2014intermittent}, was unimodal in all cases.
Similarly, the structural order parameter $\zeta$~\cite{russo14-natcomm} shows also no signs of a phase separation.

In any case, this does not exclude such a scenario, suggested in many experimental studies~\cite{angell68-jcp, angell70-jcp,kanno87-jcp,suzuki00-prl,suzuki02-jcp, bove13-jcp,suzuki13-jcp}.
It could very well be that deeper supercooling is necessary to reveal indications of such a phenomenology in simulations, especially when considering that one experimental study~\cite{bove13-jcp} locates the onset of heterogeneity at \SI{190}{\kelvin}.
Moreover, our results are consistent with the assessment that the immiscibility dome ends between  $x_\text{LiCl}=\SI{10}{\percent}$~\cite{suzuki13-jcp} and  $x_\text{LiCl}=\SI{14.3}{\percent}$~\cite{bove13-jcp}.
For instance, we find that the order parameter $\zeta$ is dominated by overlapping hydration shells when $x_\text{LiCl}$ exceeds \SI{11.1}{\percent}, a finding that is also reflected in the hydration shell statistics.
Here it is revealed that at $x_\text{LiCl}\geq\SI{11.1}{\percent}$ the majority of water molecules is part of more than one hydration shell.
We surmise that once this point is reached it is unlikely for the system to phase separate, since most water molecules are essentially trapped by the ions.

\section*{Electronic Supplementary Information}

Electronic supplementary information (ESI) available:
Data obtained for the standard Lorentz-Berthelot (LB) combination rules are compared to data  obtained for the modified combination rules (MLB).

\begin{acknowledgments}

The author thanks Johannes Bachler and John Russo for helpful discussions and comments.
Financial support by the  Austrian  Science  Fund  FWF  (Erwin  Schrödinger  Fellowship  J3811  N34)  and  the  University  of  Innsbruck  (NWF-Project  282396)  is acknowledged.
The computational results presented here have been achieved (in part) using the LEO HPC infrastructure of the University of Innsbruck.

\end{acknowledgments}

\bibliography{liq-licl.bib}

\begin{thebibliography}{4}
\expandafter\ifx\csname natexlab\endcsname\relax\def\natexlab#1{#1}\fi
\expandafter\ifx\csname bibnamefont\endcsname\relax
  \def\bibnamefont#1{#1}\fi
\expandafter\ifx\csname bibfnamefont\endcsname\relax
  \def\bibfnamefont#1{#1}\fi
\expandafter\ifx\csname citenamefont\endcsname\relax
  \def\citenamefont#1{#1}\fi
\expandafter\ifx\csname url\endcsname\relax
  \def\url#1{\texttt{#1}}\fi
\expandafter\ifx\csname urlprefix\endcsname\relax\def\urlprefix{URL }\fi
\providecommand{\bibinfo}[2]{#2}
\providecommand{\eprint}[2][]{\url{#2}}

\bibitem[{\citenamefont{Aragones et~al.}(2014)\citenamefont{Aragones, Rovere,
  Vega, and Gallo}}]{aragones14-jpcb}
\bibinfo{author}{\bibfnamefont{J.~L.} \bibnamefont{Aragones}},
  \bibinfo{author}{\bibfnamefont{M.}~\bibnamefont{Rovere}},
  \bibinfo{author}{\bibfnamefont{C.}~\bibnamefont{Vega}}, \bibnamefont{and}
  \bibinfo{author}{\bibfnamefont{P.}~\bibnamefont{Gallo}}, \bibinfo{journal}{J.
  Phys. Chem. B} \textbf{\bibinfo{volume}{118}}, \bibinfo{pages}{7680}
  (\bibinfo{year}{2014}).

\bibitem[{\citenamefont{Camisasca et~al.}(2018)\citenamefont{Camisasca,
  De~Marzio, Rovere, and Gallo}}]{camisasca18-jcp}
\bibinfo{author}{\bibfnamefont{G.}~\bibnamefont{Camisasca}},
  \bibinfo{author}{\bibfnamefont{M.}~\bibnamefont{De~Marzio}},
  \bibinfo{author}{\bibfnamefont{M.}~\bibnamefont{Rovere}}, \bibnamefont{and}
  \bibinfo{author}{\bibfnamefont{P.}~\bibnamefont{Gallo}}, \bibinfo{journal}{J.
  Chem. Phys.} \textbf{\bibinfo{volume}{148}}, \bibinfo{pages}{222829}
  (\bibinfo{year}{2018}).

\bibitem[{\citenamefont{Camisasca et~al.}(2020)\citenamefont{Camisasca,
  De~Marzio, Rovere, and Gallo}}]{camisasca20erratum}
\bibinfo{author}{\bibfnamefont{G.}~\bibnamefont{Camisasca}},
  \bibinfo{author}{\bibfnamefont{M.}~\bibnamefont{De~Marzio}},
  \bibinfo{author}{\bibfnamefont{M.}~\bibnamefont{Rovere}}, \bibnamefont{and}
  \bibinfo{author}{\bibfnamefont{P.}~\bibnamefont{Gallo}}, \bibinfo{journal}{J.
  Chem. Phys.} \textbf{\bibinfo{volume}{152}}, \bibinfo{pages}{109901}
  (\bibinfo{year}{2020}).

\bibitem[{\citenamefont{Marcus and Hefter}(2006)}]{marcus06-crev}
\bibinfo{author}{\bibfnamefont{Y.}~\bibnamefont{Marcus}} \bibnamefont{and}
  \bibinfo{author}{\bibfnamefont{G.}~\bibnamefont{Hefter}},
  \bibinfo{journal}{Chem. Rev.} \textbf{\bibinfo{volume}{106}},
  \bibinfo{pages}{4585} (\bibinfo{year}{2006}).

\end{thebibliography}


\begin{thebibliography}{86}
\expandafter\ifx\csname natexlab\endcsname\relax\def\natexlab#1{#1}\fi
\expandafter\ifx\csname bibnamefont\endcsname\relax
  \def\bibnamefont#1{#1}\fi
\expandafter\ifx\csname bibfnamefont\endcsname\relax
  \def\bibfnamefont#1{#1}\fi
\expandafter\ifx\csname citenamefont\endcsname\relax
  \def\citenamefont#1{#1}\fi
\expandafter\ifx\csname url\endcsname\relax
  \def\url#1{\texttt{#1}}\fi
\expandafter\ifx\csname urlprefix\endcsname\relax\def\urlprefix{URL }\fi
\providecommand{\bibinfo}[2]{#2}
\providecommand{\eprint}[2][]{\url{#2}}

\bibitem[{\citenamefont{Franks}(2000)}]{franks2000water}
\bibinfo{author}{\bibfnamefont{F.}~\bibnamefont{Franks}},
  \emph{\bibinfo{title}{Water: a matrix of life}} (\bibinfo{publisher}{Royal
  Society of Chemistry}, \bibinfo{year}{2000}), \bibinfo{edition}{2nd} ed.

\bibitem[{\citenamefont{Finney}(2015)}]{finney2015water}
\bibinfo{author}{\bibfnamefont{J.~L.} \bibnamefont{Finney}},
  \emph{\bibinfo{title}{Water: A very short introduction}}, vol.
  \bibinfo{volume}{440} (\bibinfo{publisher}{Oxford University Press, USA},
  \bibinfo{year}{2015}).

\bibitem[{\citenamefont{Kahler et~al.}(2020)\citenamefont{Kahler, Kamenik,
  Kraml, and Liedl}}]{kahler20-scirep}
\bibinfo{author}{\bibfnamefont{U.}~\bibnamefont{Kahler}},
  \bibinfo{author}{\bibfnamefont{A.~S.} \bibnamefont{Kamenik}},
  \bibinfo{author}{\bibfnamefont{J.}~\bibnamefont{Kraml}}, \bibnamefont{and}
  \bibinfo{author}{\bibfnamefont{K.~R.} \bibnamefont{Liedl}},
  \bibinfo{journal}{Sci. Rep.} \textbf{\bibinfo{volume}{10}},
  \bibinfo{pages}{1} (\bibinfo{year}{2020}).

\bibitem[{\citenamefont{Angell}(2002)}]{angell02-crev}
\bibinfo{author}{\bibfnamefont{C.~A.} \bibnamefont{Angell}},
  \bibinfo{journal}{Chem. Rev.} \textbf{\bibinfo{volume}{102}},
  \bibinfo{pages}{2627} (\bibinfo{year}{2002}).

\bibitem[{\citenamefont{Bachler et~al.}(2019)\citenamefont{Bachler, Handle,
  Giovambattista, and Loerting}}]{bachler19-pccp}
\bibinfo{author}{\bibfnamefont{J.}~\bibnamefont{Bachler}},
  \bibinfo{author}{\bibfnamefont{P.~H.} \bibnamefont{Handle}},
  \bibinfo{author}{\bibfnamefont{N.}~\bibnamefont{Giovambattista}},
  \bibnamefont{and} \bibinfo{author}{\bibfnamefont{T.}~\bibnamefont{Loerting}},
  \bibinfo{journal}{Phys. Chem. Chem. Phys.} \textbf{\bibinfo{volume}{21}},
  \bibinfo{pages}{23238} (\bibinfo{year}{2019}).

\bibitem[{\citenamefont{Debenedetti}(2003)}]{debenedetti03-jpcm}
\bibinfo{author}{\bibfnamefont{P.~G.} \bibnamefont{Debenedetti}},
  \bibinfo{journal}{J. Phys.: Condens. Matter} \textbf{\bibinfo{volume}{15}},
  \bibinfo{pages}{R1669} (\bibinfo{year}{2003}).

\bibitem[{\citenamefont{Mishima and Stanley}(1998)}]{mishima98-nature}
\bibinfo{author}{\bibfnamefont{O.}~\bibnamefont{Mishima}} \bibnamefont{and}
  \bibinfo{author}{\bibfnamefont{H.}~\bibnamefont{Stanley}},
  \bibinfo{journal}{Nature} \textbf{\bibinfo{volume}{396}},
  \bibinfo{pages}{329} (\bibinfo{year}{1998}).

\bibitem[{\citenamefont{Angell}(2004)}]{angell04-arpc}
\bibinfo{author}{\bibfnamefont{C.}~\bibnamefont{Angell}},
  \bibinfo{journal}{Annu. Rev. Phys. Chem.} \textbf{\bibinfo{volume}{55}},
  \bibinfo{pages}{559} (\bibinfo{year}{2004}).

\bibitem[{\citenamefont{Angell}(2008)}]{angell08-science}
\bibinfo{author}{\bibfnamefont{C.~A.} \bibnamefont{Angell}},
  \bibinfo{journal}{Science} \textbf{\bibinfo{volume}{319}},
  \bibinfo{pages}{582} (\bibinfo{year}{2008}).

\bibitem[{\citenamefont{Gallo et~al.}(2016)\citenamefont{Gallo, Amann-Winkel,
  Angell, Anisimov, Caupin, Chakravarty, Lascaris, Loerting, Panagiotopoulos,
  Russo et~al.}}]{gallo16-crev}
\bibinfo{author}{\bibfnamefont{P.}~\bibnamefont{Gallo}},
  \bibinfo{author}{\bibfnamefont{K.}~\bibnamefont{Amann-Winkel}},
  \bibinfo{author}{\bibfnamefont{C.~A.} \bibnamefont{Angell}},
  \bibinfo{author}{\bibfnamefont{M.~A.} \bibnamefont{Anisimov}},
  \bibinfo{author}{\bibfnamefont{F.}~\bibnamefont{Caupin}},
  \bibinfo{author}{\bibfnamefont{C.}~\bibnamefont{Chakravarty}},
  \bibinfo{author}{\bibfnamefont{E.}~\bibnamefont{Lascaris}},
  \bibinfo{author}{\bibfnamefont{T.}~\bibnamefont{Loerting}},
  \bibinfo{author}{\bibfnamefont{A.~Z.} \bibnamefont{Panagiotopoulos}},
  \bibinfo{author}{\bibfnamefont{J.}~\bibnamefont{Russo}},
  \bibnamefont{et~al.}, \bibinfo{journal}{Chem. Rev.}
  \textbf{\bibinfo{volume}{116}}, \bibinfo{pages}{7463} (\bibinfo{year}{2016}).

\bibitem[{\citenamefont{Handle et~al.}(2017)\citenamefont{Handle, Loerting, and
  Sciortino}}]{handle2017supercooled}
\bibinfo{author}{\bibfnamefont{P.~H.} \bibnamefont{Handle}},
  \bibinfo{author}{\bibfnamefont{T.}~\bibnamefont{Loerting}}, \bibnamefont{and}
  \bibinfo{author}{\bibfnamefont{F.}~\bibnamefont{Sciortino}},
  \bibinfo{journal}{Proc. Natl. Acad. Sci. U.S.A.}
  \textbf{\bibinfo{volume}{114}}, \bibinfo{pages}{13336}
  (\bibinfo{year}{2017}).

\bibitem[{\citenamefont{Anisimov et~al.}(2018)\citenamefont{Anisimov,
  Du{\v{s}}ka, Caupin, Amrhein, Rosenbaum, and Sadus}}]{anisimov18-prx}
\bibinfo{author}{\bibfnamefont{M.~A.} \bibnamefont{Anisimov}},
  \bibinfo{author}{\bibfnamefont{M.}~\bibnamefont{Du{\v{s}}ka}},
  \bibinfo{author}{\bibfnamefont{F.}~\bibnamefont{Caupin}},
  \bibinfo{author}{\bibfnamefont{L.~E.} \bibnamefont{Amrhein}},
  \bibinfo{author}{\bibfnamefont{A.}~\bibnamefont{Rosenbaum}},
  \bibnamefont{and} \bibinfo{author}{\bibfnamefont{R.~J.} \bibnamefont{Sadus}},
  \bibinfo{journal}{Phys. Rev. X} \textbf{\bibinfo{volume}{8}},
  \bibinfo{pages}{011004} (\bibinfo{year}{2018}).

\bibitem[{\citenamefont{Poole et~al.}(1992)\citenamefont{Poole, Sciortino,
  Essmann, and Stanley}}]{poole92-nature}
\bibinfo{author}{\bibfnamefont{P.~H.} \bibnamefont{Poole}},
  \bibinfo{author}{\bibfnamefont{F.}~\bibnamefont{Sciortino}},
  \bibinfo{author}{\bibfnamefont{U.}~\bibnamefont{Essmann}}, \bibnamefont{and}
  \bibinfo{author}{\bibfnamefont{H.}~\bibnamefont{Stanley}},
  \bibinfo{journal}{Nature} \textbf{\bibinfo{volume}{360}},
  \bibinfo{pages}{324} (\bibinfo{year}{1992}).

\bibitem[{\citenamefont{Stillinger and Rahman}(1974)}]{stillinger74-jcp}
\bibinfo{author}{\bibfnamefont{F.~H.} \bibnamefont{Stillinger}}
  \bibnamefont{and} \bibinfo{author}{\bibfnamefont{A.}~\bibnamefont{Rahman}},
  \bibinfo{journal}{J. Chem. Phys.} \textbf{\bibinfo{volume}{60}},
  \bibinfo{pages}{1545} (\bibinfo{year}{1974}).

\bibitem[{\citenamefont{Poole et~al.}(2005)\citenamefont{Poole, Saika-Voivod,
  and Sciortino}}]{poole05-jpcm}
\bibinfo{author}{\bibfnamefont{P.~H.} \bibnamefont{Poole}},
  \bibinfo{author}{\bibfnamefont{I.}~\bibnamefont{Saika-Voivod}},
  \bibnamefont{and}
  \bibinfo{author}{\bibfnamefont{F.}~\bibnamefont{Sciortino}},
  \bibinfo{journal}{J. Phys.: Condens. Matter} \textbf{\bibinfo{volume}{17}},
  \bibinfo{pages}{L431} (\bibinfo{year}{2005}).

\bibitem[{\citenamefont{Palmer et~al.}(2014)\citenamefont{Palmer, Martelli,
  Liu, Car, Panagiotopoulos, and Debenedetti}}]{palmer14-nature}
\bibinfo{author}{\bibfnamefont{J.~C.} \bibnamefont{Palmer}},
  \bibinfo{author}{\bibfnamefont{F.}~\bibnamefont{Martelli}},
  \bibinfo{author}{\bibfnamefont{Y.}~\bibnamefont{Liu}},
  \bibinfo{author}{\bibfnamefont{R.}~\bibnamefont{Car}},
  \bibinfo{author}{\bibfnamefont{A.~Z.} \bibnamefont{Panagiotopoulos}},
  \bibnamefont{and} \bibinfo{author}{\bibfnamefont{P.~G.}
  \bibnamefont{Debenedetti}}, \bibinfo{journal}{Nature}
  \textbf{\bibinfo{volume}{510}}, \bibinfo{pages}{385} (\bibinfo{year}{2014}).

\bibitem[{\citenamefont{Paschek et~al.}(2008)\citenamefont{Paschek,
  R{\"u}ppert, and Geiger}}]{paschek08-cpc}
\bibinfo{author}{\bibfnamefont{D.}~\bibnamefont{Paschek}},
  \bibinfo{author}{\bibfnamefont{A.}~\bibnamefont{R{\"u}ppert}},
  \bibnamefont{and} \bibinfo{author}{\bibfnamefont{A.}~\bibnamefont{Geiger}},
  \bibinfo{journal}{ChemPhysChem} \textbf{\bibinfo{volume}{9}},
  \bibinfo{pages}{2737} (\bibinfo{year}{2008}).

\bibitem[{\citenamefont{Corradini et~al.}(2010)\citenamefont{Corradini, Rovere,
  and Gallo}}]{corradini10-jcp}
\bibinfo{author}{\bibfnamefont{D.}~\bibnamefont{Corradini}},
  \bibinfo{author}{\bibfnamefont{M.}~\bibnamefont{Rovere}}, \bibnamefont{and}
  \bibinfo{author}{\bibfnamefont{P.}~\bibnamefont{Gallo}}, \bibinfo{journal}{J.
  Chem. Phys.} \textbf{\bibinfo{volume}{132}}, \bibinfo{pages}{134508}
  (\bibinfo{year}{2010}).

\bibitem[{\citenamefont{Abascal and Vega}(2010)}]{abascal10-jcp}
\bibinfo{author}{\bibfnamefont{J.~L.} \bibnamefont{Abascal}} \bibnamefont{and}
  \bibinfo{author}{\bibfnamefont{C.}~\bibnamefont{Vega}}, \bibinfo{journal}{J.
  Chem. Phys.} \textbf{\bibinfo{volume}{133}}, \bibinfo{pages}{234502}
  (\bibinfo{year}{2010}).

\bibitem[{\citenamefont{Wikfeldt et~al.}(2011)\citenamefont{Wikfeldt, Nilsson,
  and Pettersson}}]{wikfeldt2011spatially}
\bibinfo{author}{\bibfnamefont{K.}~\bibnamefont{Wikfeldt}},
  \bibinfo{author}{\bibfnamefont{A.}~\bibnamefont{Nilsson}}, \bibnamefont{and}
  \bibinfo{author}{\bibfnamefont{L.~G.} \bibnamefont{Pettersson}},
  \bibinfo{journal}{Phys. Chem. Chem. Phys.} \textbf{\bibinfo{volume}{13}},
  \bibinfo{pages}{19918} (\bibinfo{year}{2011}).

\bibitem[{\citenamefont{Sumi and Sekino}(2013)}]{sumi13-rscadv}
\bibinfo{author}{\bibfnamefont{T.}~\bibnamefont{Sumi}} \bibnamefont{and}
  \bibinfo{author}{\bibfnamefont{H.}~\bibnamefont{Sekino}},
  \bibinfo{journal}{RSC Adv.} \textbf{\bibinfo{volume}{3}},
  \bibinfo{pages}{12743} (\bibinfo{year}{2013}).

\bibitem[{\citenamefont{Yagasaki et~al.}(2014)\citenamefont{Yagasaki,
  Matsumoto, and Tanaka}}]{yagasaki14-pre}
\bibinfo{author}{\bibfnamefont{T.}~\bibnamefont{Yagasaki}},
  \bibinfo{author}{\bibfnamefont{M.}~\bibnamefont{Matsumoto}},
  \bibnamefont{and} \bibinfo{author}{\bibfnamefont{H.}~\bibnamefont{Tanaka}},
  \bibinfo{journal}{Phys. Rev. E} \textbf{\bibinfo{volume}{89}},
  \bibinfo{pages}{020301} (\bibinfo{year}{2014}).

\bibitem[{\citenamefont{Biddle et~al.}(2017)\citenamefont{Biddle, Singh,
  Sparano, Ricci, Gonz{\'a}lez, Valeriani, Abascal, Debenedetti, Anisimov, and
  Caupin}}]{biddle17-jcp}
\bibinfo{author}{\bibfnamefont{J.~W.} \bibnamefont{Biddle}},
  \bibinfo{author}{\bibfnamefont{R.~S.} \bibnamefont{Singh}},
  \bibinfo{author}{\bibfnamefont{E.~M.} \bibnamefont{Sparano}},
  \bibinfo{author}{\bibfnamefont{F.}~\bibnamefont{Ricci}},
  \bibinfo{author}{\bibfnamefont{M.~A.} \bibnamefont{Gonz{\'a}lez}},
  \bibinfo{author}{\bibfnamefont{C.}~\bibnamefont{Valeriani}},
  \bibinfo{author}{\bibfnamefont{J.~L.} \bibnamefont{Abascal}},
  \bibinfo{author}{\bibfnamefont{P.~G.} \bibnamefont{Debenedetti}},
  \bibinfo{author}{\bibfnamefont{M.~A.} \bibnamefont{Anisimov}},
  \bibnamefont{and} \bibinfo{author}{\bibfnamefont{F.}~\bibnamefont{Caupin}},
  \bibinfo{journal}{J. Chem. Phys.} \textbf{\bibinfo{volume}{146}},
  \bibinfo{pages}{034502} (\bibinfo{year}{2017}).

\bibitem[{\citenamefont{Handle and Sciortino}(2018)}]{handle18-jcp}
\bibinfo{author}{\bibfnamefont{P.~H.} \bibnamefont{Handle}} \bibnamefont{and}
  \bibinfo{author}{\bibfnamefont{F.}~\bibnamefont{Sciortino}},
  \bibinfo{journal}{J. Chem. Phys.} \textbf{\bibinfo{volume}{148}},
  \bibinfo{pages}{134505} (\bibinfo{year}{2018}).

\bibitem[{\citenamefont{Palmer et~al.}(2018)\citenamefont{Palmer, Poole,
  Sciortino, and Debenedetti}}]{palmer18-crev}
\bibinfo{author}{\bibfnamefont{J.~C.} \bibnamefont{Palmer}},
  \bibinfo{author}{\bibfnamefont{P.~H.} \bibnamefont{Poole}},
  \bibinfo{author}{\bibfnamefont{F.}~\bibnamefont{Sciortino}},
  \bibnamefont{and} \bibinfo{author}{\bibfnamefont{P.~G.}
  \bibnamefont{Debenedetti}}, \bibinfo{journal}{Chem. Rev.}
  \textbf{\bibinfo{volume}{118}}, \bibinfo{pages}{9129} (\bibinfo{year}{2018}).

\bibitem[{\citenamefont{Kim et~al.}(2017)\citenamefont{Kim, Sp{\"a}h, Pathak,
  Perakis, Mariedahl, Amann-Winkel, Sellberg, Lee, Kim, Park
  et~al.}}]{kim17-science}
\bibinfo{author}{\bibfnamefont{K.~H.} \bibnamefont{Kim}},
  \bibinfo{author}{\bibfnamefont{A.}~\bibnamefont{Sp{\"a}h}},
  \bibinfo{author}{\bibfnamefont{H.}~\bibnamefont{Pathak}},
  \bibinfo{author}{\bibfnamefont{F.}~\bibnamefont{Perakis}},
  \bibinfo{author}{\bibfnamefont{D.}~\bibnamefont{Mariedahl}},
  \bibinfo{author}{\bibfnamefont{K.}~\bibnamefont{Amann-Winkel}},
  \bibinfo{author}{\bibfnamefont{J.~A.} \bibnamefont{Sellberg}},
  \bibinfo{author}{\bibfnamefont{J.~H.} \bibnamefont{Lee}},
  \bibinfo{author}{\bibfnamefont{S.}~\bibnamefont{Kim}},
  \bibinfo{author}{\bibfnamefont{J.}~\bibnamefont{Park}}, \bibnamefont{et~al.},
  \bibinfo{journal}{Science} \textbf{\bibinfo{volume}{358}},
  \bibinfo{pages}{1589} (\bibinfo{year}{2017}).

\bibitem[{\citenamefont{Holten et~al.}(2017)\citenamefont{Holten, Qiu,
  Guillerm, Wilke, Ricka, Frenz, and Caupin}}]{holten2017compressibility}
\bibinfo{author}{\bibfnamefont{V.}~\bibnamefont{Holten}},
  \bibinfo{author}{\bibfnamefont{C.}~\bibnamefont{Qiu}},
  \bibinfo{author}{\bibfnamefont{E.}~\bibnamefont{Guillerm}},
  \bibinfo{author}{\bibfnamefont{M.}~\bibnamefont{Wilke}},
  \bibinfo{author}{\bibfnamefont{J.}~\bibnamefont{Ricka}},
  \bibinfo{author}{\bibfnamefont{M.}~\bibnamefont{Frenz}}, \bibnamefont{and}
  \bibinfo{author}{\bibfnamefont{F.}~\bibnamefont{Caupin}},
  \bibinfo{journal}{J. Phys. Chem. Lett.} \textbf{\bibinfo{volume}{8}},
  \bibinfo{pages}{5519} (\bibinfo{year}{2017}).

\bibitem[{\citenamefont{Mishima et~al.}(1984)\citenamefont{Mishima, Calvert,
  and Whalley}}]{mishima84-nature}
\bibinfo{author}{\bibfnamefont{O.}~\bibnamefont{Mishima}},
  \bibinfo{author}{\bibfnamefont{L.~D.} \bibnamefont{Calvert}},
  \bibnamefont{and} \bibinfo{author}{\bibfnamefont{E.}~\bibnamefont{Whalley}},
  \bibinfo{journal}{Nature} \textbf{\bibinfo{volume}{310}},
  \bibinfo{pages}{393} (\bibinfo{year}{1984}).

\bibitem[{\citenamefont{Mishima et~al.}(1985)\citenamefont{Mishima, Calvert,
  and Whalley}}]{mishima85-nature}
\bibinfo{author}{\bibfnamefont{O.}~\bibnamefont{Mishima}},
  \bibinfo{author}{\bibfnamefont{L.~D.} \bibnamefont{Calvert}},
  \bibnamefont{and} \bibinfo{author}{\bibfnamefont{E.}~\bibnamefont{Whalley}},
  \bibinfo{journal}{Nature} \textbf{\bibinfo{volume}{314}}, \bibinfo{pages}{76}
  (\bibinfo{year}{1985}).

\bibitem[{\citenamefont{Mishima}(1994)}]{mishima94-jcp}
\bibinfo{author}{\bibfnamefont{O.}~\bibnamefont{Mishima}}, \bibinfo{journal}{J.
  Chem. Phys.} \textbf{\bibinfo{volume}{100}}, \bibinfo{pages}{5910}
  (\bibinfo{year}{1994}).

\bibitem[{\citenamefont{Loerting et~al.}(2001)\citenamefont{Loerting, Salzmann,
  Kohl, Mayer, and Hallbrucker}}]{loerting01-pccp}
\bibinfo{author}{\bibfnamefont{T.}~\bibnamefont{Loerting}},
  \bibinfo{author}{\bibfnamefont{C.}~\bibnamefont{Salzmann}},
  \bibinfo{author}{\bibfnamefont{I.}~\bibnamefont{Kohl}},
  \bibinfo{author}{\bibfnamefont{E.}~\bibnamefont{Mayer}}, \bibnamefont{and}
  \bibinfo{author}{\bibfnamefont{A.}~\bibnamefont{Hallbrucker}},
  \bibinfo{journal}{Phys. Chem. Chem. Phys.} \textbf{\bibinfo{volume}{3}},
  \bibinfo{pages}{5355} (\bibinfo{year}{2001}).

\bibitem[{\citenamefont{Klotz et~al.}(2005)\citenamefont{Klotz, Str\"{a}ssle,
  Nelmes, Loveday, Hamel, Rousse, Canny, Chervin, and Saitta}}]{klotz05-prl}
\bibinfo{author}{\bibfnamefont{S.}~\bibnamefont{Klotz}},
  \bibinfo{author}{\bibfnamefont{T.}~\bibnamefont{Str\"{a}ssle}},
  \bibinfo{author}{\bibfnamefont{R.~J.} \bibnamefont{Nelmes}},
  \bibinfo{author}{\bibfnamefont{J.~S.} \bibnamefont{Loveday}},
  \bibinfo{author}{\bibfnamefont{G.}~\bibnamefont{Hamel}},
  \bibinfo{author}{\bibfnamefont{G.}~\bibnamefont{Rousse}},
  \bibinfo{author}{\bibfnamefont{B.}~\bibnamefont{Canny}},
  \bibinfo{author}{\bibfnamefont{J.~C.} \bibnamefont{Chervin}},
  \bibnamefont{and} \bibinfo{author}{\bibfnamefont{A.~M.}
  \bibnamefont{Saitta}}, \bibinfo{journal}{Phys. Rev. Lett.}
  \textbf{\bibinfo{volume}{94}}, \bibinfo{pages}{025506}
  (\bibinfo{year}{2005}).

\bibitem[{\citenamefont{Winkel et~al.}(2008)\citenamefont{Winkel, Elsaesser,
  Mayer, and Loerting}}]{winkel08-jcp}
\bibinfo{author}{\bibfnamefont{K.}~\bibnamefont{Winkel}},
  \bibinfo{author}{\bibfnamefont{M.~S.} \bibnamefont{Elsaesser}},
  \bibinfo{author}{\bibfnamefont{E.}~\bibnamefont{Mayer}}, \bibnamefont{and}
  \bibinfo{author}{\bibfnamefont{T.}~\bibnamefont{Loerting}},
  \bibinfo{journal}{J. Chem. Phys.} \textbf{\bibinfo{volume}{128}},
  \bibinfo{pages}{044510} (\bibinfo{year}{2008}).

\bibitem[{\citenamefont{Loerting et~al.}(2011)\citenamefont{Loerting, Winkel,
  Seidl, Bauer, Mitterdorfer, Handle, Salzmann, Mayer, Finney, and
  Bowron}}]{loerting11-pccp}
\bibinfo{author}{\bibfnamefont{T.}~\bibnamefont{Loerting}},
  \bibinfo{author}{\bibfnamefont{K.}~\bibnamefont{Winkel}},
  \bibinfo{author}{\bibfnamefont{M.}~\bibnamefont{Seidl}},
  \bibinfo{author}{\bibfnamefont{M.}~\bibnamefont{Bauer}},
  \bibinfo{author}{\bibfnamefont{C.}~\bibnamefont{Mitterdorfer}},
  \bibinfo{author}{\bibfnamefont{P.~H.} \bibnamefont{Handle}},
  \bibinfo{author}{\bibfnamefont{C.~G.} \bibnamefont{Salzmann}},
  \bibinfo{author}{\bibfnamefont{E.}~\bibnamefont{Mayer}},
  \bibinfo{author}{\bibfnamefont{J.~L.} \bibnamefont{Finney}},
  \bibnamefont{and} \bibinfo{author}{\bibfnamefont{D.~T.}
  \bibnamefont{Bowron}}, \bibinfo{journal}{Phys. Chem. Chem. Phys.}
  \textbf{\bibinfo{volume}{13}}, \bibinfo{pages}{8783} (\bibinfo{year}{2011}).

\bibitem[{\citenamefont{Handle and
  Loerting}(2018{\natexlab{a}})}]{handle2018experimental1}
\bibinfo{author}{\bibfnamefont{P.~H.} \bibnamefont{Handle}} \bibnamefont{and}
  \bibinfo{author}{\bibfnamefont{T.}~\bibnamefont{Loerting}},
  \bibinfo{journal}{J. Chem. Phys.} \textbf{\bibinfo{volume}{148}},
  \bibinfo{pages}{124508} (\bibinfo{year}{2018}{\natexlab{a}}).

\bibitem[{\citenamefont{Handle and
  Loerting}(2018{\natexlab{b}})}]{handle2018experimental2}
\bibinfo{author}{\bibfnamefont{P.~H.} \bibnamefont{Handle}} \bibnamefont{and}
  \bibinfo{author}{\bibfnamefont{T.}~\bibnamefont{Loerting}},
  \bibinfo{journal}{J. Chem. Phys.} \textbf{\bibinfo{volume}{148}},
  \bibinfo{pages}{124509} (\bibinfo{year}{2018}{\natexlab{b}}).

\bibitem[{\citenamefont{Mariedahl et~al.}(2018)\citenamefont{Mariedahl,
  Perakis, Sp\"{a}h, Pathak, Kim, Camisasca, Schlesinger, Benmore, Pettersson,
  Nilsson et~al.}}]{mariedahl2018x}
\bibinfo{author}{\bibfnamefont{D.}~\bibnamefont{Mariedahl}},
  \bibinfo{author}{\bibfnamefont{F.}~\bibnamefont{Perakis}},
  \bibinfo{author}{\bibfnamefont{A.}~\bibnamefont{Sp\"{a}h}},
  \bibinfo{author}{\bibfnamefont{H.}~\bibnamefont{Pathak}},
  \bibinfo{author}{\bibfnamefont{K.~H.} \bibnamefont{Kim}},
  \bibinfo{author}{\bibfnamefont{G.}~\bibnamefont{Camisasca}},
  \bibinfo{author}{\bibfnamefont{D.}~\bibnamefont{Schlesinger}},
  \bibinfo{author}{\bibfnamefont{C.}~\bibnamefont{Benmore}},
  \bibinfo{author}{\bibfnamefont{L.~G.~M.} \bibnamefont{Pettersson}},
  \bibinfo{author}{\bibfnamefont{A.}~\bibnamefont{Nilsson}},
  \bibnamefont{et~al.}, \bibinfo{journal}{J. Phys. Chem. B}
  \textbf{\bibinfo{volume}{122}}, \bibinfo{pages}{7616} (\bibinfo{year}{2018}).

\bibitem[{\citenamefont{Angell and Sare}(1968)}]{angell68-jcp}
\bibinfo{author}{\bibfnamefont{C.}~\bibnamefont{Angell}} \bibnamefont{and}
  \bibinfo{author}{\bibfnamefont{E.}~\bibnamefont{Sare}}, \bibinfo{journal}{J.
  Chem. Phys.} \textbf{\bibinfo{volume}{49}}, \bibinfo{pages}{4713}
  (\bibinfo{year}{1968}).

\bibitem[{\citenamefont{Angell and Sare}(1970)}]{angell70-jcp}
\bibinfo{author}{\bibfnamefont{C.}~\bibnamefont{Angell}} \bibnamefont{and}
  \bibinfo{author}{\bibfnamefont{E.}~\bibnamefont{Sare}}, \bibinfo{journal}{J.
  Chem. Phys.} \textbf{\bibinfo{volume}{52}}, \bibinfo{pages}{1058}
  (\bibinfo{year}{1970}).

\bibitem[{\citenamefont{Kanno}(1987)}]{kanno87-jcp}
\bibinfo{author}{\bibfnamefont{H.}~\bibnamefont{Kanno}}, \bibinfo{journal}{J.
  Chem. Phys.} \textbf{\bibinfo{volume}{91}}, \bibinfo{pages}{1967}
  (\bibinfo{year}{1987}).

\bibitem[{\citenamefont{Suzuki and Mishima}(2000)}]{suzuki00-prl}
\bibinfo{author}{\bibfnamefont{Y.}~\bibnamefont{Suzuki}} \bibnamefont{and}
  \bibinfo{author}{\bibfnamefont{O.}~\bibnamefont{Mishima}},
  \bibinfo{journal}{Phys. Rev. Lett.} \textbf{\bibinfo{volume}{85}},
  \bibinfo{pages}{1322} (\bibinfo{year}{2000}).

\bibitem[{\citenamefont{Suzuki and Mishima}(2002)}]{suzuki02-jcp}
\bibinfo{author}{\bibfnamefont{Y.}~\bibnamefont{Suzuki}} \bibnamefont{and}
  \bibinfo{author}{\bibfnamefont{O.}~\bibnamefont{Mishima}},
  \bibinfo{journal}{J. Chem. Phys.} \textbf{\bibinfo{volume}{117}},
  \bibinfo{pages}{1673} (\bibinfo{year}{2002}).

\bibitem[{\citenamefont{Mishima}(2004)}]{mishima04-jcp}
\bibinfo{author}{\bibfnamefont{O.}~\bibnamefont{Mishima}}, \bibinfo{journal}{J.
  Chem. Phys.} \textbf{\bibinfo{volume}{121}}, \bibinfo{pages}{3161}
  (\bibinfo{year}{2004}).

\bibitem[{\citenamefont{Mishima}(2005)}]{mishima05-jcp}
\bibinfo{author}{\bibfnamefont{O.}~\bibnamefont{Mishima}}, \bibinfo{journal}{J.
  Chem. Phys.} \textbf{\bibinfo{volume}{123}}, \bibinfo{pages}{154506}
  (\bibinfo{year}{2005}).

\bibitem[{\citenamefont{Mishima}(2007)}]{mishima07-jcp}
\bibinfo{author}{\bibfnamefont{O.}~\bibnamefont{Mishima}}, \bibinfo{journal}{J.
  Chem. Phys.} \textbf{\bibinfo{volume}{126}}, \bibinfo{pages}{244507}
  (\bibinfo{year}{2007}).

\bibitem[{\citenamefont{Mishima}(2011)}]{mishima11-jpcb}
\bibinfo{author}{\bibfnamefont{O.}~\bibnamefont{Mishima}}, \bibinfo{journal}{J.
  Phys. Chem. B} \textbf{\bibinfo{volume}{115}}, \bibinfo{pages}{14064}
  (\bibinfo{year}{2011}).

\bibitem[{\citenamefont{Suzuki and Tominaga}(2011)}]{suzuki11-jcp}
\bibinfo{author}{\bibfnamefont{Y.}~\bibnamefont{Suzuki}} \bibnamefont{and}
  \bibinfo{author}{\bibfnamefont{Y.}~\bibnamefont{Tominaga}},
  \bibinfo{journal}{J. Chem. Phys.} \textbf{\bibinfo{volume}{134}},
  \bibinfo{pages}{244511} (\bibinfo{year}{2011}).

\bibitem[{\citenamefont{Bove et~al.}(2011)\citenamefont{Bove, Klotz, Philippe,
  and Saitta}}]{bove11-prl}
\bibinfo{author}{\bibfnamefont{L.~E.} \bibnamefont{Bove}},
  \bibinfo{author}{\bibfnamefont{S.}~\bibnamefont{Klotz}},
  \bibinfo{author}{\bibfnamefont{J.}~\bibnamefont{Philippe}}, \bibnamefont{and}
  \bibinfo{author}{\bibfnamefont{A.~M.} \bibnamefont{Saitta}},
  \bibinfo{journal}{Phys. Rev. Lett.} \textbf{\bibinfo{volume}{106}},
  \bibinfo{pages}{125701} (\bibinfo{year}{2011}).

\bibitem[{\citenamefont{Winkel et~al.}(2011)\citenamefont{Winkel, Seidl,
  Loerting, Bove, Imberti, Molinero, Bruni, Mancinelli, and
  Ricci}}]{winkel11-jcp}
\bibinfo{author}{\bibfnamefont{K.}~\bibnamefont{Winkel}},
  \bibinfo{author}{\bibfnamefont{M.}~\bibnamefont{Seidl}},
  \bibinfo{author}{\bibfnamefont{T.}~\bibnamefont{Loerting}},
  \bibinfo{author}{\bibfnamefont{L.}~\bibnamefont{Bove}},
  \bibinfo{author}{\bibfnamefont{S.}~\bibnamefont{Imberti}},
  \bibinfo{author}{\bibfnamefont{V.}~\bibnamefont{Molinero}},
  \bibinfo{author}{\bibfnamefont{F.}~\bibnamefont{Bruni}},
  \bibinfo{author}{\bibfnamefont{R.}~\bibnamefont{Mancinelli}},
  \bibnamefont{and} \bibinfo{author}{\bibfnamefont{M.}~\bibnamefont{Ricci}},
  \bibinfo{journal}{J. Chem. Phys.} \textbf{\bibinfo{volume}{134}},
  \bibinfo{pages}{024515} (\bibinfo{year}{2011}).

\bibitem[{\citenamefont{Bove et~al.}(2013)\citenamefont{Bove, Dreyfus, Torre,
  and Pick}}]{bove13-jcp}
\bibinfo{author}{\bibfnamefont{L.}~\bibnamefont{Bove}},
  \bibinfo{author}{\bibfnamefont{C.}~\bibnamefont{Dreyfus}},
  \bibinfo{author}{\bibfnamefont{R.}~\bibnamefont{Torre}}, \bibnamefont{and}
  \bibinfo{author}{\bibfnamefont{R.}~\bibnamefont{Pick}}, \bibinfo{journal}{J.
  Chem. Phys.} \textbf{\bibinfo{volume}{139}}, \bibinfo{pages}{044501}
  (\bibinfo{year}{2013}).

\bibitem[{\citenamefont{Suzuki and Mishima}(2013)}]{suzuki13-jcp}
\bibinfo{author}{\bibfnamefont{Y.}~\bibnamefont{Suzuki}} \bibnamefont{and}
  \bibinfo{author}{\bibfnamefont{O.}~\bibnamefont{Mishima}},
  \bibinfo{journal}{J. Chem. Phys.} \textbf{\bibinfo{volume}{138}},
  \bibinfo{pages}{084507} (\bibinfo{year}{2013}).

\bibitem[{\citenamefont{Ruiz et~al.}(2014)\citenamefont{Ruiz, Bove, Corti, and
  Loerting}}]{ruiz14-pccp}
\bibinfo{author}{\bibfnamefont{G.}~\bibnamefont{Ruiz}},
  \bibinfo{author}{\bibfnamefont{L.}~\bibnamefont{Bove}},
  \bibinfo{author}{\bibfnamefont{H.~R.} \bibnamefont{Corti}}, \bibnamefont{and}
  \bibinfo{author}{\bibfnamefont{T.}~\bibnamefont{Loerting}},
  \bibinfo{journal}{Phys. Chem. Chem. Phys.} \textbf{\bibinfo{volume}{16}},
  \bibinfo{pages}{18553} (\bibinfo{year}{2014}).

\bibitem[{\citenamefont{Ruiz et~al.}(2018)\citenamefont{Ruiz, Amann-Winkel,
  Bove, Corti, and Loerting}}]{ruiz18-pccp}
\bibinfo{author}{\bibfnamefont{G.~N.} \bibnamefont{Ruiz}},
  \bibinfo{author}{\bibfnamefont{K.}~\bibnamefont{Amann-Winkel}},
  \bibinfo{author}{\bibfnamefont{L.~E.} \bibnamefont{Bove}},
  \bibinfo{author}{\bibfnamefont{H.~R.} \bibnamefont{Corti}}, \bibnamefont{and}
  \bibinfo{author}{\bibfnamefont{T.}~\bibnamefont{Loerting}},
  \bibinfo{journal}{Phys. Chem. Chem. Phys.} \textbf{\bibinfo{volume}{20}},
  \bibinfo{pages}{6401} (\bibinfo{year}{2018}).

\bibitem[{\citenamefont{Camisasca et~al.}(2018)\citenamefont{Camisasca,
  De~Marzio, Rovere, and Gallo}}]{camisasca18-jcp}
\bibinfo{author}{\bibfnamefont{G.}~\bibnamefont{Camisasca}},
  \bibinfo{author}{\bibfnamefont{M.}~\bibnamefont{De~Marzio}},
  \bibinfo{author}{\bibfnamefont{M.}~\bibnamefont{Rovere}}, \bibnamefont{and}
  \bibinfo{author}{\bibfnamefont{P.}~\bibnamefont{Gallo}}, \bibinfo{journal}{J.
  Chem. Phys.} \textbf{\bibinfo{volume}{148}}, \bibinfo{pages}{222829}
  (\bibinfo{year}{2018}).

\bibitem[{\citenamefont{Xu et~al.}(2005)\citenamefont{Xu, Kumar, Buldyrev,
  Chen, Poole, Sciortino, and Stanley}}]{xu2005relation}
\bibinfo{author}{\bibfnamefont{L.}~\bibnamefont{Xu}},
  \bibinfo{author}{\bibfnamefont{P.}~\bibnamefont{Kumar}},
  \bibinfo{author}{\bibfnamefont{S.~V.} \bibnamefont{Buldyrev}},
  \bibinfo{author}{\bibfnamefont{S.-H.} \bibnamefont{Chen}},
  \bibinfo{author}{\bibfnamefont{P.~H.} \bibnamefont{Poole}},
  \bibinfo{author}{\bibfnamefont{F.}~\bibnamefont{Sciortino}},
  \bibnamefont{and} \bibinfo{author}{\bibfnamefont{H.~E.}
  \bibnamefont{Stanley}}, \bibinfo{journal}{Proc. Natl. Acad. Sci. U.S.A.}
  \textbf{\bibinfo{volume}{102}}, \bibinfo{pages}{16558}
  (\bibinfo{year}{2005}).

\bibitem[{\citenamefont{Le and Molinero}(2011)}]{le2011nanophase}
\bibinfo{author}{\bibfnamefont{L.}~\bibnamefont{Le}} \bibnamefont{and}
  \bibinfo{author}{\bibfnamefont{V.}~\bibnamefont{Molinero}},
  \bibinfo{journal}{J. Phys. Chem. A} \textbf{\bibinfo{volume}{115}},
  \bibinfo{pages}{5900} (\bibinfo{year}{2011}).

\bibitem[{\citenamefont{Molinero and Moore}(2009)}]{molinero2009water}
\bibinfo{author}{\bibfnamefont{V.}~\bibnamefont{Molinero}} \bibnamefont{and}
  \bibinfo{author}{\bibfnamefont{E.~B.} \bibnamefont{Moore}},
  \bibinfo{journal}{J. Phys. Chem. B} \textbf{\bibinfo{volume}{113}},
  \bibinfo{pages}{4008} (\bibinfo{year}{2009}).

\bibitem[{\citenamefont{Aragones et~al.}(2014)\citenamefont{Aragones, Rovere,
  Vega, and Gallo}}]{aragones14-jpcb}
\bibinfo{author}{\bibfnamefont{J.~L.} \bibnamefont{Aragones}},
  \bibinfo{author}{\bibfnamefont{M.}~\bibnamefont{Rovere}},
  \bibinfo{author}{\bibfnamefont{C.}~\bibnamefont{Vega}}, \bibnamefont{and}
  \bibinfo{author}{\bibfnamefont{P.}~\bibnamefont{Gallo}}, \bibinfo{journal}{J.
  Phys. Chem. B} \textbf{\bibinfo{volume}{118}}, \bibinfo{pages}{7680}
  (\bibinfo{year}{2014}).

\bibitem[{\citenamefont{Horn et~al.}(2004)\citenamefont{Horn, Swope, Pitera,
  Madura, Dick, Hura, and Head-Gordon}}]{horn04-jcp}
\bibinfo{author}{\bibfnamefont{H.~W.} \bibnamefont{Horn}},
  \bibinfo{author}{\bibfnamefont{W.~C.} \bibnamefont{Swope}},
  \bibinfo{author}{\bibfnamefont{J.~W.} \bibnamefont{Pitera}},
  \bibinfo{author}{\bibfnamefont{J.~D.} \bibnamefont{Madura}},
  \bibinfo{author}{\bibfnamefont{T.~J.} \bibnamefont{Dick}},
  \bibinfo{author}{\bibfnamefont{G.~L.} \bibnamefont{Hura}}, \bibnamefont{and}
  \bibinfo{author}{\bibfnamefont{T.}~\bibnamefont{Head-Gordon}},
  \bibinfo{journal}{J. Chem. Phys.} \textbf{\bibinfo{volume}{120}},
  \bibinfo{pages}{9665} (\bibinfo{year}{2004}).

\bibitem[{\citenamefont{Joung and Cheatham~III}(2008)}]{joung08-jpcb}
\bibinfo{author}{\bibfnamefont{I.~S.} \bibnamefont{Joung}} \bibnamefont{and}
  \bibinfo{author}{\bibfnamefont{T.~E.} \bibnamefont{Cheatham~III}},
  \bibinfo{journal}{J. Phys. Chem. B} \textbf{\bibinfo{volume}{112}},
  \bibinfo{pages}{9020} (\bibinfo{year}{2008}).

\bibitem[{\citenamefont{Abascal and Vega}(2005)}]{abascal05-jcp}
\bibinfo{author}{\bibfnamefont{J.~L.} \bibnamefont{Abascal}} \bibnamefont{and}
  \bibinfo{author}{\bibfnamefont{C.}~\bibnamefont{Vega}}, \bibinfo{journal}{J.
  Chem. Phys.} \textbf{\bibinfo{volume}{123}}, \bibinfo{pages}{234505}
  (\bibinfo{year}{2005}).

\bibitem[{\citenamefont{Vega and Abascal}(2011)}]{vega11-pccp}
\bibinfo{author}{\bibfnamefont{C.}~\bibnamefont{Vega}} \bibnamefont{and}
  \bibinfo{author}{\bibfnamefont{J.~L.} \bibnamefont{Abascal}},
  \bibinfo{journal}{Phys. Chem. Chem. Phys.} \textbf{\bibinfo{volume}{13}},
  \bibinfo{pages}{19663} (\bibinfo{year}{2011}).

\bibitem[{\citenamefont{Wong et~al.}(2015)\citenamefont{Wong, Jahn, and
  Giovambattista}}]{wong15-jcp}
\bibinfo{author}{\bibfnamefont{J.}~\bibnamefont{Wong}},
  \bibinfo{author}{\bibfnamefont{D.~A.} \bibnamefont{Jahn}}, \bibnamefont{and}
  \bibinfo{author}{\bibfnamefont{N.}~\bibnamefont{Giovambattista}},
  \bibinfo{journal}{J. Chem. Phys.} \textbf{\bibinfo{volume}{143}},
  \bibinfo{pages}{074501} (\bibinfo{year}{2015}).

\bibitem[{\citenamefont{Handle et~al.}(2019)\citenamefont{Handle, Sciortino,
  and Giovambattista}}]{handle19-jcp}
\bibinfo{author}{\bibfnamefont{P.~H.} \bibnamefont{Handle}},
  \bibinfo{author}{\bibfnamefont{F.}~\bibnamefont{Sciortino}},
  \bibnamefont{and}
  \bibinfo{author}{\bibfnamefont{N.}~\bibnamefont{Giovambattista}},
  \bibinfo{journal}{J. Chem. Phys.} \textbf{\bibinfo{volume}{150}},
  \bibinfo{pages}{244506} (\bibinfo{year}{2019}).

\bibitem[{\citenamefont{Mou\v{c}ka et~al.}(2012)\citenamefont{Mou\v{c}ka,
  L{\'\i}sal, and Smith}}]{moucka12-jpcb}
\bibinfo{author}{\bibfnamefont{F.}~\bibnamefont{Mou\v{c}ka}},
  \bibinfo{author}{\bibfnamefont{M.}~\bibnamefont{L{\'\i}sal}},
  \bibnamefont{and} \bibinfo{author}{\bibfnamefont{W.~R.} \bibnamefont{Smith}},
  \bibinfo{journal}{J. Phys. Chem. B} \textbf{\bibinfo{volume}{116}},
  \bibinfo{pages}{5468} (\bibinfo{year}{2012}).

\bibitem[{\citenamefont{Monnin et~al.}(2002)\citenamefont{Monnin, Dubois,
  Papaiconomou, and Simonin}}]{monnin02-jced}
\bibinfo{author}{\bibfnamefont{C.}~\bibnamefont{Monnin}},
  \bibinfo{author}{\bibfnamefont{M.}~\bibnamefont{Dubois}},
  \bibinfo{author}{\bibfnamefont{N.}~\bibnamefont{Papaiconomou}},
  \bibnamefont{and} \bibinfo{author}{\bibfnamefont{J.-P.}
  \bibnamefont{Simonin}}, \bibinfo{journal}{J. Chem. Eng. Data}
  \textbf{\bibinfo{volume}{47}}, \bibinfo{pages}{1331} (\bibinfo{year}{2002}).

\bibitem[{\citenamefont{Van Der~Spoel et~al.}(2005)\citenamefont{Van Der~Spoel,
  Lindahl, Hess, Groenhof, Mark, and Berendsen}}]{vanderspoel05-jcc}
\bibinfo{author}{\bibfnamefont{D.}~\bibnamefont{Van Der~Spoel}},
  \bibinfo{author}{\bibfnamefont{E.}~\bibnamefont{Lindahl}},
  \bibinfo{author}{\bibfnamefont{B.}~\bibnamefont{Hess}},
  \bibinfo{author}{\bibfnamefont{G.}~\bibnamefont{Groenhof}},
  \bibinfo{author}{\bibfnamefont{A.~E.} \bibnamefont{Mark}}, \bibnamefont{and}
  \bibinfo{author}{\bibfnamefont{H.~J.} \bibnamefont{Berendsen}},
  \bibinfo{journal}{J. Comput. Chem.} \textbf{\bibinfo{volume}{26}},
  \bibinfo{pages}{1701} (\bibinfo{year}{2005}).

\bibitem[{\citenamefont{Nos{\'e}}(1984)}]{nose84-mp}
\bibinfo{author}{\bibfnamefont{S.}~\bibnamefont{Nos{\'e}}},
  \bibinfo{journal}{Mol. Phys.} \textbf{\bibinfo{volume}{52}},
  \bibinfo{pages}{255} (\bibinfo{year}{1984}).

\bibitem[{\citenamefont{Hoover}(1985)}]{hoover85-pra}
\bibinfo{author}{\bibfnamefont{W.~G.} \bibnamefont{Hoover}},
  \bibinfo{journal}{Phys. Rev. A} \textbf{\bibinfo{volume}{31}},
  \bibinfo{pages}{1695} (\bibinfo{year}{1985}).

\bibitem[{\citenamefont{Parrinello and Rahman}(1981)}]{parrinello81-jap}
\bibinfo{author}{\bibfnamefont{M.}~\bibnamefont{Parrinello}} \bibnamefont{and}
  \bibinfo{author}{\bibfnamefont{A.}~\bibnamefont{Rahman}},
  \bibinfo{journal}{J. Appl. Phys.} \textbf{\bibinfo{volume}{52}},
  \bibinfo{pages}{7182} (\bibinfo{year}{1981}).

\bibitem[{\citenamefont{Essmann et~al.}(1995)\citenamefont{Essmann, Perera,
  Berkowitz, Darden, Lee, and Pedersen}}]{essmann95-jcp}
\bibinfo{author}{\bibfnamefont{U.}~\bibnamefont{Essmann}},
  \bibinfo{author}{\bibfnamefont{L.}~\bibnamefont{Perera}},
  \bibinfo{author}{\bibfnamefont{M.~L.} \bibnamefont{Berkowitz}},
  \bibinfo{author}{\bibfnamefont{T.}~\bibnamefont{Darden}},
  \bibinfo{author}{\bibfnamefont{H.}~\bibnamefont{Lee}}, \bibnamefont{and}
  \bibinfo{author}{\bibfnamefont{L.~G.} \bibnamefont{Pedersen}},
  \bibinfo{journal}{J. Chem. Phys.} \textbf{\bibinfo{volume}{103}},
  \bibinfo{pages}{8577} (\bibinfo{year}{1995}).

\bibitem[{\citenamefont{Hess}(2008)}]{hess08-jctc}
\bibinfo{author}{\bibfnamefont{B.}~\bibnamefont{Hess}}, \bibinfo{journal}{J.
  Chem. Theory Comput.} \textbf{\bibinfo{volume}{4}}, \bibinfo{pages}{116}
  (\bibinfo{year}{2008}).

\bibitem[{\citenamefont{Testard et~al.}(2014)\citenamefont{Testard, Berthier,
  and Kob}}]{testard2014intermittent}
\bibinfo{author}{\bibfnamefont{V.}~\bibnamefont{Testard}},
  \bibinfo{author}{\bibfnamefont{L.}~\bibnamefont{Berthier}}, \bibnamefont{and}
  \bibinfo{author}{\bibfnamefont{W.}~\bibnamefont{Kob}}, \bibinfo{journal}{J.
  Chem. Phys.} \textbf{\bibinfo{volume}{140}}, \bibinfo{pages}{164502}
  (\bibinfo{year}{2014}).

\bibitem[{\citenamefont{Russo and Tanaka}(2014)}]{russo14-natcomm}
\bibinfo{author}{\bibfnamefont{J.}~\bibnamefont{Russo}} \bibnamefont{and}
  \bibinfo{author}{\bibfnamefont{H.}~\bibnamefont{Tanaka}},
  \bibinfo{journal}{Nat. Commun.} \textbf{\bibinfo{volume}{5}},
  \bibinfo{pages}{3556} (\bibinfo{year}{2014}).

\bibitem[{\citenamefont{Tanaka et~al.}(2019)\citenamefont{Tanaka, Tong, Shi,
  and Russo}}]{tanaka2019revealing}
\bibinfo{author}{\bibfnamefont{H.}~\bibnamefont{Tanaka}},
  \bibinfo{author}{\bibfnamefont{H.}~\bibnamefont{Tong}},
  \bibinfo{author}{\bibfnamefont{R.}~\bibnamefont{Shi}}, \bibnamefont{and}
  \bibinfo{author}{\bibfnamefont{J.}~\bibnamefont{Russo}},
  \bibinfo{journal}{Nat. Rev. Phys.} \textbf{\bibinfo{volume}{1}},
  \bibinfo{pages}{333} (\bibinfo{year}{2019}).

\bibitem[{\citenamefont{Errington and
  Debenedetti}(2001)}]{errington2001relationship}
\bibinfo{author}{\bibfnamefont{J.~R.} \bibnamefont{Errington}}
  \bibnamefont{and} \bibinfo{author}{\bibfnamefont{P.~G.}
  \bibnamefont{Debenedetti}}, \bibinfo{journal}{Nature}
  \textbf{\bibinfo{volume}{409}}, \bibinfo{pages}{318} (\bibinfo{year}{2001}).

\bibitem[{\citenamefont{Saika-Voivod et~al.}(2000)\citenamefont{Saika-Voivod,
  Sciortino, and Poole}}]{saika2000computer}
\bibinfo{author}{\bibfnamefont{I.}~\bibnamefont{Saika-Voivod}},
  \bibinfo{author}{\bibfnamefont{F.}~\bibnamefont{Sciortino}},
  \bibnamefont{and} \bibinfo{author}{\bibfnamefont{P.~H.} \bibnamefont{Poole}},
  \bibinfo{journal}{Phys. Rev. E} \textbf{\bibinfo{volume}{63}},
  \bibinfo{pages}{011202} (\bibinfo{year}{2000}).

\bibitem[{\citenamefont{Cuthbertson and Poole}(2011)}]{cuthbertson11-prl}
\bibinfo{author}{\bibfnamefont{M.~J.} \bibnamefont{Cuthbertson}}
  \bibnamefont{and} \bibinfo{author}{\bibfnamefont{P.~H.} \bibnamefont{Poole}},
  \bibinfo{journal}{Phys. Rev. Lett.} \textbf{\bibinfo{volume}{106}},
  \bibinfo{pages}{115706} (\bibinfo{year}{2011}).

\bibitem[{\citenamefont{Martelli}(2019)}]{martelli19-jcp}
\bibinfo{author}{\bibfnamefont{F.}~\bibnamefont{Martelli}}, \bibinfo{journal}{J
  Chem. Phys.} \textbf{\bibinfo{volume}{150}}, \bibinfo{pages}{094506}
  (\bibinfo{year}{2019}).

\bibitem[{\citenamefont{Camisasca et~al.}(2020)\citenamefont{Camisasca,
  De~Marzio, Rovere, and Gallo}}]{camisasca20erratum}
\bibinfo{author}{\bibfnamefont{G.}~\bibnamefont{Camisasca}},
  \bibinfo{author}{\bibfnamefont{M.}~\bibnamefont{De~Marzio}},
  \bibinfo{author}{\bibfnamefont{M.}~\bibnamefont{Rovere}}, \bibnamefont{and}
  \bibinfo{author}{\bibfnamefont{P.}~\bibnamefont{Gallo}}, \bibinfo{journal}{J.
  Chem. Phys.} \textbf{\bibinfo{volume}{152}}, \bibinfo{pages}{109901}
  (\bibinfo{year}{2020}).

\bibitem[{\citenamefont{Kob and Andersen}(1995)}]{kob95-pre}
\bibinfo{author}{\bibfnamefont{W.}~\bibnamefont{Kob}} \bibnamefont{and}
  \bibinfo{author}{\bibfnamefont{H.~C.} \bibnamefont{Andersen}},
  \bibinfo{journal}{Phys. Rev. E} \textbf{\bibinfo{volume}{51}},
  \bibinfo{pages}{4626} (\bibinfo{year}{1995}).

\bibitem[{\citenamefont{Testard}(2011)}]{testard2011etude}
\bibinfo{author}{\bibfnamefont{V.}~\bibnamefont{Testard}}, Ph.D. thesis,
  \bibinfo{school}{Universit{\'e} Montpellier 2} (\bibinfo{year}{2011}).

\bibitem[{\citenamefont{Shi et~al.}(2018)\citenamefont{Shi, Russo, and
  Tanaka}}]{shi18-jcp}
\bibinfo{author}{\bibfnamefont{R.}~\bibnamefont{Shi}},
  \bibinfo{author}{\bibfnamefont{J.}~\bibnamefont{Russo}}, \bibnamefont{and}
  \bibinfo{author}{\bibfnamefont{H.}~\bibnamefont{Tanaka}},
  \bibinfo{journal}{J. Chem. Phys.} \textbf{\bibinfo{volume}{149}},
  \bibinfo{pages}{224502} (\bibinfo{year}{2018}).

\bibitem[{\citenamefont{Marcus and Hefter}(2006)}]{marcus06-crev}
\bibinfo{author}{\bibfnamefont{Y.}~\bibnamefont{Marcus}} \bibnamefont{and}
  \bibinfo{author}{\bibfnamefont{G.}~\bibnamefont{Hefter}},
  \bibinfo{journal}{Chem. Rev.} \textbf{\bibinfo{volume}{106}},
  \bibinfo{pages}{4585} (\bibinfo{year}{2006}).

\bibitem[{\citenamefont{Omta et~al.}(2003)\citenamefont{Omta, Kropman,
  Woutersen, and Bakker}}]{omta03-science}
\bibinfo{author}{\bibfnamefont{A.~W.} \bibnamefont{Omta}},
  \bibinfo{author}{\bibfnamefont{M.~F.} \bibnamefont{Kropman}},
  \bibinfo{author}{\bibfnamefont{S.}~\bibnamefont{Woutersen}},
  \bibnamefont{and} \bibinfo{author}{\bibfnamefont{H.~J.}
  \bibnamefont{Bakker}}, \bibinfo{journal}{Science}
  \textbf{\bibinfo{volume}{301}}, \bibinfo{pages}{347} (\bibinfo{year}{2003}).

\bibitem[{\citenamefont{Marcus}(2009)}]{marcus09-crev}
\bibinfo{author}{\bibfnamefont{Y.}~\bibnamefont{Marcus}},
  \bibinfo{journal}{Chem. Rev.} \textbf{\bibinfo{volume}{109}},
  \bibinfo{pages}{1346} (\bibinfo{year}{2009}).

\end{thebibliography}

\end{document}


\title{Electronic Supplementary Information for:\\ ``Simulations of LiCl--Water: The Effects of Concentration and Supercooling"\\
}

\author{Philip H. Handle}
\affiliation{%
 Institute of Physical Chemistry,  University of Innsbruck, Innsbruck, Austria
}
\affiliation{%
Institute of General, Inorganic and Theoretical Chemistry, University of Innsbruck, Innsbruck, Austria
}

\date{\today}

\maketitle

In this document we provide additional data and figures to supplement the discussion of the main manuscript.
In particular, we compare data obtained for the standard Lorentz-Berthelot (LB) combination rules, as well as the modified (MLB) combination rules as introduced by Aragones et al.~\cite{aragones14-jpcb}.
The systems contain 1000 water molecules and the amount of LiCl is altered to match the desired concentration (see Tab.~\ref{tab:thd}).
In addition, we also present data for smaller systems containing 480 water molecules.
These systems were simulated at \SI{298}{\kelvin} and \SI{1}{\bar} for four molar fractions ($x_\text{LiCl}=\SIlist{2.4;5.9;11.1;14.3}{\percent}$).
This enables a direct comparison with data from literature, where such system sizes were studied~\cite{aragones14-jpcb, camisasca18-jcp, camisasca20erratum}.
Moreover, the study of two system sizes allows to asses size effects.

The structure of this document follows the structure of the main document.
That is, we present supplementary information regarding thermodynamics (Section~\ref{sec:thd}), diffusion coefficients (Section~\ref{sec:diff}), structure (Section~\ref{sec:rdf}), the coarse-grained density field (Section~\ref{sec:cg}), the structural order parameter $\zeta$ (Section~\ref{sec:rt}), and the overlapping hydration shells (Section~\ref{sec:ohs}).

\newpage

\section{Thermodynamics}
\label{sec:thd}

The studied thermodynamic quantities of all considered systems is summarised in Tab.~\ref{tab:thd}.
For both water models the change from the LB combination rules to the MLB combination rules has little effect on $\left<U\right>$, but it provokes an increase of $\left<\rho\right>$ at high concentrations.
The system size has no significant influence on the results and the results obtained here agree with the values reported by Aragones et al.~\cite{aragones14-jpcb} and Camisasca et al.~\cite{camisasca18-jcp, camisasca20erratum} (see Tab.~\ref{tab:thd}).

\begin{table}[b!]

\begin{threeparttable}[b]

\caption{Summary of the obtained averages of the potential energy $\left<U\right>$ and the density $\left<\rho\right>$ for all simulated systems.
$\left<U\right>$ is given per mole of particles ($N_\text{Li}+N_\text{Cl}+N_{\text{H}_2\text{O}}$).
The concentrations are given as mole fractions of LiCl $x_\text{LiCl}$ and the ratio $R$ of water molecules per LiCl. The employed water models TIP4P/2005 and TIP4P-Ew are abbreviated as 2005 and Ew, respectively.
The abbreviations LB and MLB encode the Lorentz-Berthelot and modified Lorentz-Berthelot combination rules, respectively.
Where available values from literature are also reproduced.}
\label{tab:thd}
\begin{tabular}{@{\extracolsep{2pt}}rrcrrrrrrrrrrc@{}}
\multicolumn{5}{c}{}&
\multicolumn{4}{c}{$\left<U\right>$ / $\left(\si{\kilo\joule\per\mole}\right)$}&
\multicolumn{4}{c}{$\left<\rho\right>$ / $\left(\si{\gram\per\cubic\centi\meter}\right)$}\\[0.25em]
\cline{6-9}\cline{10-13}\\[-0.5em]
\multicolumn{5}{c}{}&
\multicolumn{2}{c}{\SI{240}{\kelvin}}&
\multicolumn{2}{c}{\SI{298}{\kelvin}}&
\multicolumn{2}{c}{\SI{240}{\kelvin}}&
\multicolumn{2}{c}{\SI{298}{\kelvin}}&\\[0.25em]
\cline{6-7}\cline{8-9}\cline{10-11}\cline{12-13}\\[-0.5em]
\multicolumn{1}{c}{$x_\text{LiCl}$ / \si{\percent}} &
\multicolumn{1}{c}{$R$} &
Water &
\multicolumn{1}{c}{$N_{\text{H}_2\text{O}}$} &
\multicolumn{1}{c}{$N_\text{LiCl}$} &
\multicolumn{1}{c}{LB} &
\multicolumn{1}{c}{MLB} &
\multicolumn{1}{c}{LB} &
\multicolumn{1}{c}{MLB} &
\multicolumn{1}{c}{LB} &
\multicolumn{1}{c}{MLB} &
\multicolumn{1}{c}{LB} &
\multicolumn{1}{c}{MLB} &
Source\\[0.25em]
\hline\\[-0.5em]
\num{ 0.0} & \multicolumn{1}{c}{$\infty$} &   Ew & 1000 &   0 &               & \num{ -50.66} & \num{ -46.52} & \num{ -46.52} &             & \num{0.988} & \num{0.995} & \num{0.995} & This work\\ 
           &              & 2005 & 1000 &   0 &               & \num{ -51.96} & \num{ -47.87} & \num{ -47.87} &             & \num{0.985} & \num{0.997} & \num{0.997} & This work\\ 

\num{ 0.1} & \num{1000.0} &   Ew & 1000 &   1 &               & \num{ -51.45} & \num{ -47.33} & \num{ -47.33} &             & \num{0.990} & \num{0.996} & \num{0.996} & This work\\ 
           &              & 2005 & 1000 &   1 &               & \num{ -52.74} & \num{ -48.67} & \num{ -48.67} &             & \num{0.987} & \num{0.998} & \num{0.998} & This work\\ 

\num{ 0.2} & \num{ 500.0} &   Ew & 1000 &   2 &               & \num{ -52.24} & \num{ -48.14} & \num{ -48.14} &             & \num{0.991} & \num{0.998} & \num{0.998} & This work\\ 
           &              & 2005 & 1000 &   2 &               & \num{ -53.52} & \num{ -49.47} & \num{ -49.47} &             & \num{0.988} & \num{0.999} & \num{0.999} & This work\\ 

\num{ 0.4} & \num{ 250.0} &   Ew & 1000 &   4 &               & \num{ -53.81} & \num{ -49.74} & \num{ -49.74} &             & \num{0.994} & \num{1.000} & \num{1.000} & This work\\ 
           &              & 2005 & 1000 &   4 &               & \num{ -55.08} & \num{ -51.07} & \num{ -51.07} &             & \num{0.992} & \num{1.002} & \num{1.002} & This work\\ 

\num{ 0.8} & \num{ 125.0} &   Ew & 1000 &   8 &               & \num{ -56.91} & \num{ -52.92} & \num{ -52.91} &             & \num{1.001} & \num{1.005} & \num{1.005} & This work\\ 
           &              & 2005 & 1000 &   8 &               & \num{ -58.16} & \num{ -54.21} & \num{ -54.21} &             & \num{0.998} & \num{1.006} & \num{1.006} & This work\\ 

\num{ 1.6} & \num{  62.5} &   Ew & 1000 &  16 &               & \num{ -63.00} & \num{ -59.10} & \num{ -59.09} &             & \num{1.012} & \num{1.013} & \num{1.013} & This work\\ 
           &              & 2005 & 1000 &  16 &               & \num{ -64.19} & \num{ -60.34} & \num{ -60.33} &             & \num{1.010} & \num{1.015} & \num{1.014} & This work\\ 

\num{ 2.4} & \num{  40.0} &   Ew &  480 &  12 &               &               & \num{ -65.81} & \num{ -65.78} &             &             & \num{1.022} & \num{1.021} & \cite{aragones14-jpcb}\\ 
           &              &      &      &     &               &               & \num{ -65.94} & \num{ -65.92} &             &             & \num{1.022} & \num{1.021} & This work\\ 
           &              &      & 1000 &  25 & \num{ -69.69} & \num{ -69.69} & \num{ -65.93} & \num{ -65.92} & \num{1.024} & \num{1.024} & \num{1.023} & \num{1.022} & This work\\ 
           &              & 2005 &  480 &  12 &               &               & \num{ -67.81} &               &             &             & \num{1.024} &           & \cite{aragones14-jpcb}\\ 
           &              &      &      &     &               &               & \num{ -67.12} & \num{ -67.10} &             &             & \num{1.024} & \num{1.023} & This work\\ 
           &              &      & 1000 &  25 & \num{ -70.83} & \num{ -70.82} & \num{ -67.11} & \num{ -67.09} & \num{1.022} & \num{1.022} & \num{1.024} & \num{1.023} & This work\\ 

\num{ 5.9} & \num{  16.0} &   Ew &  480 &  30 &               &               & \num{ -91.59} & \num{ -91.49} &             &             & \num{1.056} & \num{1.055} & \cite{aragones14-jpcb}\\ 
           &              &      &      &     &               &               & \num{ -91.63} & \num{ -91.53} &             &             & \num{1.057} & \num{1.056} & This work\\ 
           &              &      & 1000 &  62 &               & \num{ -95.36} & \num{ -92.03} & \num{ -91.94} &             & \num{1.066} & \num{1.058} & \num{1.056} & This work\\ 
           &              & 2005 &  480 &  30 &               &               & \num{ -92.57} & \num{ -92.49} &             &             & \num{1.057} & \num{1.055} & This work\\ 
           &              &      & 1000 &  62 &               & \num{ -96.31} & \num{ -92.97} & \num{ -92.88} &             & \num{1.063} & \num{1.057} & \num{1.056} & This work\\ 

\num{11.1} & \num{   8.0} &   Ew &  480 &  60 &               &               & \num{-127.47} & \num{-127.28} &             &             & \num{1.101} & \num{1.101} & \cite{aragones14-jpcb}\\ 
           &              &      &      &     &               &               & \num{-127.62} & \num{-127.39} &             &             & \num{1.103} & \num{1.101} & This work\\ 
           &              &      & 1000 & 125 &               & \num{-130.48} & \num{-127.61} & \num{-127.38} &             & \num{1.118} & \num{1.102} & \num{1.101} & This work\\ 
           &              & 2005 &  480 &  60 &               &               & \num{-128.20} & \num{-128.02} &             &             & \num{1.100} & \num{1.099} & This work\\ 
           &              &      & 1000 & 125 &               & \num{-131.11} & \num{-128.20} & \num{-128.00} &             & \num{1.113} & \num{1.100} & \num{1.099} & This work\\ 

\num{14.3} & \num{   6.0} &   Ew &  480 &  80 &               &               & \num{-147.43} & \num{-147.24} &             &             & \num{1.126} & \num{1.127} & \cite{aragones14-jpcb}\\                              
           &              &      &      &     &               &               & \num{-147.55} & \num{-147.30} &             &             & \num{1.127} & \num{1.128} & This work\\ 
           &              &      & 1000 & 166 & \num{-150.69} & \num{-150.43} & \num{-147.77} & \num{-147.54} & \num{1.146} & \num{1.148} & \num{1.127} & \num{1.128} & This work\\ 
           &              & 2005 &  480 &  80 &               &               & \num{-147.88} &               &             &             & \num{1.124} &        & \cite{aragones14-jpcb}\\
           &              &      &      &     &               & \num{-150.69} &               & \num{-147.70}\tnote{$\dagger$} &             & \num{1.143} &             & \num{1.124}\tnote{$\dagger$} & \cite{camisasca18-jcp, camisasca20erratum}\\
           &              &      &      &     &               &               & \num{-147.94} & \num{-147.77} &             &             & \num{1.124} & \num{1.125} & This work\\ 
           &              &      & 1000 & 166 & \num{-151.06} & \num{-150.90} & \num{-148.14} & \num{-147.99} & \num{1.140} & \num{1.142} & \num{1.124} & \num{1.125} & This work\\ 

\num{25.0} & \num{   3.0} &   Ew & 1000 & 333 &               & \num{-209.73} & \num{-206.37} & \num{-207.20} &             & \num{1.237} & \num{1.205} & \num{1.213} & This work\\ 
           &              & 2005 & 1000 & 333 &               & \num{-209.88} & \num{-206.40} & \num{-207.29} &             & \num{1.231} & \num{1.201} & \num{1.208} & This work\\ 

\num{33.3} & \num{   2.0} &   Ew & 1000 & 500 &               & \num{-248.35} & \num{-244.08} & \num{-246.35} &             & \num{1.292} & \num{1.255} & \num{1.272} & This work\\ 
           &              & 2005 & 1000 & 500 &               & \num{-248.54} & \num{-244.16} & \num{-246.31} &             & \num{1.289} & \num{1.253} & \num{1.267} & This work\\ 
\end{tabular}
\begin{tablenotes}
\item [$\dagger$] The data stem from simulations conducted at \SI{300}{\kelvin}.
\end{tablenotes}
\end{threeparttable}
\end{table}

\newpage

\section{Diffusion Coefficients}
\label{sec:diff}

\begin{figure}[h!]
\begin{center}
\includegraphics[width=0.49\textwidth]{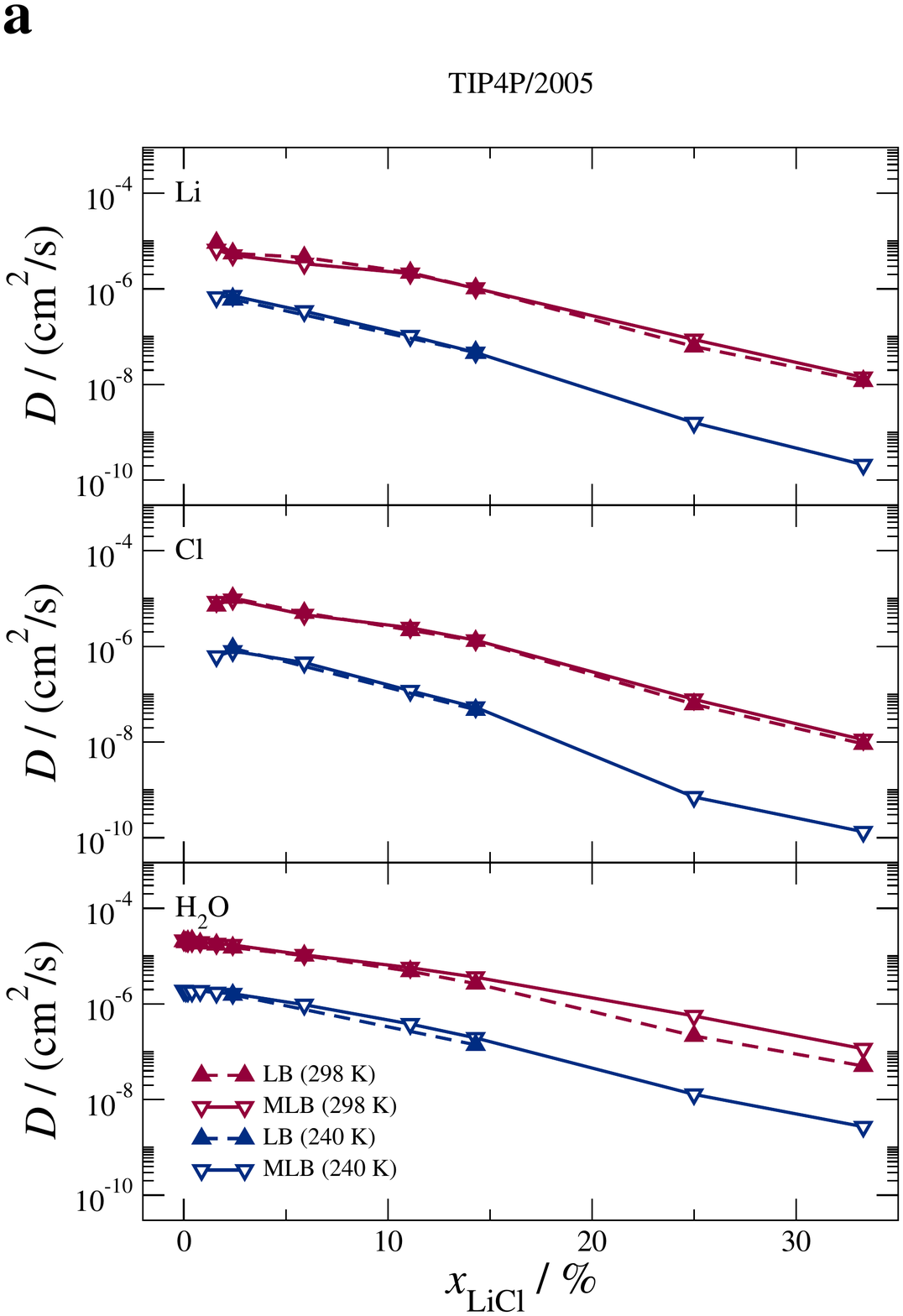}
\includegraphics[width=0.49\textwidth]{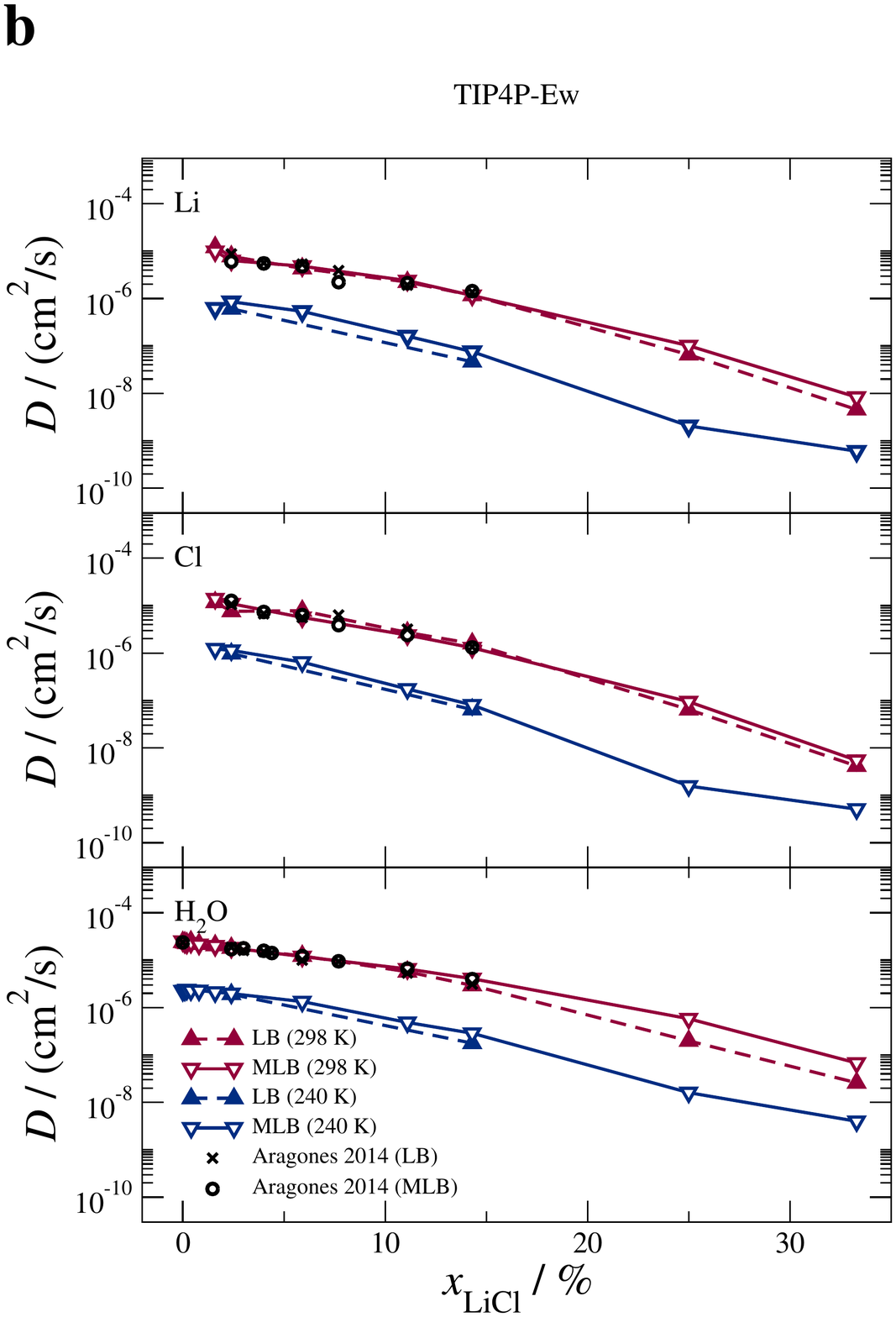}
 \caption{Diffusion coefficients $D$ for Li (top panel), Cl (middle panel) and H$_2$O (bottom panel) as a function of LiCl mole fraction $x_\text{LiCl}$ for both temperatures and combination rules (LB and MLB) studied.
  The filled upward triangles show the data for the LB combination rules, the open downward triangles the data for the MLB combination rules.
  Part a shows the results for TIP4P/2005 and part b  shows the results for TIP4P-Ew.
  Blue data are obtained for \SI{240}{\kelvin} and red data are obtained for \SI{298}{\kelvin}.
   For the ions only $x_\text{LiCl}\geq\SI{1.6}{\percent}$ allow for statistically significant data.
 All data are obtained from the large systems ($N_{\text{H}_2\text{O}}=1000$).
 The black open circles indicate the data  for the TIP4P-Ew water model as reported by Aragones et al.~\cite{aragones14-jpcb}.
 }
\label{fig:diff}
\end{center}
\end{figure}

The change from the LB to the MLB combination rules enhances the water diffusivity at the highest two concentrations (see Fig.~\ref{fig:diff}).
A smaller enhancement is visible for the ions, which is slightly more pronounced in TIP4P-Ew.
It was pointed out earlier that the change from the LB to the MLB combination rules enhances the diffusivity of the ions~\cite{aragones14-jpcb}.
However, the enhancement in the water diffusivity visible in Fig.~\ref{fig:diff} occurs at concentrations not studied by Aragones et al.~\cite{aragones14-jpcb}.
We rationalise this, through the tendency of the ions to cluster when described using the MLB combination rules.
Therefore, it is likely that the MLB combination rules free some water molecules from the ions' immediate surroundings enabling them to diffuse.

\newpage

\section{Structure}
\label{sec:rdf}

\begin{figure}[h!]
\begin{center}
\includegraphics[width=0.49\textwidth]{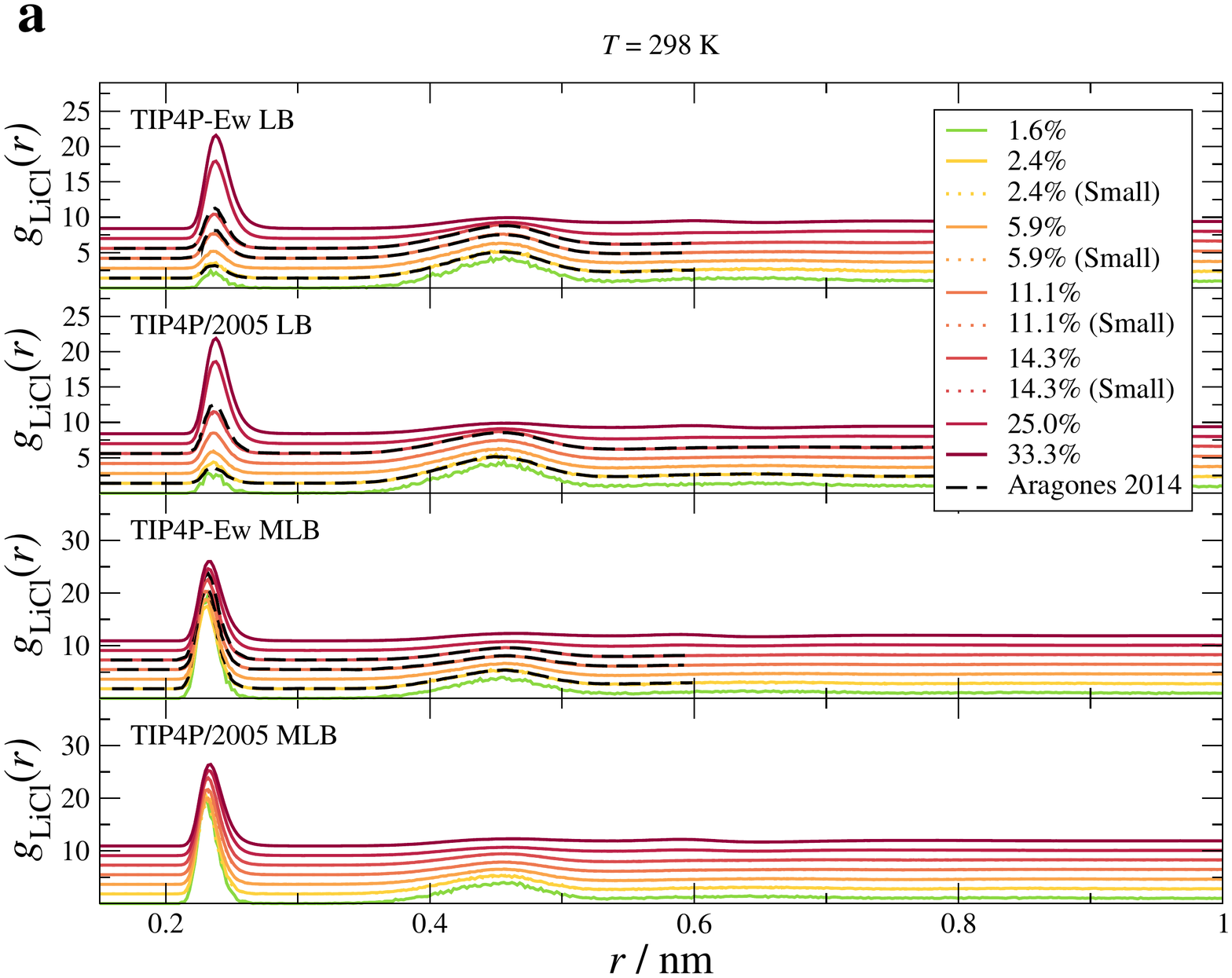}
\includegraphics[width=0.49\textwidth]{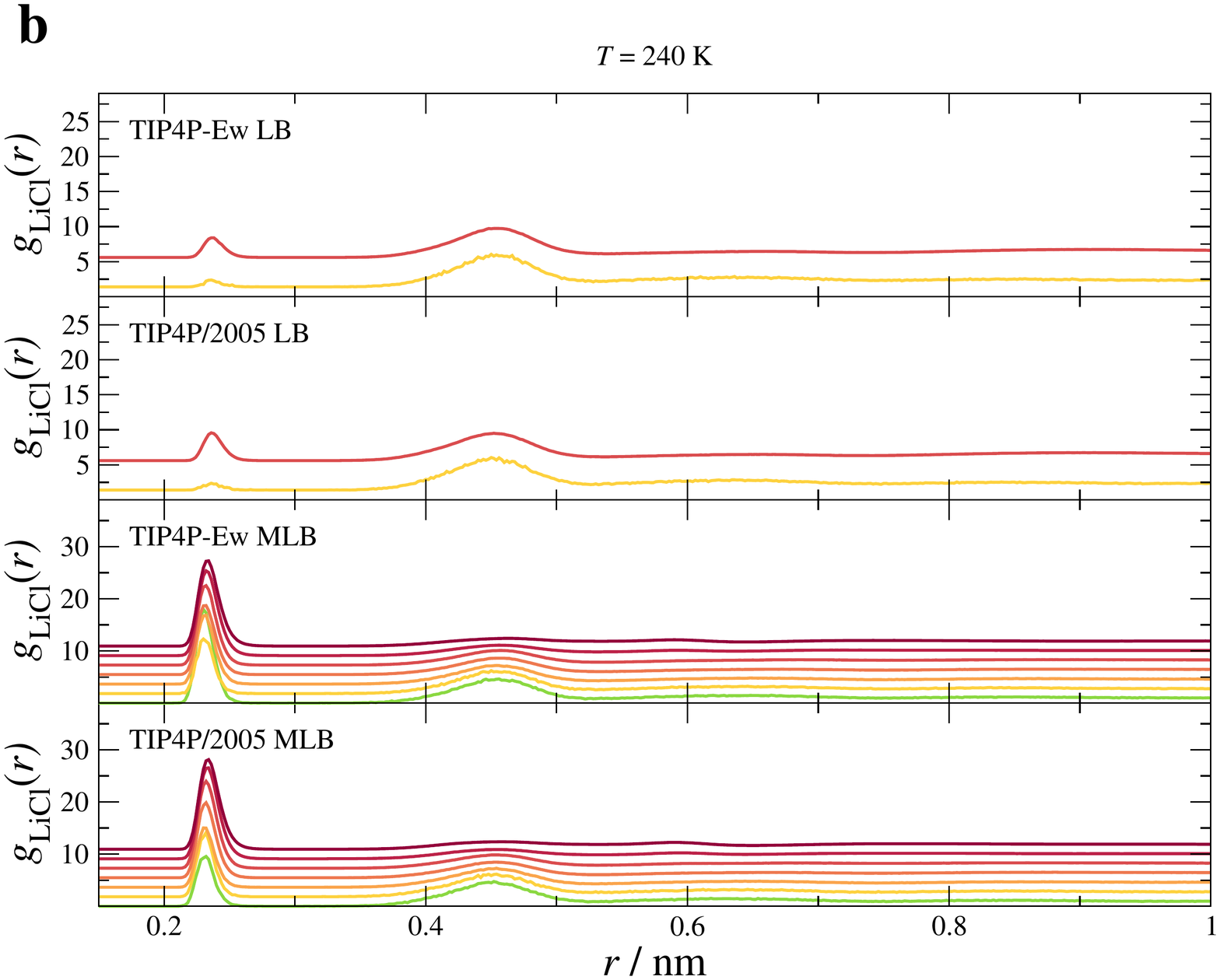}
\caption{Li-Cl RDFs as a function of concentration. Part a shows the data for \SI{298}{\kelvin}, part b for \SI{240}{\kelvin}.
The two upper panels show the behaviour obtained for the LB combination rules for both studied water models. The bottom two panels show the behaviour obtained for the MLB combination rules for both studied water models. The dotted data were obtained for the small systems ($N_{\text{H}_2\text{O}}=480$). The dashed black lines indicate the RDFs as reported by Aragones et al.~\cite{aragones14-jpcb}.}
\label{fig:licl}
\end{center}
\end{figure}

\begin{figure}[h!]
\begin{center}
\includegraphics[width=0.49\textwidth]{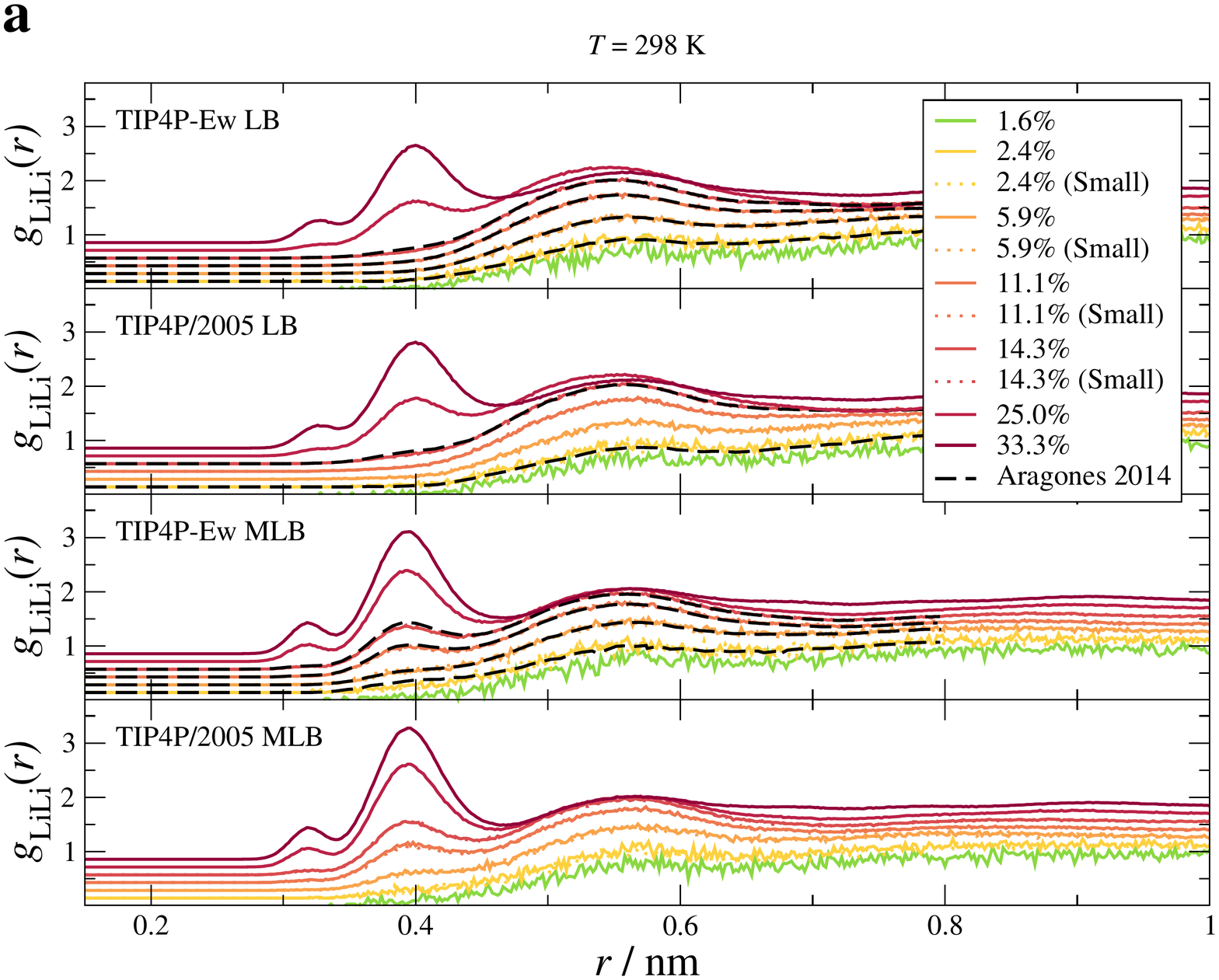}
\includegraphics[width=0.49\textwidth]{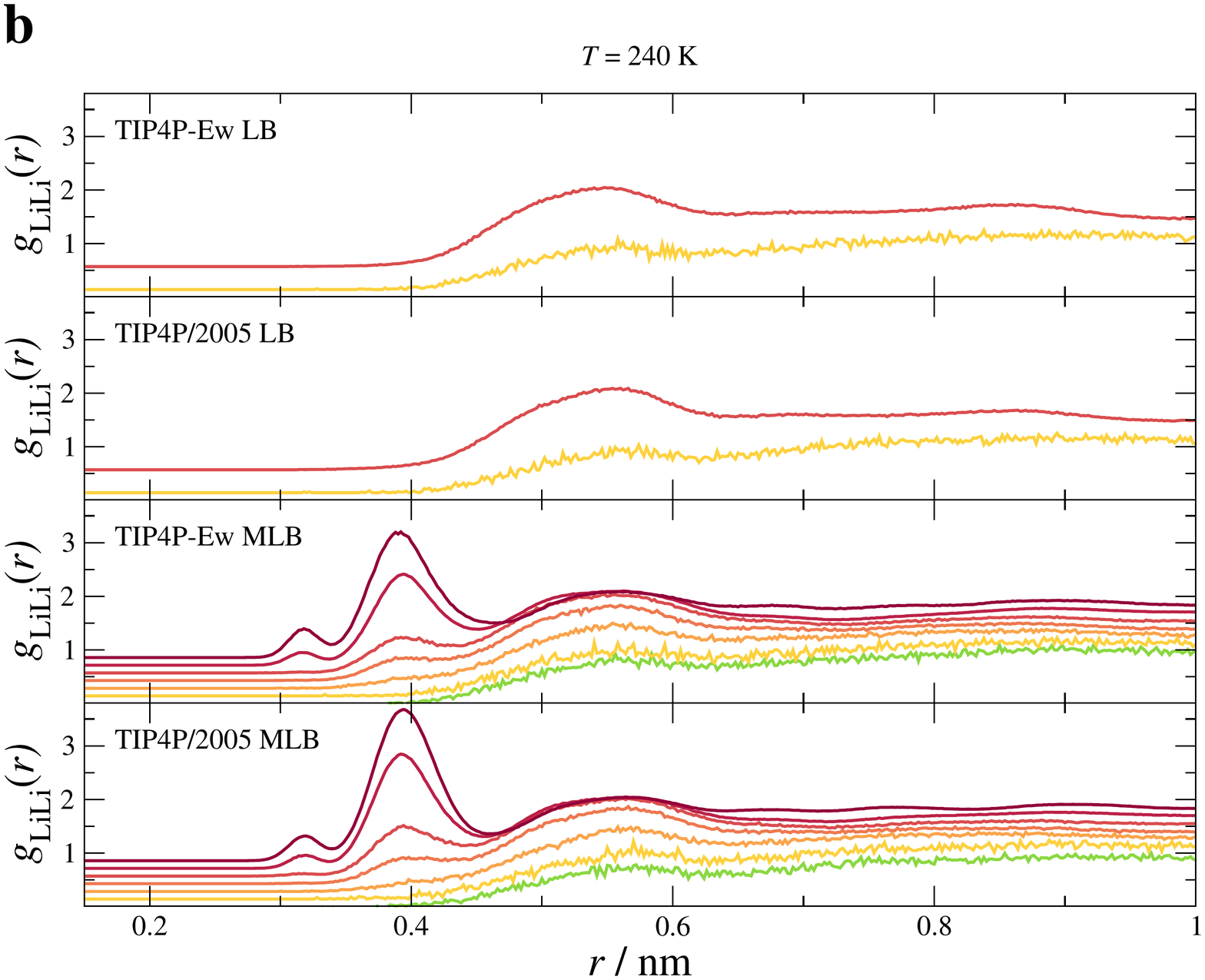}
\caption{Li-Li RDFs as a function of concentration. Part a shows the data for \SI{298}{\kelvin}, part b for \SI{240}{\kelvin}.
The two upper panels show the behaviour obtained for the LB combination rules for both studied water models. The bottom two panels show the behaviour obtained for the MLB combination rules for both studied water models. The dotted data were obtained for the small systems ($N_{\text{H}_2\text{O}}=480$). The dashed black lines indicate the RDFs as reported by Aragones et al.~\cite{aragones14-jpcb}.}
\label{fig:lili}
\end{center}
\end{figure}

For the Li-Cl RDFs shown in Fig.~\ref{fig:licl} the use of the MLB combination rules enhances the first peak in the Li-Cl RDFs at both studied temperatures when compared to the LB case.
This is actually the effect that the MLB combination rules were designed to achieve~\cite{aragones14-jpcb}.
Also, for the Li-Li and the Cl-Cl RDFs we find that the MLB combination rules enhance the pre-peaks at both studied temperatures (cf. Figs.~\ref{fig:lili} and \ref{fig:licl}).

The hydration structure of Li is hardly affected by the choice of combination rule as can be seen in Figs.~\ref{fig:lio} and ~\ref{fig:lih}.
For the Cl hydration the combination rules influence the shoulder in the Cl-O RDFs and the third maximum at $r\approx\SI{0.42}{\nano\meter}$ in the Cl-H RDFs.
Both features become enhanced when the MLB combination rules are used (cf. Figs.~\ref{fig:clo} and ~\ref{fig:clh}).
This again corroborates the link between these two features discussed in the main document.

As is visible in Figs.~\ref{fig:oo}-\ref{fig:hh} the water-water structure shows almost no change when different combination rules are used.

\begin{figure}[h!]
\begin{center}
\includegraphics[width=0.49\textwidth]{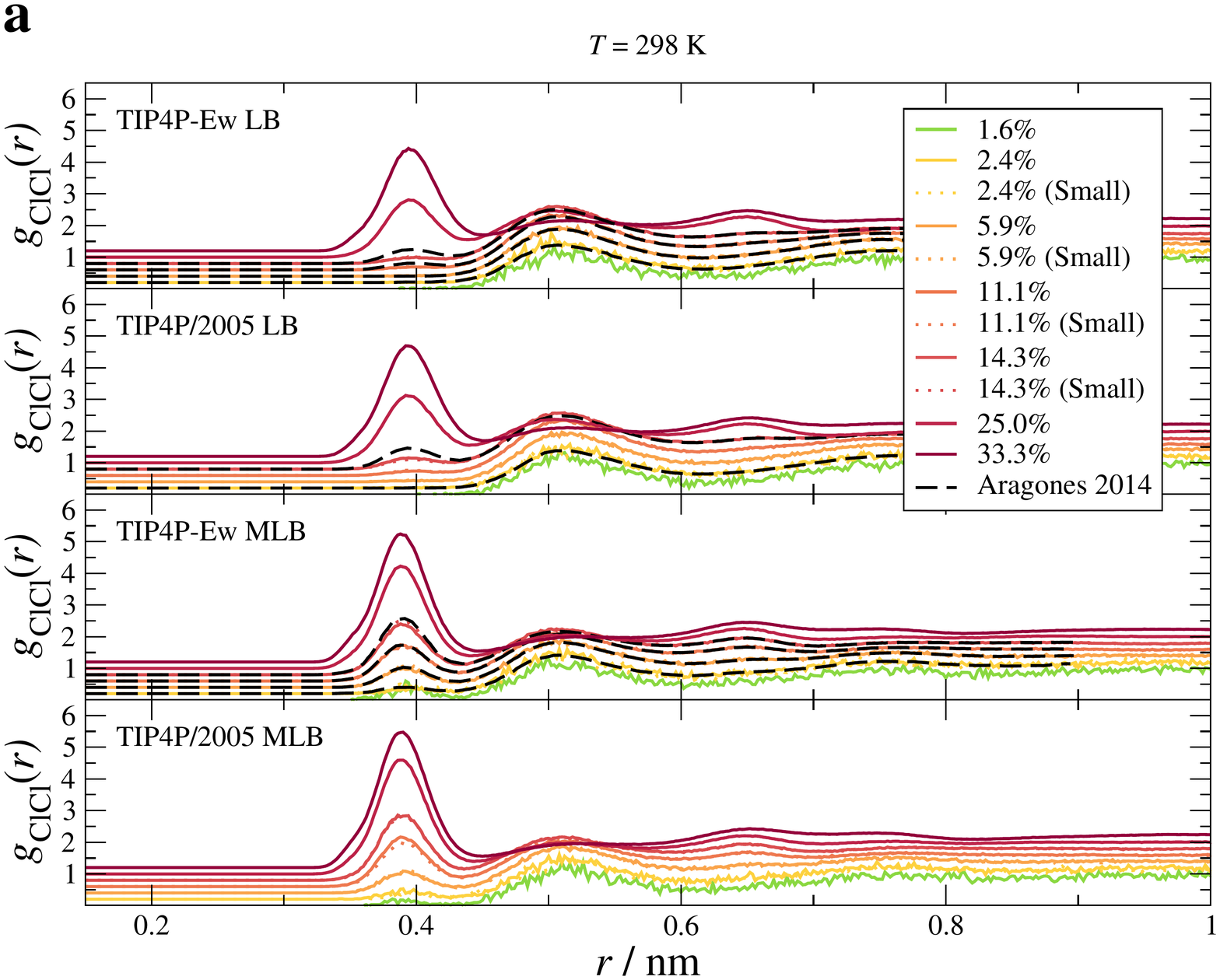}
\includegraphics[width=0.49\textwidth]{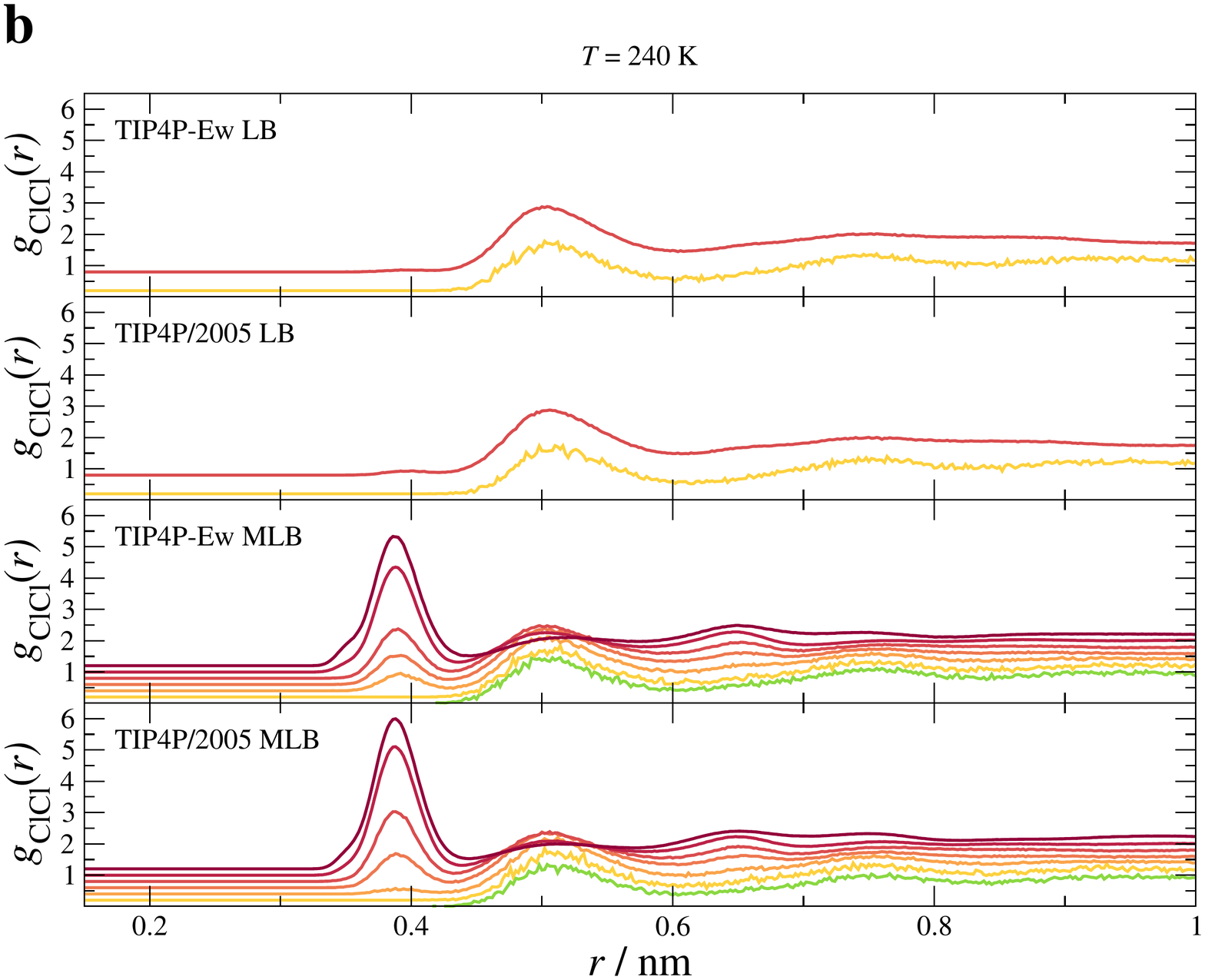}
\caption{Cl-Cl RDFs as a function of concentration. Part a shows the data for \SI{298}{\kelvin}, part b for \SI{240}{\kelvin}.
The two upper panels show the behaviour obtained for the LB combination rules for both studied water models. The bottom two panels show the behaviour obtained for the MLB combination rules for both studied water models. The dotted data were obtained for the small systems ($N_{\text{H}_2\text{O}}=480$). The dashed black lines indicate the RDFs as reported by Aragones et al.~\cite{aragones14-jpcb}.}
\label{fig:clcl}
\end{center}
\end{figure}

\begin{figure}[h!]
\begin{center}
\includegraphics[width=0.49\textwidth]{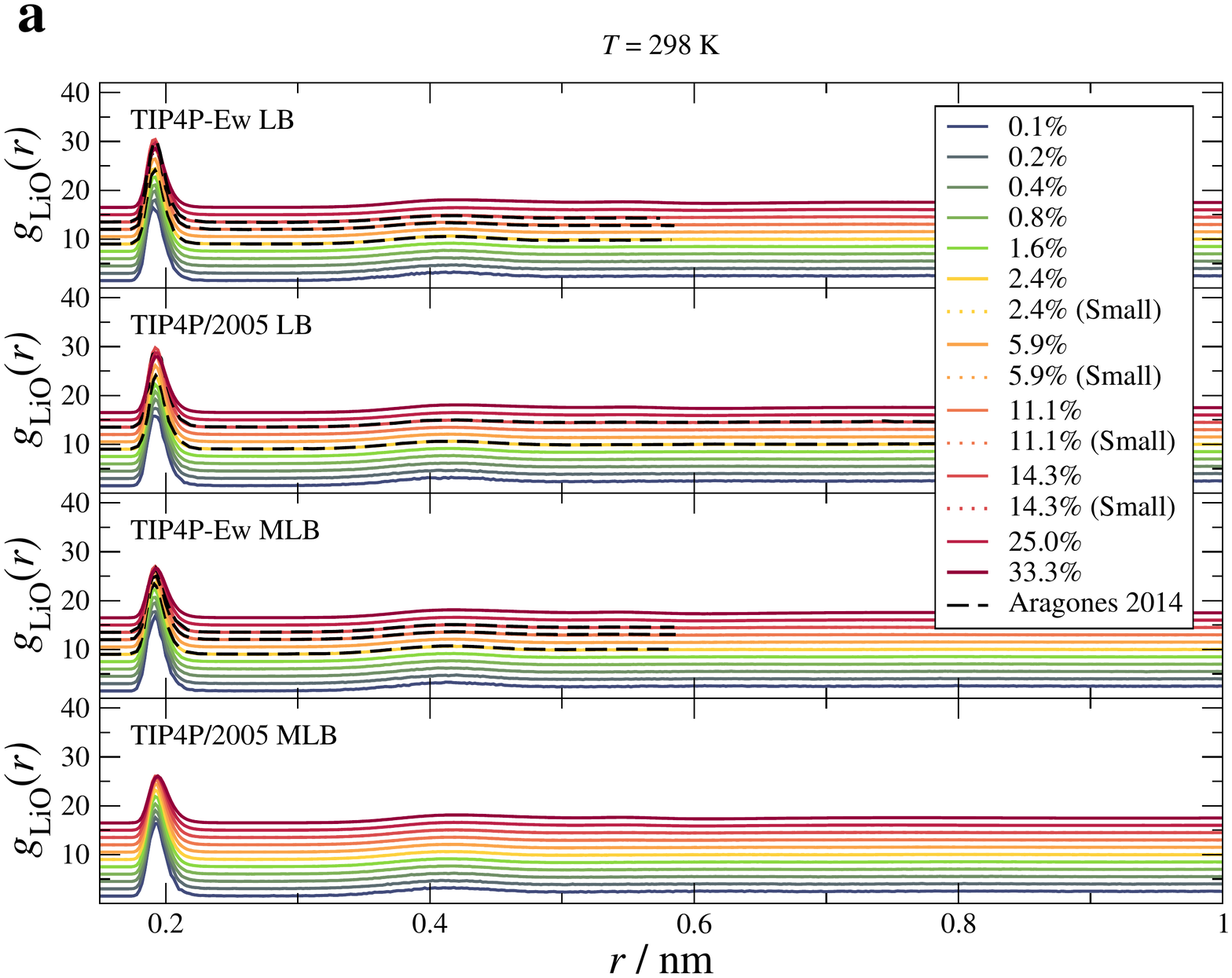}
\includegraphics[width=0.49\textwidth]{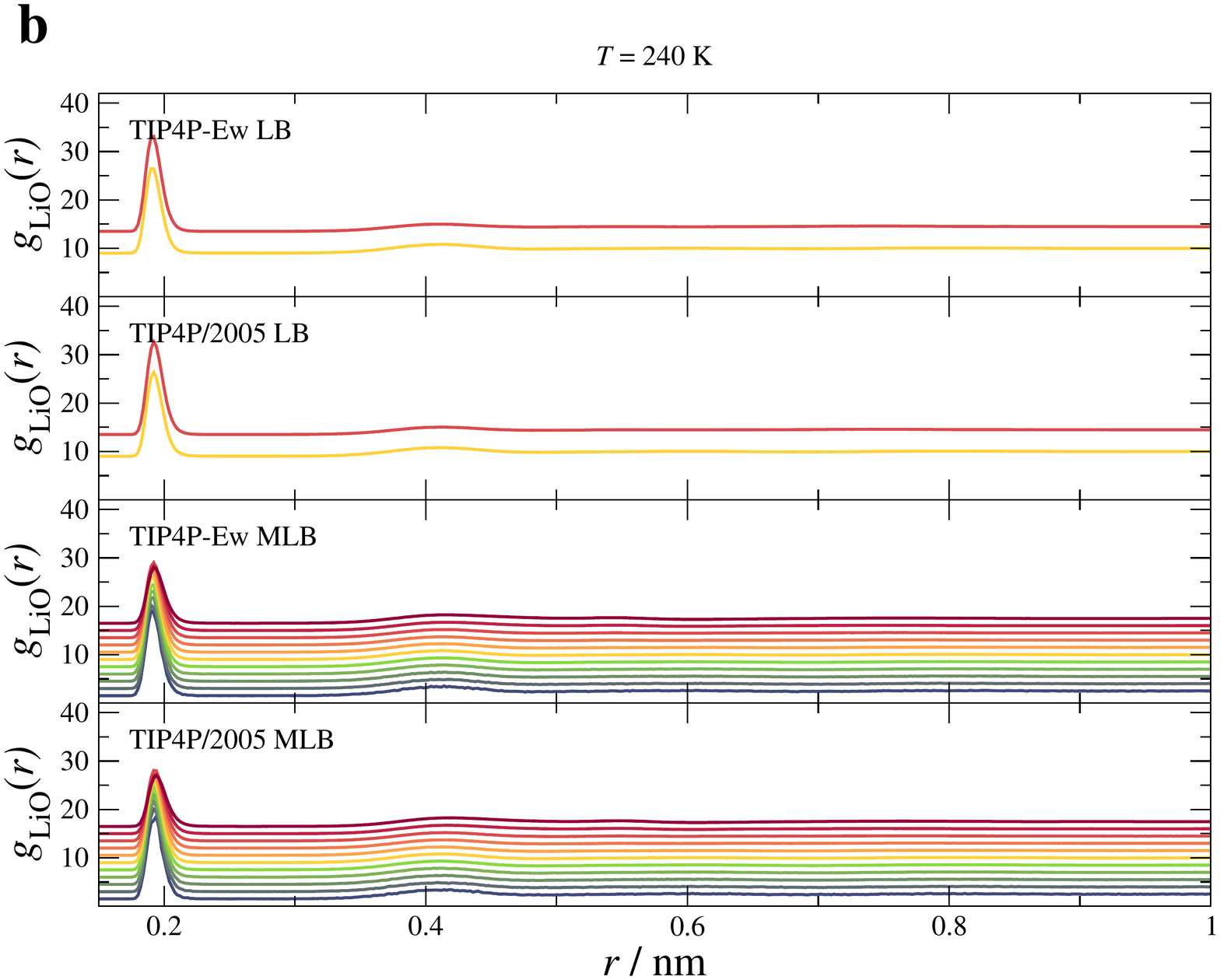}
\caption{Li-O RDFs as a function of concentration. Part a shows the data for \SI{298}{\kelvin}, part b for \SI{240}{\kelvin}.
The two upper panels show the behaviour obtained for the LB combination rules for both studied water models. The bottom two panels show the behaviour obtained for the MLB combination rules for both studied water models. The dotted data were obtained for the small systems ($N_{\text{H}_2\text{O}}=480$). The dashed black lines indicate the RDFs as reported by Aragones et al.~\cite{aragones14-jpcb}.}
\label{fig:lio}
\end{center}
\end{figure}

\begin{figure}[h!]
\begin{center}
\includegraphics[width=0.49\textwidth]{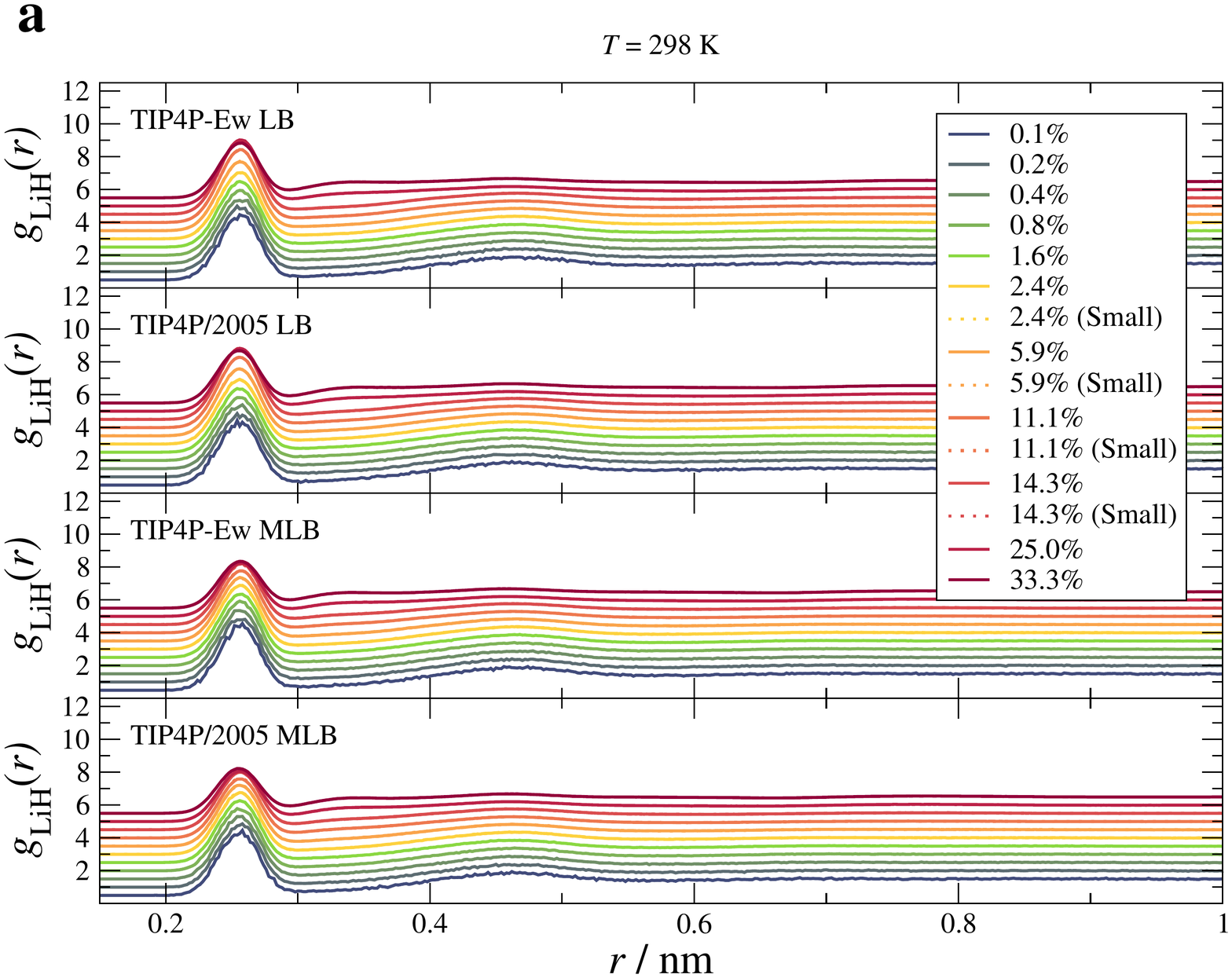}
\includegraphics[width=0.49\textwidth]{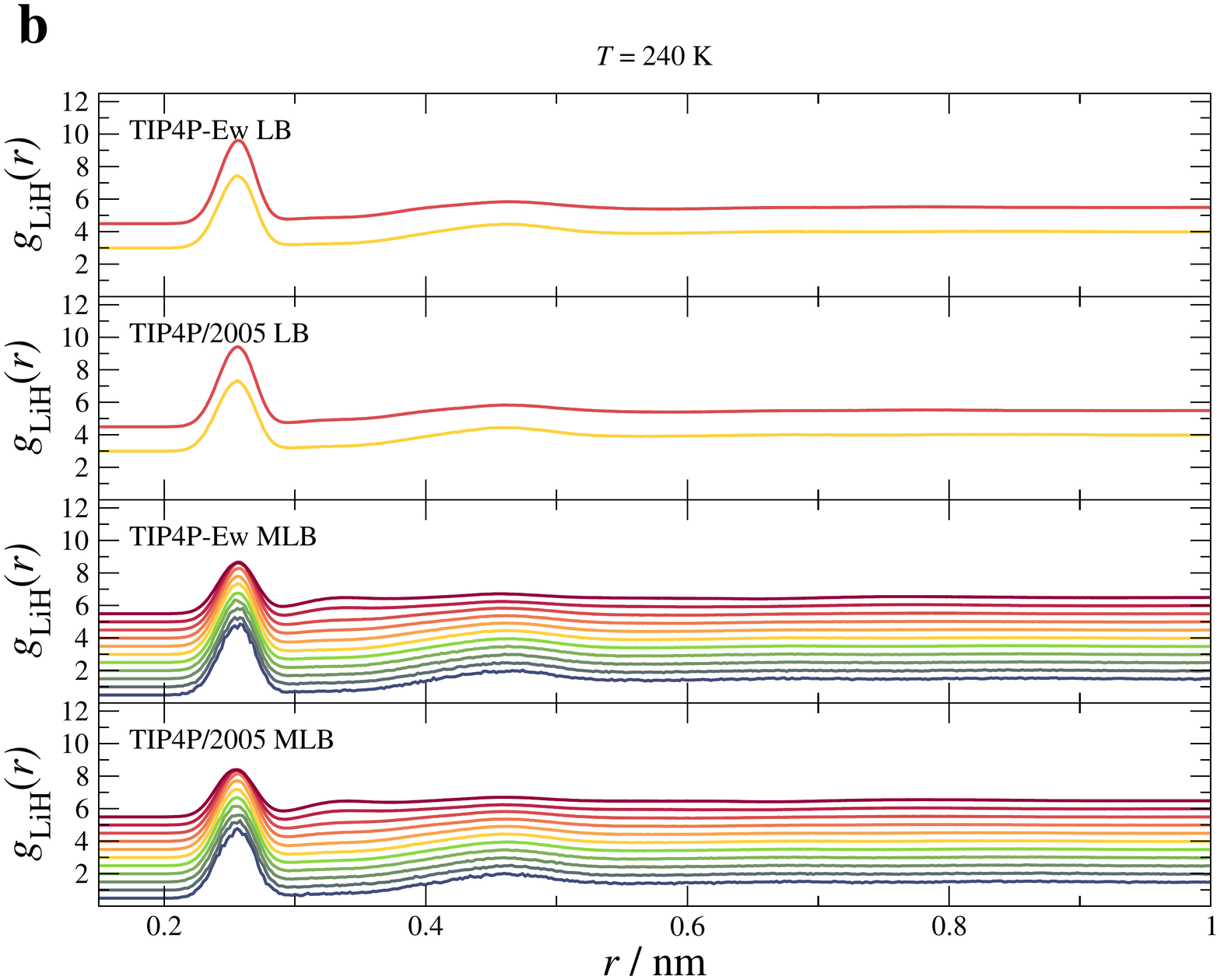}
\caption{Li-H RDFs as a function of concentration. Part a shows the data for \SI{298}{\kelvin}, part b for \SI{240}{\kelvin}.
The two upper panels show the behaviour obtained for the LB combination rules for both studied water models. The bottom two panels show the behaviour obtained for the MLB combination rules for both studied water models. The dotted data were obtained for the small systems ($N_{\text{H}_2\text{O}}=480$). The dashed black lines indicate the RDFs as reported by Aragones et al.~\cite{aragones14-jpcb}.}
\label{fig:lih}
\end{center}
\end{figure}

\begin{figure}[h!]
\begin{center}
\includegraphics[width=0.49\textwidth]{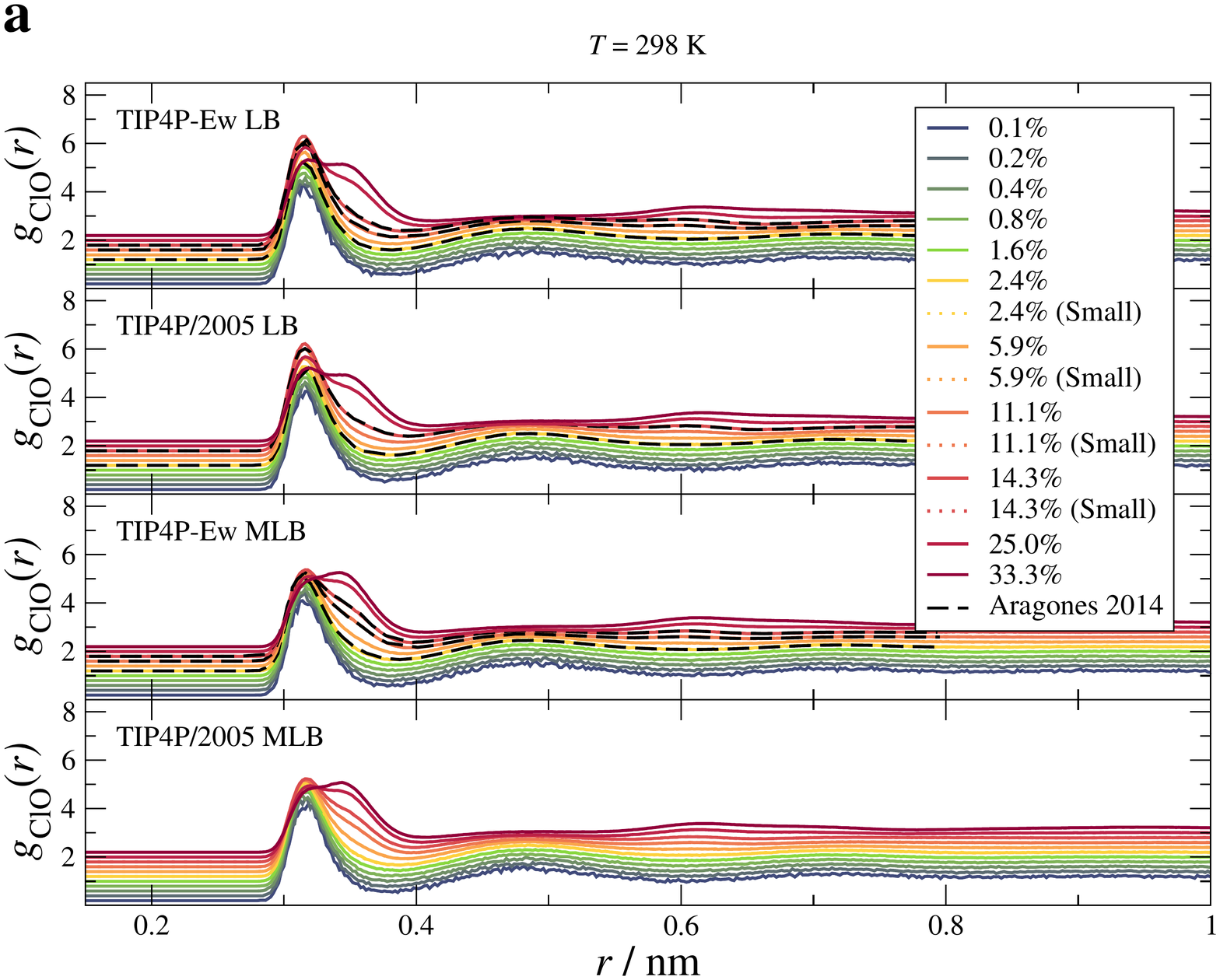}
\includegraphics[width=0.49\textwidth]{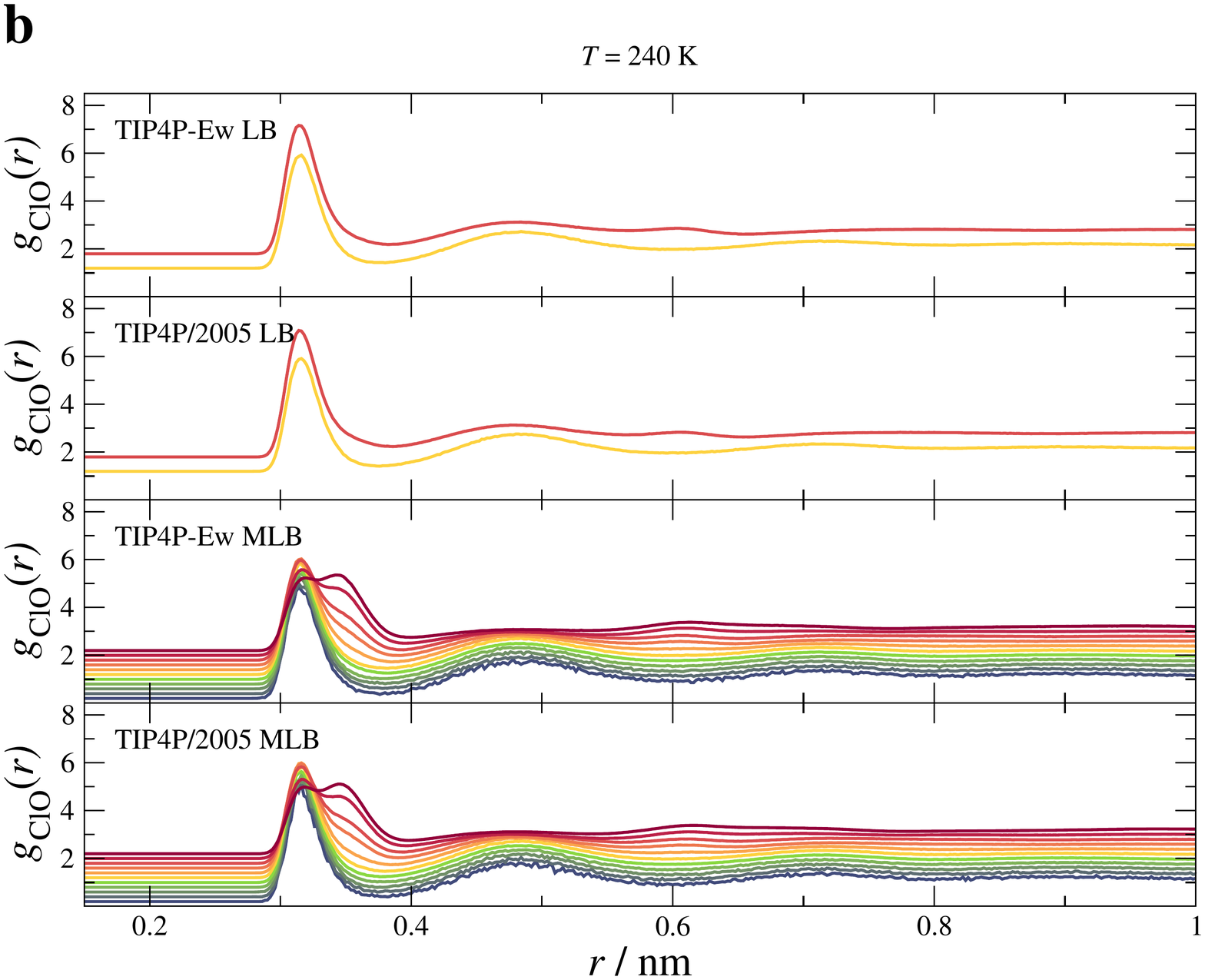}
\caption{Cl-O RDFs as a function of concentration. Part a shows the data for \SI{298}{\kelvin}, part b for \SI{240}{\kelvin}.
The two upper panels show the behaviour obtained for the LB combination rules for both studied water models. The bottom two panels show the behaviour obtained for the MLB combination rules for both studied water models. The dotted data were obtained for the small systems ($N_{\text{H}_2\text{O}}=480$). The dashed black lines indicate the RDFs as reported by Aragones et al.~\cite{aragones14-jpcb}.}
\label{fig:clo}
\end{center}
\end{figure}

\begin{figure}[h!]
\begin{center}
\includegraphics[width=0.49\textwidth]{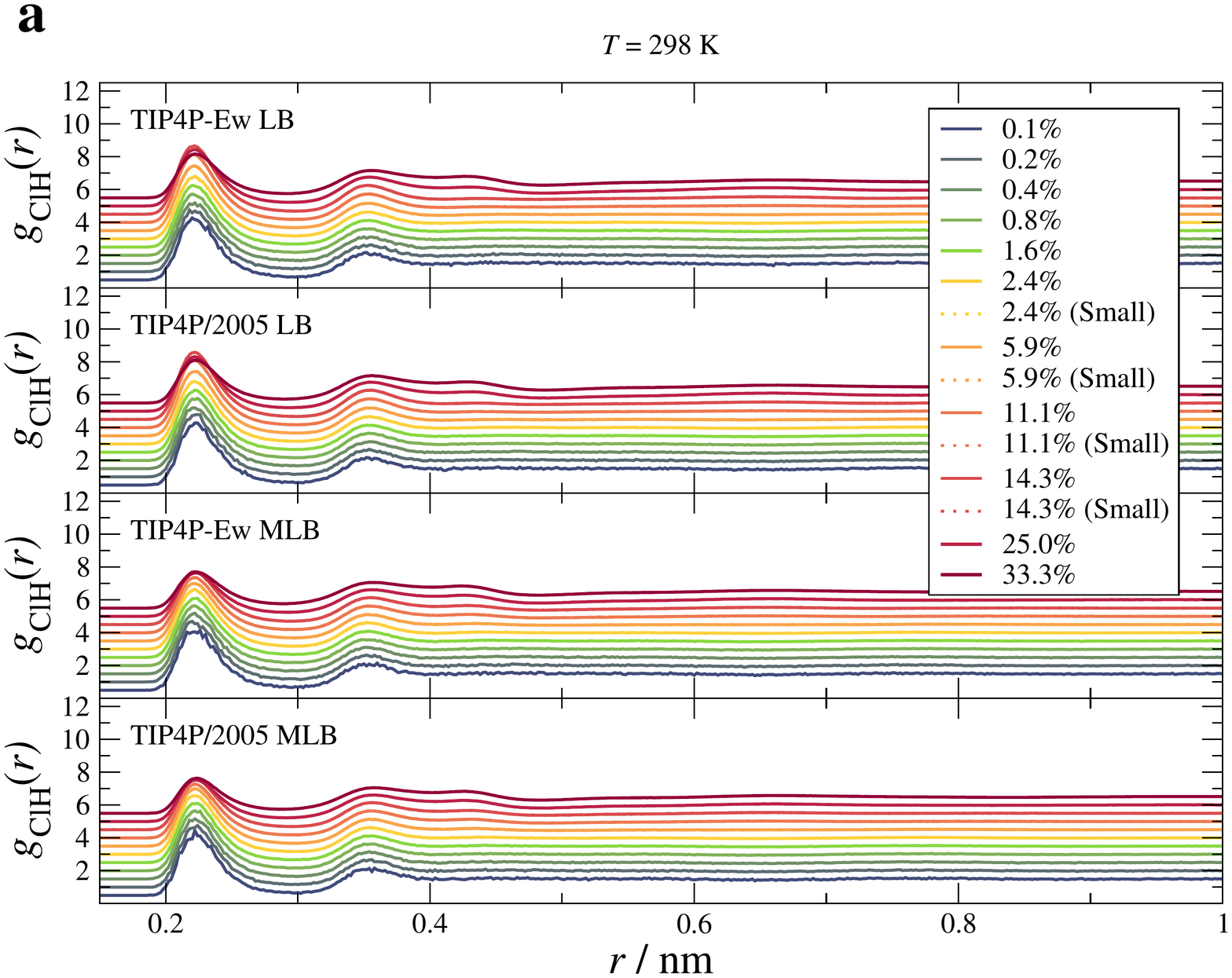}
\includegraphics[width=0.49\textwidth]{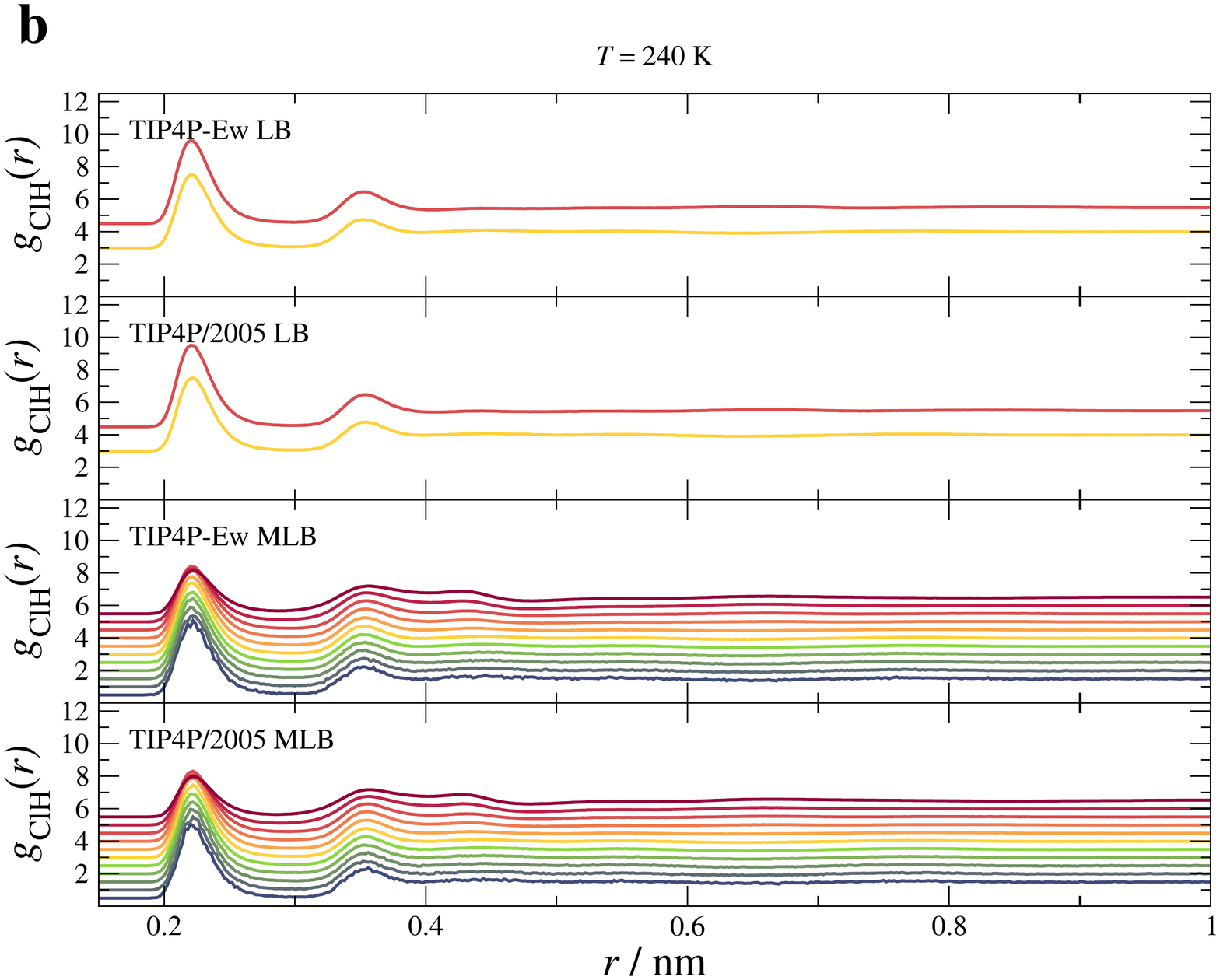}
\caption{Cl-H RDFs as a function of concentration. Part a shows the data for \SI{298}{\kelvin}, part b for \SI{240}{\kelvin}.
The two upper panels show the behaviour obtained for the LB combination rules for both studied water models. The bottom two panels show the behaviour obtained for the MLB combination rules for both studied water models. The dotted data were obtained for the small systems ($N_{\text{H}_2\text{O}}=480$). The dashed black lines indicate the RDFs as reported by Aragones et al.~\cite{aragones14-jpcb}.}
\label{fig:clh}
\end{center}
\end{figure}

\begin{figure}[h!]
\begin{center}
\includegraphics[width=0.49\textwidth]{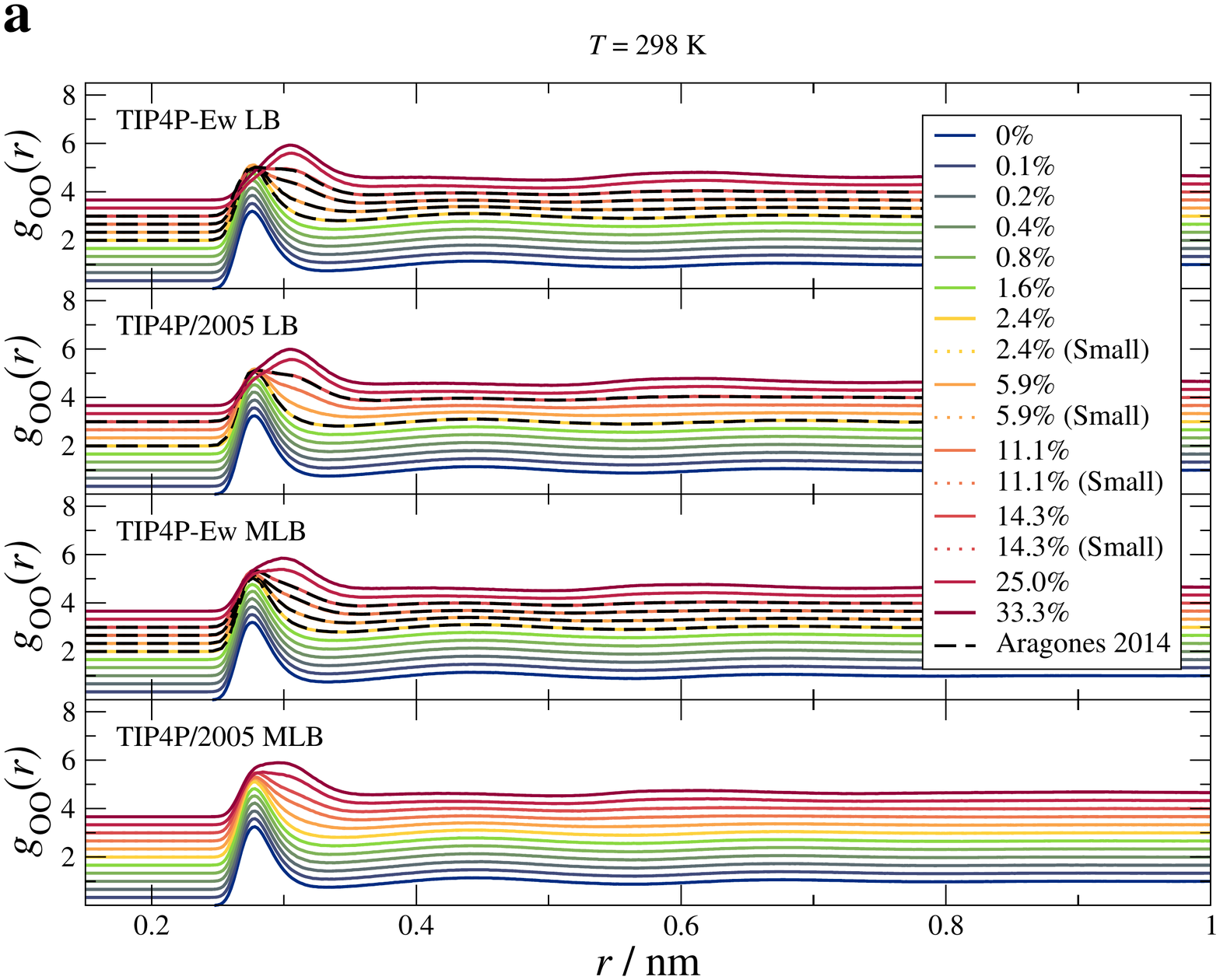}
\includegraphics[width=0.49\textwidth]{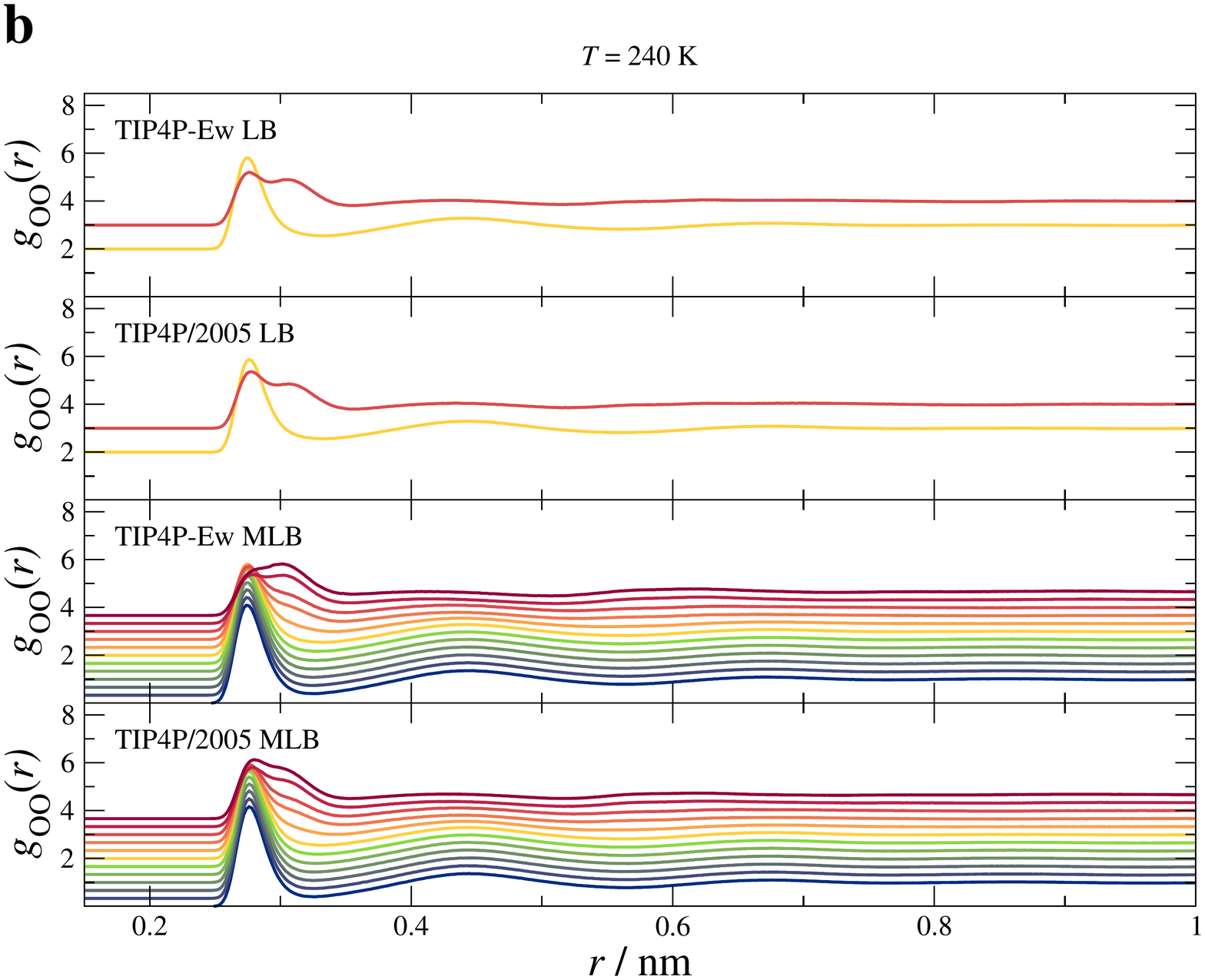}
\caption{O-O RDFs as a function of concentration. Part a shows the data for \SI{298}{\kelvin}, part b for \SI{240}{\kelvin}.
The two upper panels show the behaviour obtained for the LB combination rules for both studied water models. The bottom two panels show the behaviour obtained for the MLB combination rules for both studied water models. The dotted data were obtained for the small systems ($N_{\text{H}_2\text{O}}=480$). The dashed black lines indicate the RDFs as reported by Aragones et al.~\cite{aragones14-jpcb}.}
\label{fig:oo}
\end{center}
\end{figure}

\begin{figure}[h!]
\begin{center}
\includegraphics[width=0.49\textwidth]{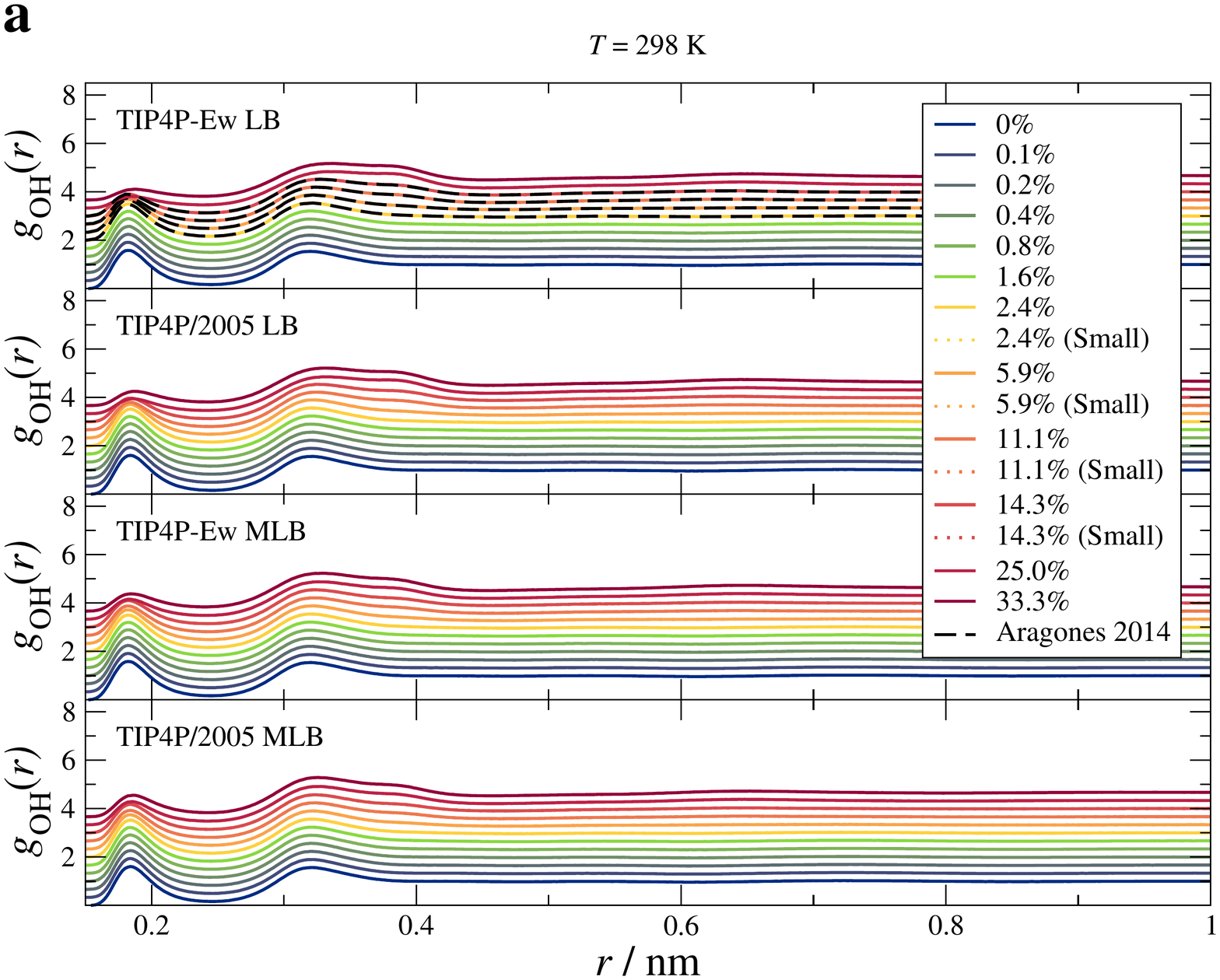}
\includegraphics[width=0.49\textwidth]{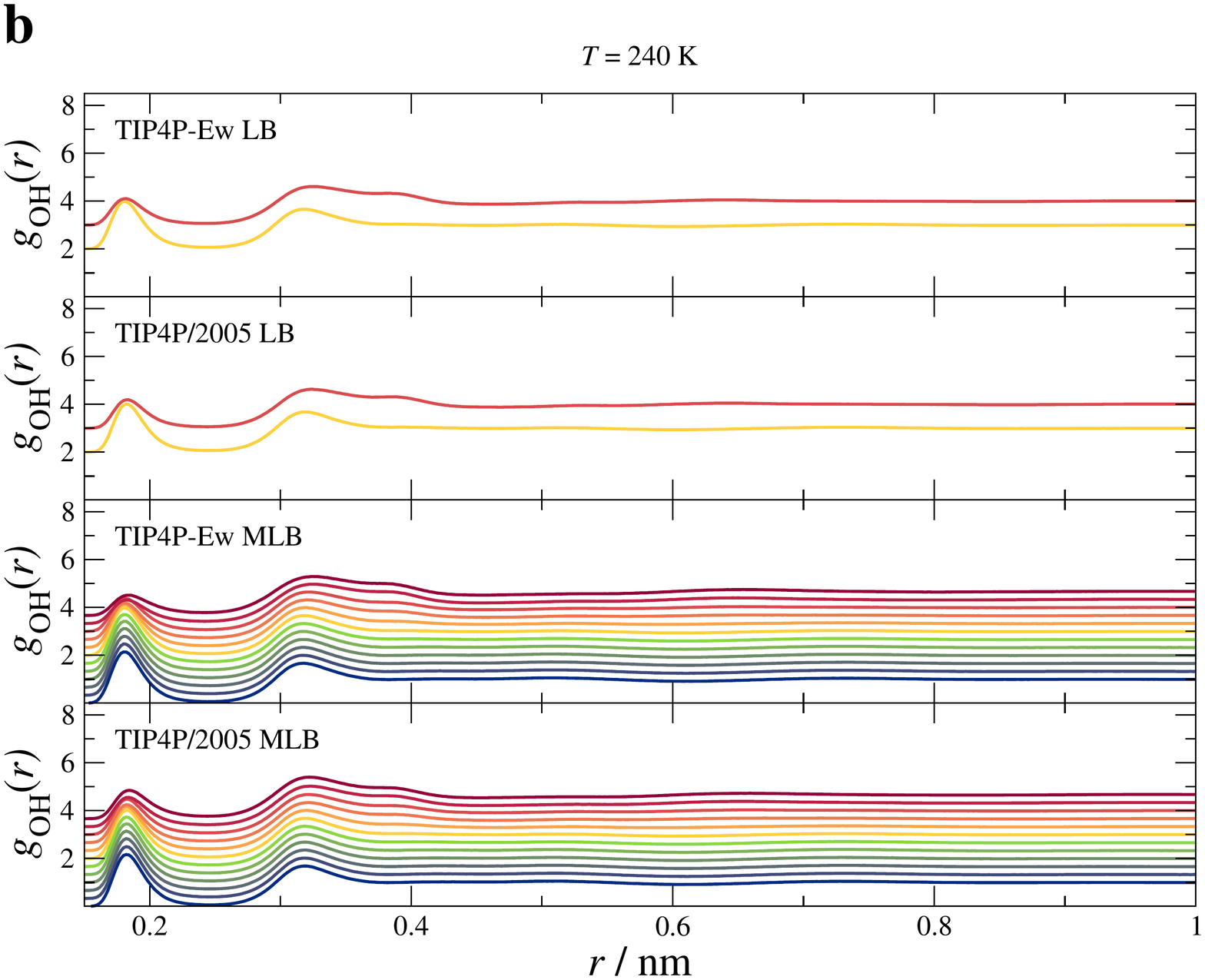}
\caption{O-H RDFs as a function of concentration. Part a shows the data for \SI{298}{\kelvin}, part b for \SI{240}{\kelvin}.
The two upper panels show the behaviour obtained for the LB combination rules for both studied water models. The bottom two panels show the behaviour obtained for the MLB combination rules for both studied water models. The dotted data were obtained for the small systems ($N_{\text{H}_2\text{O}}=480$). The dashed black lines indicate the RDFs as reported by Aragones et al.~\cite{aragones14-jpcb}.}
\label{fig:oh}
\end{center}
\end{figure}

\begin{figure}[h!]
\begin{center}
\includegraphics[width=0.49\textwidth]{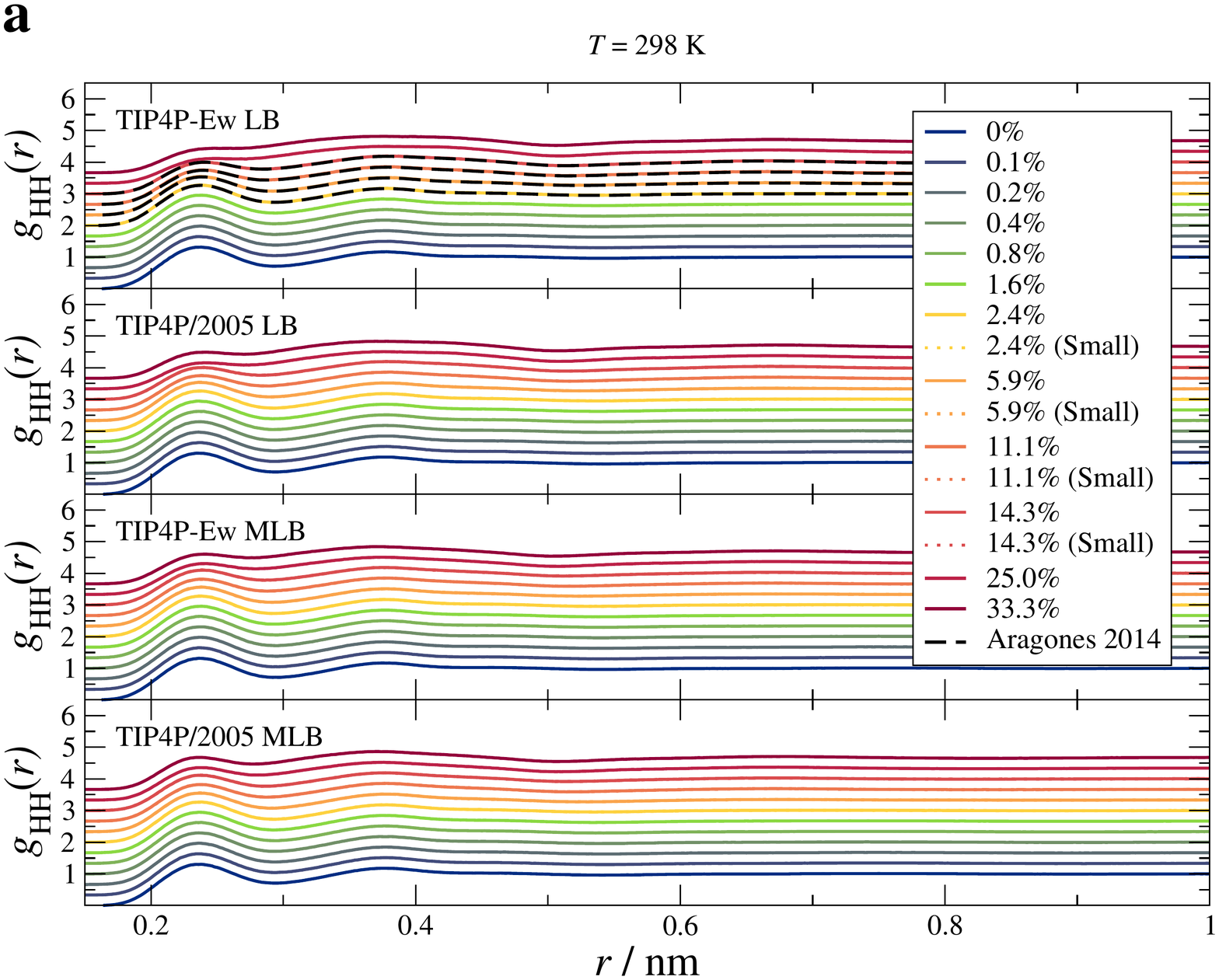}
\includegraphics[width=0.49\textwidth]{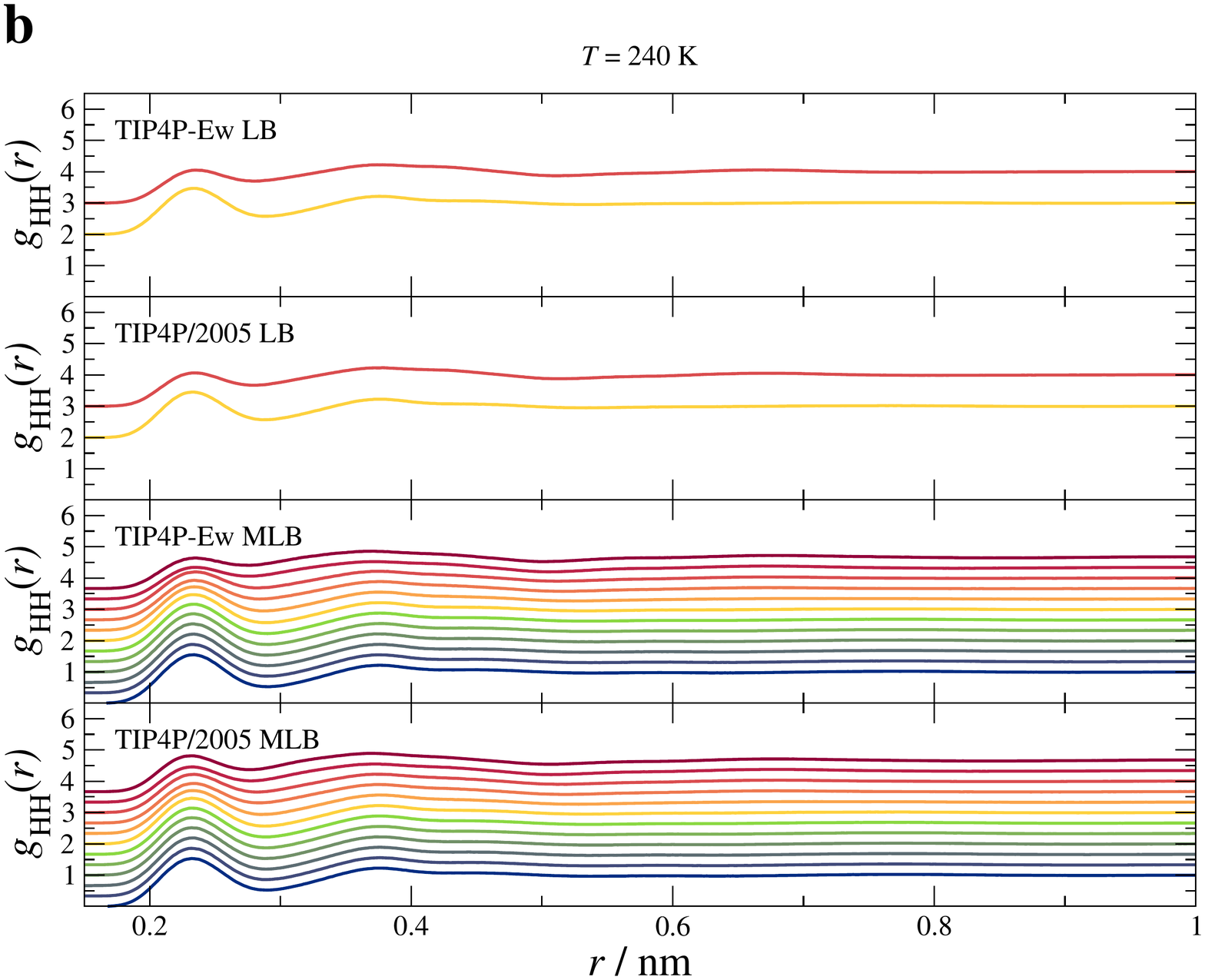}
\caption{H-H RDFs as a function of concentration. Part a shows the data for \SI{298}{\kelvin}, part b for \SI{240}{\kelvin}.
The two upper panels show the behaviour obtained for the LB combination rules for both studied water models. The bottom two panels show the behaviour obtained for the MLB combination rules for both studied water models. The dotted data were obtained for the small systems ($N_{\text{H}_2\text{O}}=480$). The dashed black lines indicate the RDFs as reported by Aragones et al.~\cite{aragones14-jpcb}.}
\label{fig:hh}
\end{center}
\end{figure}

We note that the system size has no significant influence on the RDFs.
The RDFs of the small systems ($N_{\text{H}_2\text{O}}=480$) are shown as dotted lines in Figs.~\ref{fig:licl}--\ref{fig:hh} and they coincide with the data obtained for the larger systems ($N_{\text{H}_2\text{O}}=1000$) shown as full lines.
Furthermore, we compared our RDFs to the data of Aragones et al.~\cite{aragones14-jpcb} where possible (dashed black lines).
Again we find that the two data sets agree very well.
Aragones et al.~\cite{aragones14-jpcb} studied systems with $N_{\text{H}_2\text{O}}=480$.

\newpage

\newpage

\newpage

\newpage

\newpage

\newpage

\section{Coarse-Grained Density}
\label{sec:cg}

\begin{figure}[h!]
\begin{center}
\includegraphics[width=0.49\textwidth]{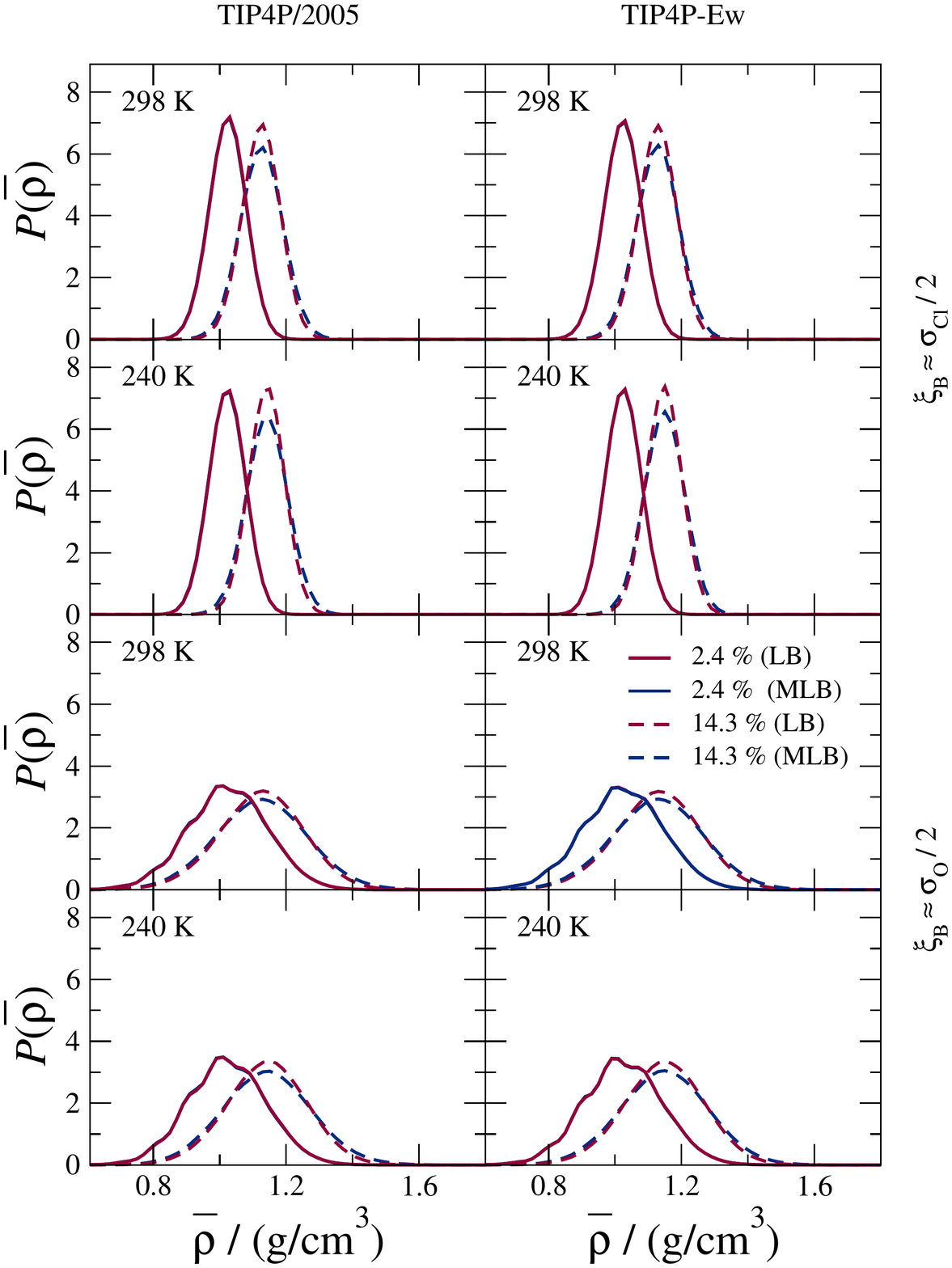}
\caption{Probability distribution of the coarse-grained density for both water models and both temperatures studied. 
The effect of the combination rules is illustrated for selected concentrations (see legend).
The left column shows the data for TIP4P/2005 and the right column for TIP4P-Ew. The top two rows show the results for $\xi_\text{b}\approx\sigma_\text{Cl}/2$ and the bottom two rows show the results for $\xi_\text{b}\approx\sigma_\text{O}/2$. The different colours indicate the different concentrations as indicated in the legend.
All data are obtained from the large systems ($N_{\text{H}_2\text{O}}=1000$).}
\label{fig:cg}
\end{center}
\end{figure}

Fig.~\ref{fig:cg} shows that the choice of combination rule has little influence on the coarse-grained density field.

\newpage

\section{Structural Order Parameter}
\label{sec:rt}

\begin{figure}[h!]
\begin{center}
\includegraphics[width=0.49\textwidth]{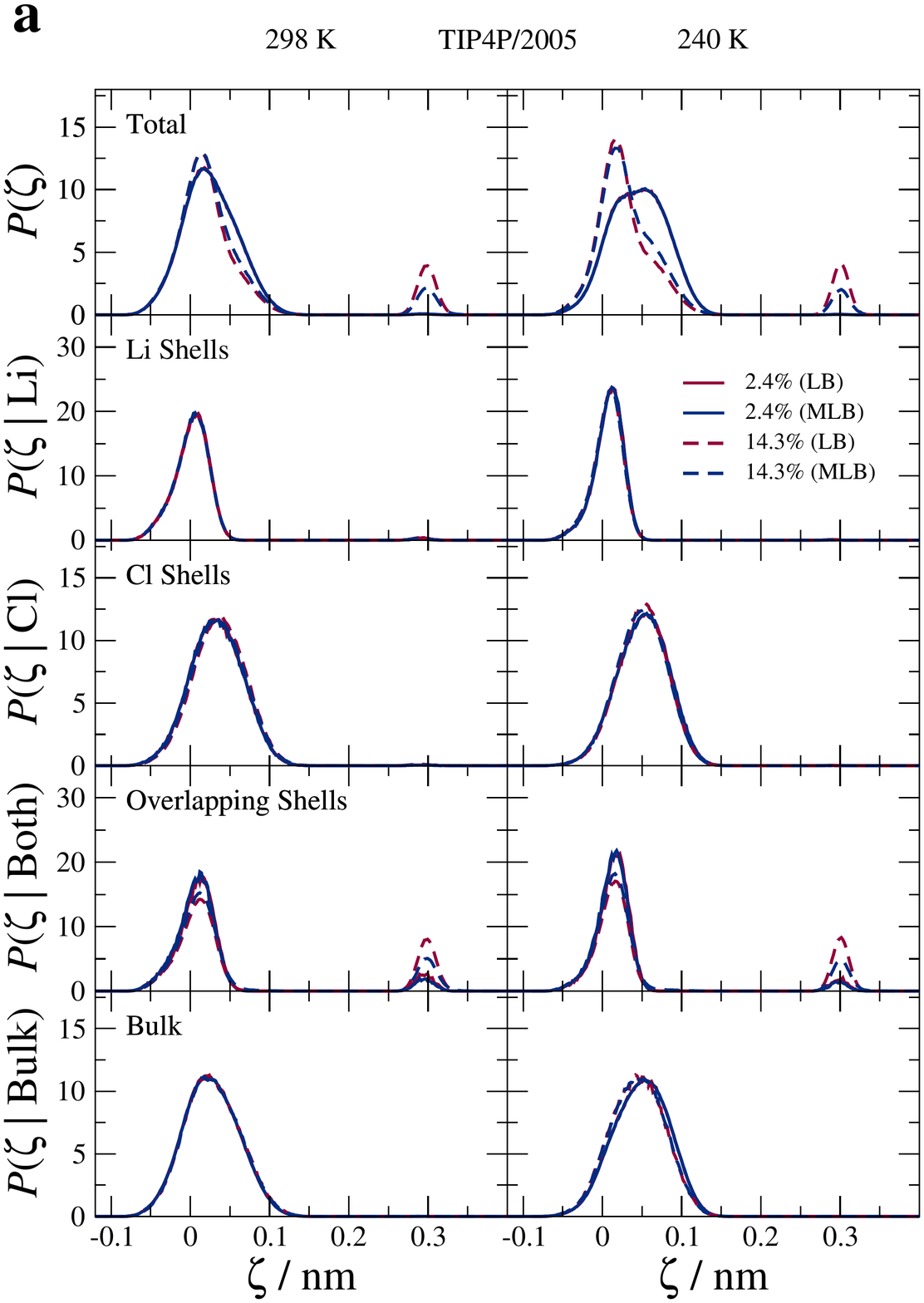}
\includegraphics[width=0.49\textwidth]{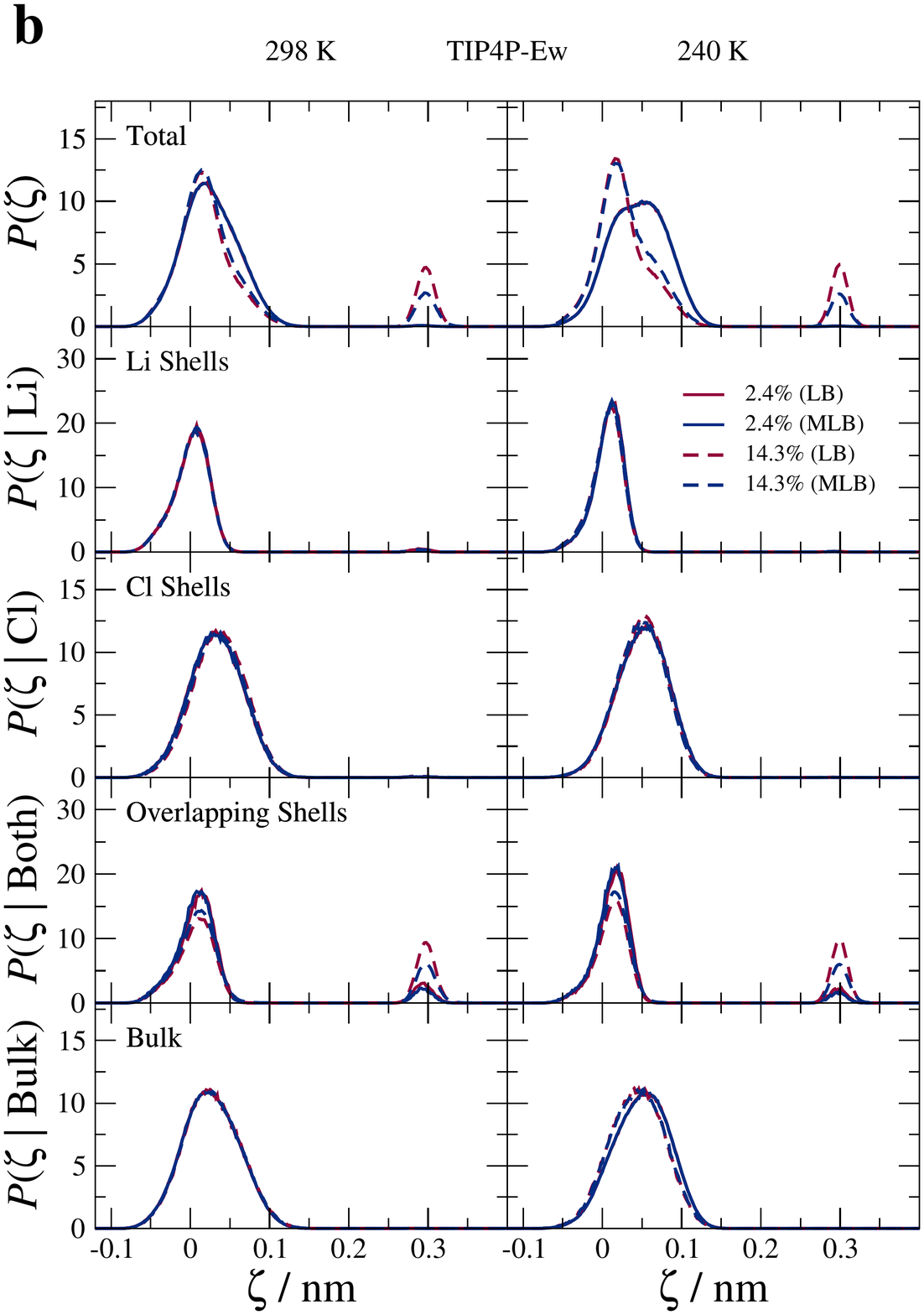}
\caption{Distributions of the structural order parameter for TIP4P/2005 (a) and TIP4P-Ew (b).
The effect of the combination rules is illustrated for selected concentrations (see legend).
 In all figures the left column shows the data for \SI{298}{\kelvin} and the right column for \SI{240}{\kelvin}.
  From top to bottom the rows show the data for all water molecules, molecules in first hydratisation shells of Li only, molecules in first hydratisation shells of Cl only, molecules in first hydratisation shells of both Li and Cl, and the bulk contribution, i.e, water not part of any first hydration shell.
  All data are obtained from the large systems ($N_{\text{H}_2\text{O}}=1000$).}
\label{fig:rt}
\end{center}
\end{figure}

The choice combination rules has little influence on the structural order parameter for both water models and both studied temperatures.
This is documented in Fig.~\ref{fig:rt}.
It is only noticeable that the non-HB peak at $\zeta\approx\SI{0.3}{\nano\meter}$ shrinks if the MLB combination rules are used.
This is in line with the ions' enhanced tendency to  cluster in the MLB variant. 

\newpage

\section{Overlapping Hydration Shells}
\label{sec:ohs}

\begin{figure*}[b!]
 \centering
    \includegraphics[width=\textwidth]{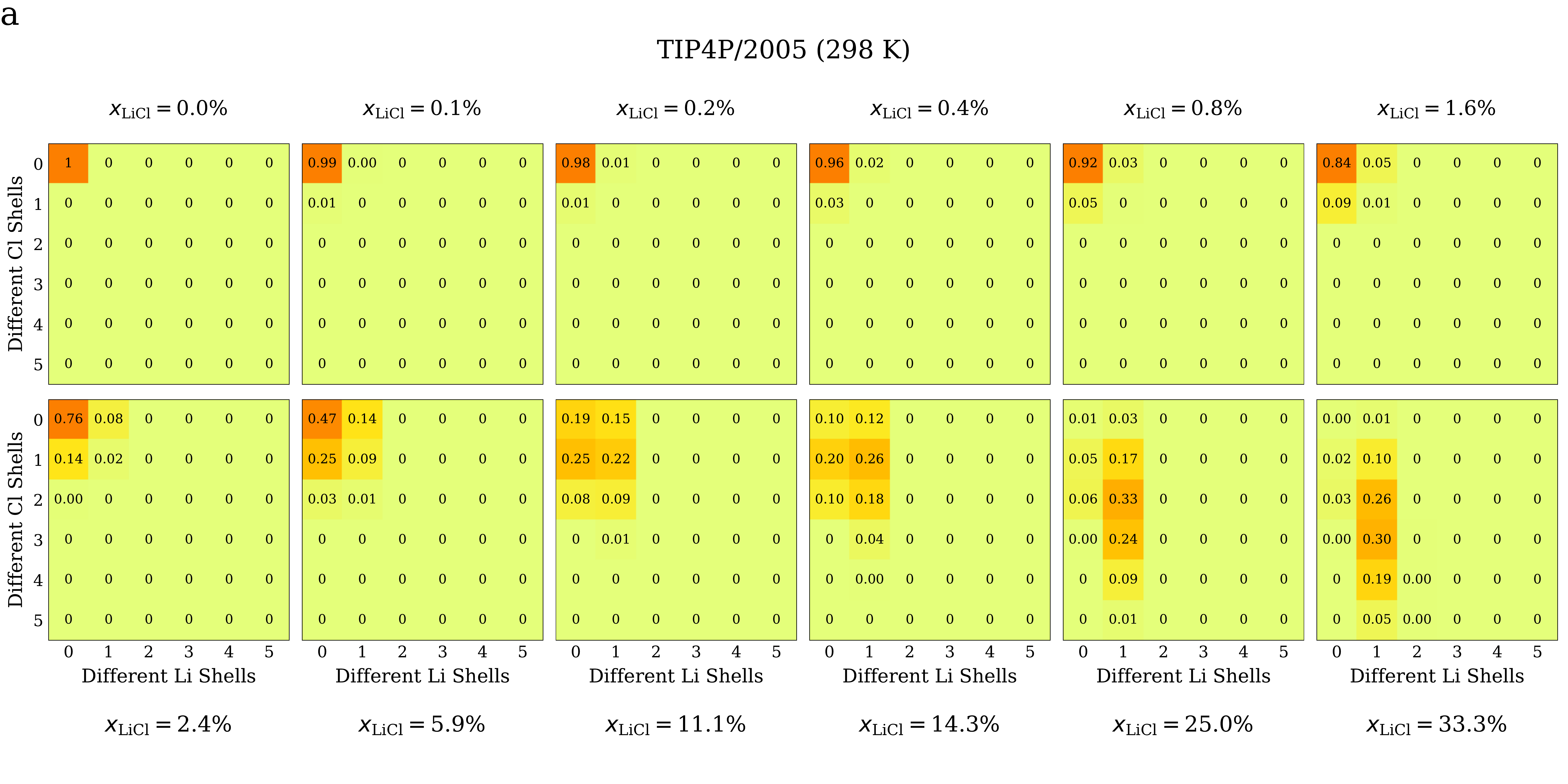}
    \includegraphics[width=\textwidth]{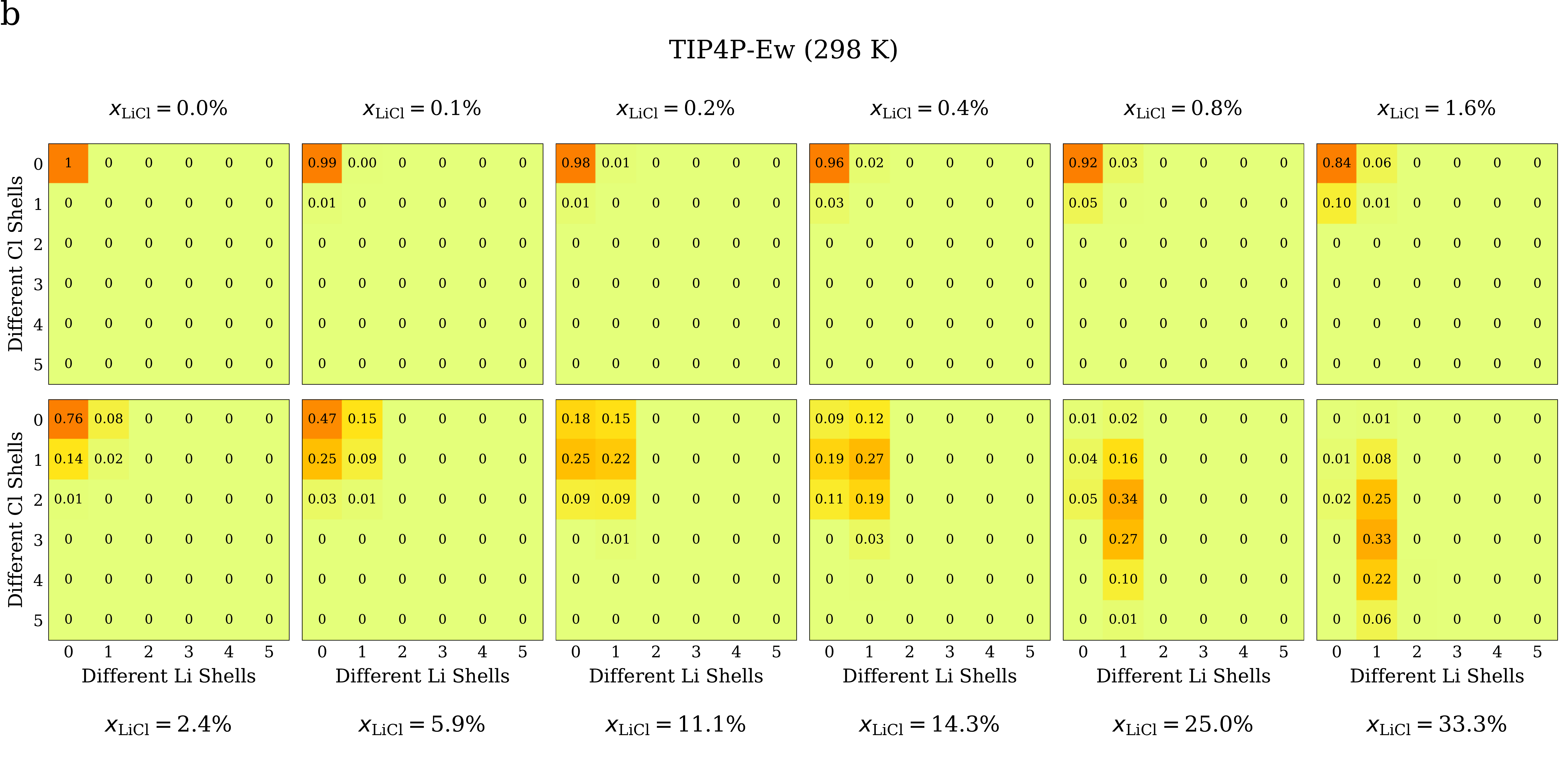}
 \caption{Relative amount of water molecules being part of different first hydration shells at \SI{298}{\kelvin} using the LB combination rules. The number of different first Li shells increases along the rows, the number of different Cl shells along the columns.
 Each matrix shows the results for a given concentration. Part a shows the results for the TIP4P/2005 water model and part b for the TIP4P-Ew water model.
  All data are obtained from the large systems ($N_{\text{H}_2\text{O}}=1000$).}
 \label{fig:shells-298}
\end{figure*}

\begin{figure*}
 \centering
    \includegraphics[width=0.55\textwidth]{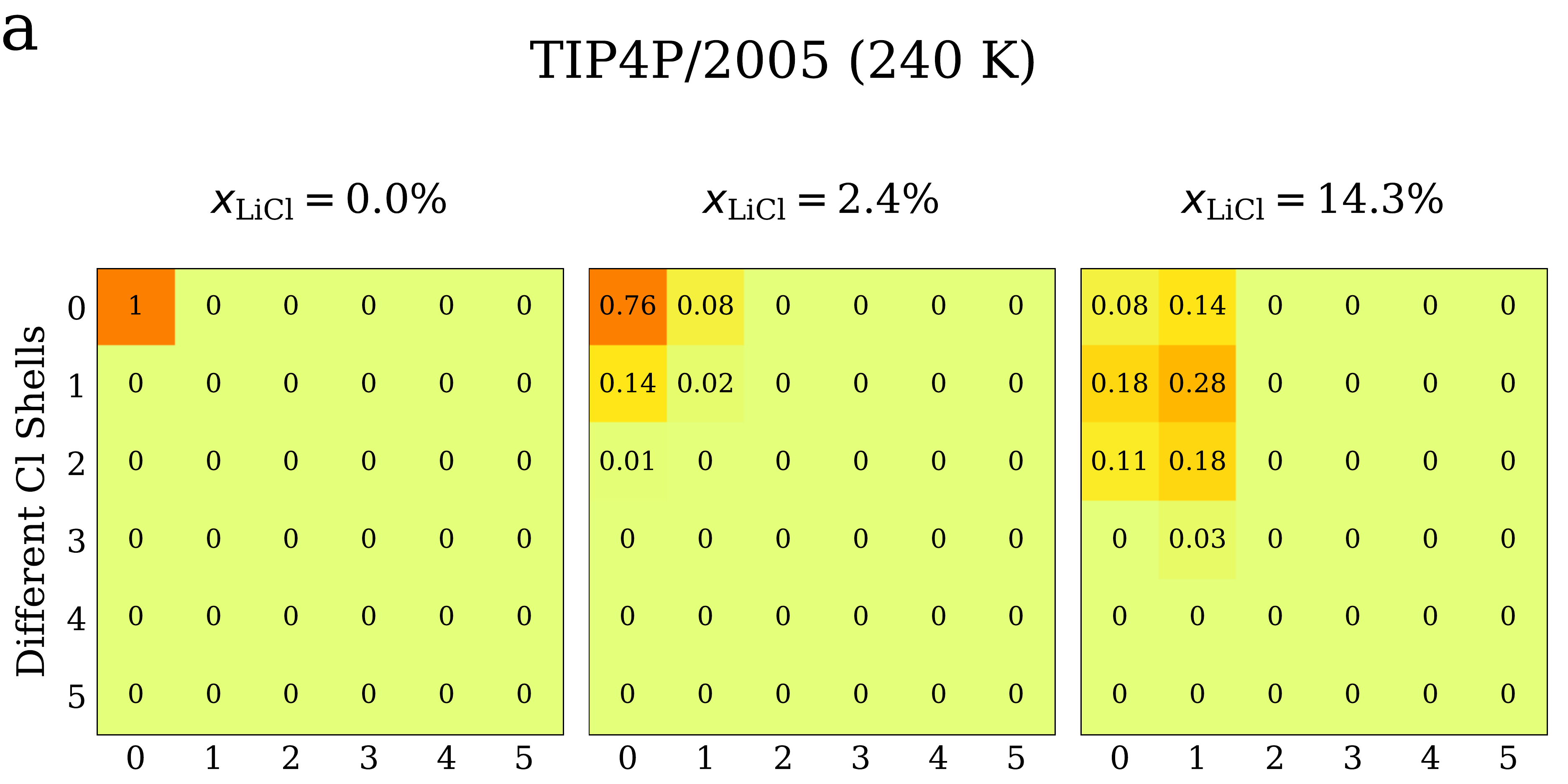}
    \includegraphics[width=0.55\textwidth]{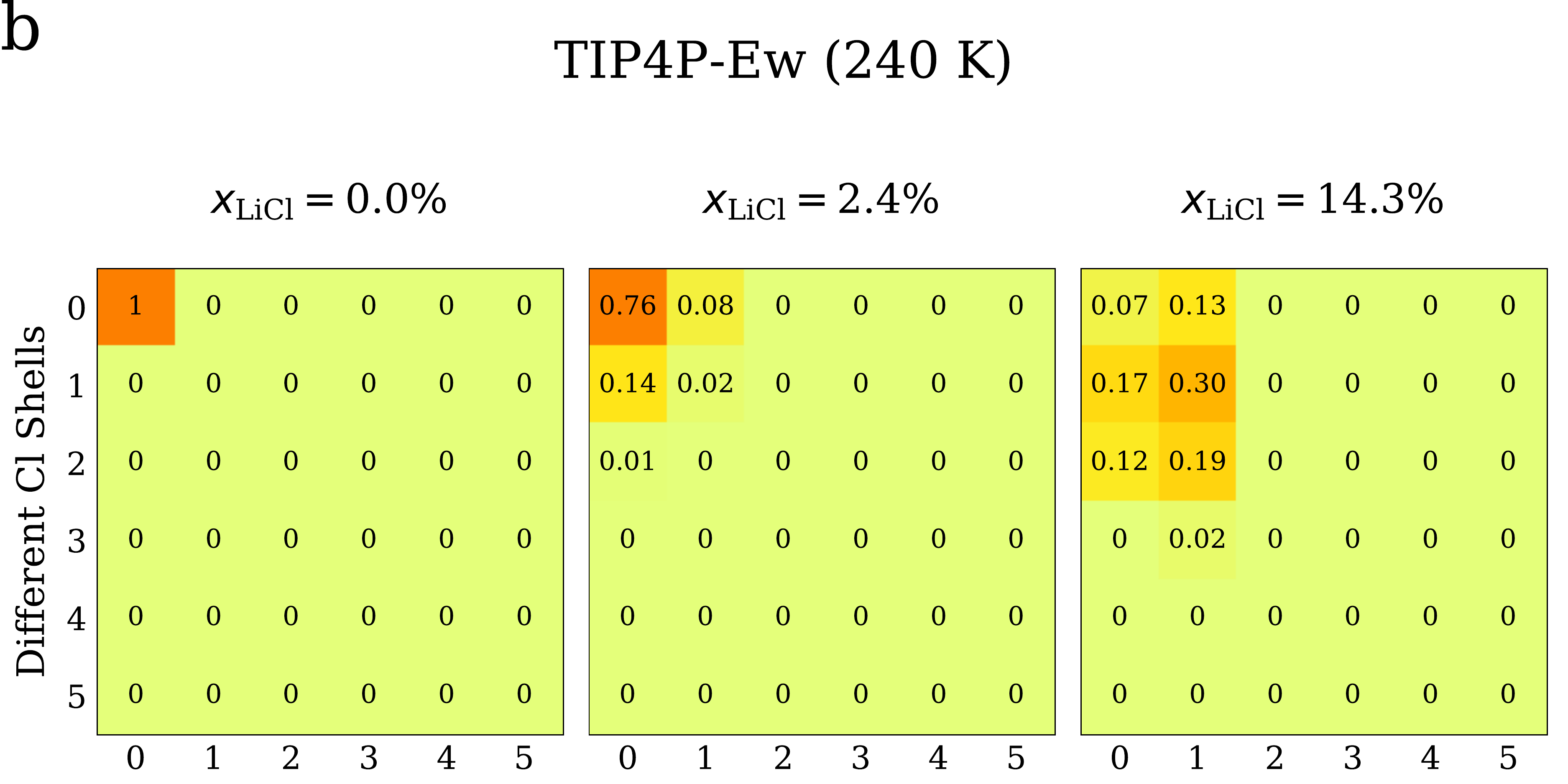}
 \caption{Relative amount of water molecules being part of different first hydration shells at \SI{240}{\kelvin} using the LB combination rules. The number of different first Li shells increases along the rows, the number of different Cl shells along the columns.
 Each matrix shows the results for a given concentration. Part a shows the results for the TIP4P/2005 water.
All data are obtained from the large systems ($N_{\text{H}_2\text{O}}=1000$).}
 \label{fig:shells-240}
\end{figure*}

Figs.~\ref{fig:shells-298} and \ref{fig:shells-240} show the hydration shell statistics when using the LB combination rules.
When compared to Figs.~12 and 13 of the main document it becomes clear that a change from the LB to the MLB rules decreases the fraction of water molecules in overlapping hydration shells at higher concentrations.
Given the increase in the first peak of the Li-Cl RDFs when the MLB rules are used (cf. Fig.~\ref{fig:licl}) this indicates that in the MLB rules direct ion contacts are favoured compared to solvent shared or solvent separated contacts (cf. Ref.~\onlinecite{marcus06-crev}).

\clearpage

\bibliography{../liq-licl.bib}